\documentclass[acmsmall, screen, nonacm]{acmart}

\AtBeginDocument{%
  }

\setcopyright{acmlicensed}
\copyrightyear{2025}
\acmYear{2025}
\acmDOI{XXXXXXX.XXXXXXX}

\acmJournal{CSUR}
\acmVolume{57}
\acmNumber{2}
\acmArticle{111}
\acmMonth{2}

\usepackage{pgfplotstable}
\usepackage{threeparttable}
\usepackage{tabularx}
\usepackage{booktabs}
\usepackage{multirow}
\usepackage{pifont}

\usepackage{lscape}
\usepackage{float}
\usepackage{stfloats}

\usepackage{fontawesome}

\pgfplotsset{compat=1.18}

\begin{document}

\title{A Survey on Privacy-Preserving Computing in the Automotive Domain}

\author{Nergiz Yuca}
\email{nergiz.yuca@uni-passau.de}
\orcid{0000-0002-2964-4673}
\affiliation{%
  \institution{University of Passau}
  \department{Faculty of Computer Science and Mathematics}
  \city{Passau}
  \country{Germany}
}

\author{Nikolay Matyunin}
\email{nikolay.matyunin@honda-ri.de}
\orcid{0000-0001-8974-3078}
\affiliation{%
  \institution{Honda Research Institute Europe GmbH}
  \city{Offenbach am Main}
  \country{Germany}}

\author{Ektor Arzoglou}
\email{ektor.arzoglou@aalto.fi}
\orcid{0000-0001-8664-1885}
\affiliation{%
  \institution{Aalto University}
  \department{School of Electrical Engineering}
  \city{Espoo}
  \country{Finland}
}

\author{Nikolaos Athanasios Anagnostopoulos}
\email{nikolaos.anagnostopoulos@uni-passau.de}
\orcid{0000-0003-0243-8594}
\affiliation{%
  \institution{University of Passau}
  \department{Faculty of Computer Science and Mathematics}
  \city{Passau}
  \country{Germany}
}

\author{Stefan Katzenbeisser}
\email{stefan.katzenbeisser@uni-passau.de}
\orcid{0009-0005-3608-874X}
\affiliation{%
 \institution{University of Passau}
 \department{Faculty of Computer Science and Mathematics}
 \city{Passau}
 \country{Germany}}

\renewcommand{\shortauthors}{Yuca et al.}

\begin{abstract}
 As vehicles become increasingly connected and autonomous, they accumulate and manage various personal data, thereby presenting a key challenge in preserving privacy during data sharing and processing. This survey reviews applications of Secure Multi-Party Computation (MPC) and Homomorphic Encryption (HE) that address these privacy concerns in the automotive domain.
First, we identify the scope of privacy-sensitive use cases for these technologies, by surveying existing works that address privacy issues in different automotive contexts, such as location-based services, mobility infrastructures, traffic management, etc.
Then, we review recent works that employ MPC and HE as solutions for these use cases in detail. 
 Our survey highlights the applicability of these privacy-preserving technologies in the automotive context, while also identifying challenges and gaps in the current research landscape. This work aims to provide a clear and comprehensive overview of this emerging field and to encourage further research in this domain.

\end{abstract}


\maketitle

\definecolor{nmblack}{RGB}{0,0,0} %
\definecolor{nmcolor}{RGB}{0,0,200} %
\definecolor{spacecolor}{RGB}{0,200,0} %

\newcommand{\nmold}[1]{\begingroup\color{nmblack}#1\endgroup}
\newcommand{\nm}[1]{\begingroup\color{nmblack}#1\endgroup}
\newcommand{\spac}[1]{\begingroup\color{spacecolor}#1\endgroup}

\newcommand\varPapersTotal{$230$}
\newcommand\varPapersMPC{$62$}

\section{Introduction}
\label{section:introduction}
In recent years, modern automotive architectures have evolved into highly connected, software-defined systems that aggregate substantial amounts of data. This shift has been driven by advancements in Intelligent Transportation Systems (ITSs) and Software-Defined Vehicles (SDVs), which integrate connectivity and smart data processing capabilities. Autonomous driving technologies further contribute to this data collection, with vehicles equipped with numerous sensors to continuously monitor the environment and vehicular performance. It is estimated that connected autonomous vehicles alone could generate exabytes of data each month, representing a significant increase over current data volumes~\cite{visualcapitalist2023}.
These advancements are further developed into concepts of fully connected vehicular networks, such as Vehicular Ad-hoc Networks (VANETs) and the Internet of Vehicles (IoV), which aim to facilitate extensive data exchange across vehicles, servers, and infrastructure, supporting services for enhanced safety, assistance, and infotainment.

However, as vehicles become more connected, they gather
increasing amounts of sensitive data.
The development of services
that utilize this information presents a challenge in ensuring privacy during data sharing and processing.
One of the primary privacy concerns centers on location data, as the collection and analysis of user locations, travel routes, and behavioral patterns may reveal sensitive information such as user interests, profiles, and habits.
These challenges have led research to the application of Privacy-Enhancing Technologies (PETs) in the automotive domain.

In this survey, we review the application of PETs based on two key technologies of privacy-preserving computation: (1) Secure Multi-Party Computation (MPC), allowing several parties to jointly compute a function over their inputs while keeping these inputs private from each other, and (2) Homomorphic Encryption (HE), a form of encryption that allows computations to be carried out on encrypted data without requiring access to the corresponding plaintexts. \textcolor{black}{Applications of MPC and HE have been studied in various domains, such as healthcare~\cite{aziz2019privacy}, blockchain~\cite{surveyblockchain}, and deep learning~\cite{Zhang2021PrivacyPreservingDL}. However, there has been limited work to holistically survey their applications within the automotive domain. This survey aims to fill this gap by exploring the role of these technologies in enhancing privacy for automotive use cases.}

\textcolor{black}{While other privacy-preserving approaches such as Differential Privacy (DP) have shown promise in automotive domains such as vehicular crowdsourcing and query services in IoV, they typically achieve privacy by injecting calibrated noise into computation outputs. Zhao et al.~\cite{dpsurvey} concluded that applying local DP to IoV presents several challenges. For instance, the high-dimensional nature of vehicle data
makes it difficult to apply LDP, as it leads to high global query sensitivity, introduces excessive noise, and makes it challenging to preserve correlations among attributes after independent perturbation. Furthermore, the privacy-accuracy trade-off introduced by DP might be unsuitable for automotive scenarios that require privacy/accuracy results, such as the computation of ride-sharing matches or some practical ITS applications (e.g. eTolling fees)~\cite{dpits}. In contrast, MPC and HE allow for exact computation over sensitive data, making them particularly relevant for secure and precise processing
of vehicle data.
While we acknowledge that MPC and HE incur performance overhead in terms of computation and communication costs, they remain promising solutions for automotive use cases where accuracy is critical and performance requirements can be optimized through system-level design. For this reason, and given that DP has already been extensively surveyed in \cite{dpsurvey, IntroducingDP}, our work focuses specifically on MPC and HE approaches.
}

To identify the scope of scenarios suitable for applying these technologies, we first survey {\varPapersTotal} recent papers to understand the current privacy challenges and solutions in the automotive domain and to identify specific use cases in vehicular systems where privacy is a major concern.
Second, we aim for completeness of MPC/HE solutions by closely examining {\varPapersMPC} studies that specifically apply these technologies to the identified use cases. Third, we compare these works in terms of their setups, functions, protocols used, security model, and evaluation datasets. Finally, we also document the cases in which MPC or HE is combined with other privacy-preserving technologies (e.g., Differential Privacy or Federated~Learning).

Our survey reveals that privacy-preserving computing technologies are applicable in various settings within the automotive domain. However, we also note several challenges in their application, identify gaps in current research, and suggest areas where further work is needed.
To the best of our knowledge, this is the first comprehensive survey that studies in detail and compares existing works applying MPC and HE to privacy-sensitive use cases specifically in the automotive domain.

We summarize the main contributions of this work as follows: %
\begin{itemize}
    \item First, we review 26 privacy-sensitive use cases across four key areas in the automotive domain: location-based services, mobility infrastructure, vehicular data analysis, and dynamic traffic management.
    
    \item Second, we conduct a comprehensive analysis of 62 state-of-the-art works that apply MPC and HE to the identified use cases,  and categorize these works based on their computational setting (client-to-client, client-to-server, distributed).

    \item  Finally, we draw conclusions about the applicability of MPC and HE in the automotive domain and identify directions for future research.

\end{itemize}

\section{Cryptographic Background}
\label{section:background}

In this section, we provide an overview of the cryptographic primitives, focusing on MPC and HE. We briefly introduce the fundamental principles of these technologies here and discuss in the next sections how they offer strong security guarantees to address vehicle privacy concerns.

\subsection{Secure Multi-Party Computation}

Secure Multi-Party Computation (MPC) is a cryptographic method that enables multiple parties to compute a function without revealing their inputs to the other parties~\cite{goldreich1998secure}. MPC has been a widely studied topic in cryptography for more than two decades, beginning with Yao's seminal work, where he introduced the millionaire problem~\cite{yao1982}. The problem involves two millionaires, Alice and Bob, who want to determine who is wealthier without disclosing their actual wealth. This scenario is an instance of a broader problem involving two numbers, $a$ and $b$, where the objective is to establish whether the inequality $a \geq b$ is true or false without revealing the values of $a$ and $b$. 

A Garbled Circuit (GC) is a generic approach for secure two-party computation ~\cite{Yao1986HowTG}.
In this protocol, party $A$ (the garbler) generates a garbled version of a Boolean circuit that represents the function to be computed and sends it to party $B$ (the evaluator). $B$ uses its inputs to evaluate the GC. To securely obtain encrypted inputs from $A$, $B$ employs the Oblivious Transfer (OT)~\cite{Lindell2011SecureTC} protocol, where $A$ holds two messages $m_0$ and $m_1$, and $B$ holds a choice bit $b \in \{0, 1\}$. The protocol ensures that $B$ learns only $m_b$ (the message corresponding to its choice), while $A$ learns nothing about $B$'s choice $b$. The Beaver-Micali-Rogaway (BMR)~\cite{bmr} scheme adapts the main idea of Yao's GC to a multi-party setting, allowing each party to independently evaluate the GC.

Another fundamental primitive for MPC is Secret Sharing (SS)~\cite{Shamir1979HowTS,Blakley1899SafeguardingCK}.
A $t$-out-of-$n$ secret sharing allows a secret to be divided into $n$ shares, where any $t$ or more shares can reconstruct the secret, but any fewer than $t$ shares reveal no information about it. There are two main secret sharing schemes: (1) Additive Secret Sharing (ASS), applicable only in the specific case of $t=n$, and (2) Shamir's Secret Sharing (SSS), which is applicable to any positive $t\leq n$. Both schemes satisfy linearity, such that the sum of two secret shares is equivalent to the share of the sum.

One key sub-area of MPC emphasizes specific functionalities such as the Private Set Intersection (PSI) method~\cite{Freedman2004EfficientPM}. It enables two parties to securely compute the intersection of their respective private datasets, such that no information beyond the common elements is disclosed to either party. Threshold PSI (TPSI)~\cite{Pinkas2014FasterPS} is a variant of PSI that allows the parties to compute the intersection of their sets only if
it exceeds a defined threshold.
A Private Equality Test (PEQT) enables two parties to compare their private values to determine if they are equal without revealing any information to each other if the values are not equal~\cite{Pinkas2014FasterPS}. Finally,  a Private Information Retrieval (PIR) allows a client to obtain a specific item from a server without disclosing which item was retrieved~\cite{Chor1995PrivateIR}.

\subsection{Homomorphic Encryption}

Homomorphic Encryption (HE) is a cryptographic technique that
enables arithmetic operations to be performed directly on encrypted data without requiring decryption~\cite{Rivest1978AMF}. It can be categorized into four main types based on the number of operations allowed on the encrypted data: (1) Additive Homomorphic Encryption (AHE), \textcolor{black}{(2) Multiplicative Homomorphic Encryption (MHE)}, (3) Somewhat Homomorphic Encryption (SHE), and (4) Fully Homomorphic Encryption (FHE). AHE schemes, such as the Paillier cryptosystem~\cite{Paillier1999PublicKeyCB}, are suitable only for specific applications where algorithms involve predominantly addition operations. AHE schemes, such as the Paillier cryptosystem [67], allow computations where the encrypted result of two ciphertexts corresponds to the sum of their original plaintexts. \textcolor{black}{In contrast, MHE schemes, such as the RSA cryptosystem~\cite{Lin2005AnES}, support computations where the encrypted result corresponds to the multiplication of plaintexts, although RSA without padding is not semantically secure.} SHE schemes, such as the BGN cryptosystem~\cite{bgn}, support both addition and a limited number of multiplications. 
Finally, FHE schemes allow an unlimited number of both addition and multiplications as well as the evaluation of arbitrary functions (such as searching, sorting, computation of max or min, etc.) over ciphertexts~\cite{acar2018survey}.

Both MPC and HE enable privacy-preserving computations over private data but differ in their approaches. MPC relies on distributed computation among multiple parties, which may require communication during the computation. Specifically, SS-based MPC, while promising, requires multiple parties to be online for the computation. In outsourcing MPC, clients can reduce communication overhead by distributing their data as secret shares to servers who perform computation on their behalf and are readily available. As a natural primitive for outsourcing scenarios, HE allows a data owner to outsource computation to an untrusted server,
minimizing communication during the computation phase. However, HE schemes have several drawbacks. First, the computational cost is high, as many privacy-preserving computation protocols require encryption of individual bits. Second, FHE relies on a key technique called bootstrapping to periodically reduce ciphertext noise, which can drastically reduce the system’s efficiency. Third, there is considerable storage overhead, as ciphertexts can be several times larger than plaintexts.
Finally, a Trusted Authority may be needed to generate and distribute public and private keys for all parties~\cite{Yang2019ACS}.

\section{Related Work}
\label{section:related-work}

\begin{table*}[t]
\scriptsize
\centering
\caption{Overview of existing related surveys}
\label{table:survey}
\scriptsize
\begin{tabularx}{\textwidth}{p{0.8cm} p{0.5cm} p{4.4cm} p{7.0cm}}
\hline
\textbf{Survey} & \textbf{Year}  & \textbf{Topic of the survey} & \textbf{
Comparison to our survey's contributions} \\ \hline

\addlinespace
\multicolumn{4}{l}{\textit{I. Surveys on MPC and HE in non-automotive applications}} \\
\addlinespace  \hline

\cite{Yang2019ACS} & 2019   & Security threats and requirements for secure outsourced computation and its applications  & Our work focuses on the automotive domain applications of MPC and HE, and not on generic secure outsourcing schemes only. \\ \hline

\cite{Zhang2021PrivacyPreservingDL,Zhou2024SecureMC} & 2021, 2024 & Secure Multi-Party Computation-based machine learning & Our scope includes the overall automotive domain, and we do not consider only machine-learning applications.
\\ \hline

\cite{acar2018survey} & 2018 & Homomorphic Encryption Schemes  & We survey MPC and HE applications in the automotive domain instead of focusing on generic HE implementations.\\
\hline
\cite{Li2023ASO} & 2023  & Trusted execution environments-based secure computation protocols & Our work covers secure computing in automotive use cases using MPC and HE, and does not focus on TEE-based protocols only. \\ \hline

\addlinespace
\multicolumn{4}{l}{\textit{II. Surveys on privacy-preserving technologies in the automotive domain}} \\
\addlinespace  \hline

\cite{sun2020} & 2020   & Homomorphic Encryption applications in VANETs & We extend their focus with MPC applications and give a holistic analysis of specific automotive use cases. \\ \hline

\cite{Mundhe2021ACS} & 2021   & Security and privacy requirements, architectures, and cryptographic schemes in VANETs & MPC and HE applications in automotive use cases, which form the focus of our work, have not been covered in their survey. \\ \hline

\cite{Jiang2021LocationPM} & 2021 & Privacy-preserving location-based services  & Unlike their focus on privacy-preserving LBSes, our work examines MPC/HE solutions across a broader scope of automotive applications.
\\ \hline

\cite{Hataba2022SecurityAP} & 2022   & Security and privacy issues in autonomous vehicles & Our automotive use case examination and analysis of MPC and HE are more comprehensive than their work. \\ \hline

\cite{Zafar2022CarpoolingIC} & 2022   & Solutions and future directions for carpooling in CAVs & Our survey considers various automotive use cases, and does not focus only on carpooling. \\ \hline

\cite{Yoshizawa2022ASO} & 2022   & Security and privacy concerns in vehicular communication systems  & Our work focuses on privacy-preserving solutions for automotive applications using MPC and HE, whereas their work centers on analyzing security and privacy gaps in V2X communication standards.\\ \hline

\cite{Sutradhar2024ASO} & 2024  & Cryptographic authentication techniques for secure vehicular communication & Unlike their specific focus on cryptographic authentication in vehicular communication, our work specifically addresses privacy-preserving computing in automotive use cases, emphasizing MPC and HE solutions. \\ \hline

\end{tabularx}
\end{table*}

In this section, we review surveys relevant to privacy-preserving computing in the automotive domain, with a particular focus on those addressing MPC and HE. We distinguish between works that cover these technologies in generic application domains and works that specifically focus on automotive applications. In Table~\ref{table:survey}, we further highlight distinctions between our work and existing surveys and discuss the enhancements in our survey.
 
Several works survey PETs in a broader, non-domain-specific context. Acar et al.~\cite{acar2018survey} survey HE schemes and recent developments. Yang et al.~\cite{Yang2019ACS} provide a technical review and comparison of secure outsourcing schemes, focusing on various secure computation methods, including MPC and HE, and conclude with an analysis of security, performance, and future directions in the field. Zhang et al.~\cite{Zhang2021PrivacyPreservingDL} and Zhou et al.~\cite{Zhou2024SecureMC} review the state of the art in privacy-preserving machine learning, focusing on secure multi-party computation and categorizing techniques used during the training and inference phases. Li et al.~\cite{Li2023ASO} give a comparison of secure computation protocols based on Trusted Execution Environments (TEEs), offering a taxonomy and assessment criteria to evaluate various protocols and highlighting their applicability to both general-purpose and specialized tasks, such as privacy-preserving machine learning and encrypted database queries.

Additionally, domain-specific surveys explore privacy risks, attacks, and other solutions in ITSs, vehicular networks, such as VANETs, and autonomous vehicles. To this end, Sun et al.~\cite{sun2020} survey HE in VANETs, detailing the relevant framework, security issues, and data handling. 
Zafar et al.~\cite{Zafar2022CarpoolingIC} present a comprehensive survey on carpooling in autonomous and connected vehicles, discussing the relevant architecture, components, and solutions, and addressing existing challenges in carpooling. Yoshizawa et al.~\cite{Yoshizawa2022ASO} identify and analyze security and privacy concerns in vehicle-to-everything communication standards and provide recommendations to address these issues for the improvement of security and privacy in vehicular networks. Sudrathar et al.~\cite{Sutradhar2024ASO}  classify and analyze various cryptographic authentication techniques for secure vehicular communication, discussing their properties, advantages, and limitations.

On the contrary, our survey focuses on the application of MPC and HE in the overall automotive domain, clearly distinguishing itself from existing surveys that either concern privacy challenges and cryptographic techniques in general or consider only a very limited area of the automotive domain. In this way, we address substantial gaps left in the current literature, particularly regarding the practical application of MPC and HE in automotive privacy scenarios, which has not so far been the central focus of another similar study. Our article highlights the role of MPC and HE in solving privacy issues in the context of connected and autonomous vehicles, providing a significant contribution to the ongoing discussion on ways of improving data privacy and security in this field.

\section{Methodology}
\label{section:methodology}

We conduct a comprehensive literature review in two steps. First, we begin by identifying existing privacy-preserving use cases in the automotive domain to understand the scope of relevant applications.
We focus on selected scenarios that are identified as privacy-sensitive.
Second, we specifically examine how MPC and HE are applied to privacy-sensitive use cases.

To conduct a comprehensive investigation, we employ keyword searches in the IEEE Xplore and Google Scholar databases, 
chosen for their broad scientific coverage. In the first step, to retrieve a relevant and exhaustive set of works, we develop two sets of keywords. The first set includes domain-specific keywords,
while the second set includes privacy-related keywords that indicate relevance to privacy concerns or solutions. A complete list of the keywords used is provided in Table~\ref{table:keywords}. We generate all possible combinations of domain- and privacy-related keywords and construct search queries by combining them with logical operators.
Specifically, we use the \texttt{OR} operator within each set of keywords to include synonyms and related terms, and the \texttt{AND} operator between the domain and privacy keyword sets to ensure that the results are relevant to both areas. The queries are constructed in the form of: \texttt{(``domain keyword'' OR ``domain synonym'') AND (``privacy keyword'' OR ``privacy synonym'')}. For example, a search query would be \texttt{(``Vehicle'' OR ``Car'') AND (``Privacy-preserving'' OR ``Data privacy'')}. Due to the potential overlap with papers on Model Predictive Control, we avoided the acronym ``MPC'' and used the explicit term ``Secure Multiparty Computation'' or ``Secure Multi-Party Computation'' in our search.

Our aim is to collect research works that address selected scenarios in the automotive domain as privacy-sensitive.
In this approach, we thus do not expect the complete coverage of all possible automotive use cases, and to not aim to discover \textit{new} privacy-sensitive use cases, but rather scope existing use cases and determine if the surveyed technologies address them or leave gaps. 

\begin{table}[t!]
\centering
\caption{Keywords Used in Literature Search}
\label{table:keywords}
\scriptsize
\begin{tabular}{|p{6.2cm}|p{7.1cm}|}
\hline
\textbf{Domain Keywords} & \textbf{Privacy Keywords} \\
\hline
vehicle, car, VANETs, location-based services, autonomous vehicles, connected cars, intelligent transportation, vehicular networks, mobility, traffic management, vehicular data analysis
&
privacy-preserving, privacy-sensitive, secure multi-party computation, homomorphic encryption, data privacy, privacy protection, secure computation, cryptographic protocols, privacy-enhancing technologies, secret sharing, garbled circuits, oblivious transfer, private set intersection \\
\hline
\end{tabular}
\end{table}

As a result of the first step, we identified a total of $243$ papers. Subsequently, we refined this pool by selecting only publications written in English, published within the last five years, and with a citation count of at least ten.
In addition, we filtered out irrelevant work, e.g., papers focused on security rather than on privacy. We also excluded papers that address privacy in communication protocols (e.g., specific authentication schemes in VANETs). Finally, we excluded $48$ papers
where the applied
methods were unclear or lacked substantial details.
Recognizing the potential existence of recent relevant works that may not fully meet the aforementioned criteria but still be important (e.g., very recent works with only a few citations), we conduct an additional screening of the works in the original pool of
papers.
This results in the inclusion of $16$ more
papers with direct relevance to the scope of our survey, despite not initially meeting the set criteria.
As a result, our final analysis dataset consists of {\varPapersTotal} papers, 
selected through criteria-based filtering and our targeted search.

In the second step, for the detailed analysis in Section~\ref{section:mpc}, we filter the collected papers to select works that apply MPC or HE as a privacy-preserving solution. To ensure comprehensive coverage, we perform an additional verification step by conducting targeted searches combining each reviewed use case (Section~\ref{section:use-cases})  with MPC and HE technologies. These searches use queries in the form of: \texttt{(``Use Case Name'') AND (``Technology Name'')}. The final dataset consists of {\varPapersMPC} papers.

\section{Use Cases}
\label{section:use-cases}
In this section, we present a set of use cases in the automotive domain based on the first step of our literature review. Each use case illustrates specific scenarios that involve the processing, analysis, or sharing of privacy-sensitive data among multiple parties.
 We cluster use cases
 into four categories: location-based services, mobility infrastructures, vehicular data analysis, and dynamic traffic management and V2X communication. 
 While we identify specific groups for different use cases, it is important to note that there are natural overlaps among them. For example, traffic signal control use case (Section ~\ref{section:traffic signal control}) can be classified to vehicular data analysis and traffic management groups. However, we categorize it under traffic management because its primary focus is on optimizing road traffic flow.
 Later in Section~\ref{section:mpc}, we analyze in detail the concrete applications of MPC and HE in these use cases.

\subsection{Privacy in Location-Based Services}
\label{subsection:location privacy}
In this section, we discuss privacy concerns associated with Location-Based Services (LBSs). LBSs include various services
that utilize the geographic location of a user as input to specific functions, such as recommending nearby points of interest, or connecting users to the taxi drivers.
Service providers, typically centralized, 
require users to 
disclose their location to access the desired services.
This centralization leaves users unable to verify whether their data is being used as intended,
and poses risks of
tracking and profiling. Therefore, finding a way to offer users personalized LBSs without compromising their location is an essential concern.
Below, we summarize concrete use cases representing LBSs with works addressing privacy issues in these use cases.

\subsubsection{Points of Interest.}
\label{section:POI}
Points of Interest (POI) are specific locations that are beneficial for drivers, such as restaurants, gas stations, and landmarks. Vehicles often request these locations to improve navigation and provide access to nearby amenities. In this scenario, vehicles query for POIs by sending requests to a server. The server processes these queries and returns relevant information without determining the exact locations of the vehicles~\cite{Tan2018PrivateIR,Zhou2020ATQ}.

\subsubsection{\textcolor{black}{Navigation and Route Planning}}
\label{section:navigation}
\textcolor{black}{
 One of the prominent services in vehicular networks is providing drivers with optimal routes and real-time navigation assistance based on traffic conditions and data collected from other vehicles. This service requires users to share
 their current location, travel routes, and destinations, which might be  exploited for user movement profiling~\cite{Tiausas2023HPRoPHP, Zhou2021EPNSEP}.}

\subsubsection{Localization.}
\label{section:localization}
Vehicle localization is the process of precisely determining a vehicle's location, speed, and direction using GPS, sensors, maps, and communication systems. Constant data collection and transmission may expose vehicle location history, travel patterns, and sensitive user information, highlighting the need for privacy-preserving measures~\cite{woo2018localization,siam2018}.

\subsubsection{Vehicular Crowdsourcing.}
\label{section:spatial crowdsourcing}

This use case involves outsourcing tasks related to specific locations to a group of mobile workers. Task requesters register through a centralized server and publish tasks with target locations or spatial routes. Available workers are considered for task assignment and are responsible for reporting their locations to the server~\cite{chenxi2022,junwei2021,Xu2023PriTAECPT}. 

\subsubsection{Ride-Sharing.}
\label{section:ride-sharing}
Ride-sharing services match drivers with passengers in order to share a journey. To facilitate this coordination, ride-sharing systems collect vast amounts of sensitive data, including pick-up and drop-off locations, the identities of  riders and drivers, and specific timings. Therefore, the server or any entity accessing this data can infer sensitive information about riders' activities by monitoring their locations~\cite{Avodji2016MeetingPI,ORide2017,pRidePR2019,Hallgren2017,Luo2023P2RidePA,marie2018,toppool2019,yuan2018,yuha2021,yu2021}.

\subsubsection{Vehicle Sharing.}
\label{section:vehicle sharing}

Vehicle sharing is a smart mobility service that provides users with access to vehicles for short-term use, often on an as-needed basis. It utilizes in-vehicle telematics and portable devices, such as smartphones, and allows vehicle owners to distribute temporary digital keys or access tokens to other users, enabling them to access the vehicle~\cite{iraklis2017,iraklis2022}.

\subsection{Privacy in Mobility Infrastructure}
\label{subsection:mobility privacy}
In this section, we discuss the privacy use cases associated with mobility infrastructure. This category focuses on services where vehicles interact with physical infrastructure, such as toll stations or charging stations. The primary privacy risks stem from the sharing of vehicle and user data
with centralized systems or third-party infrastructure providers.

\subsubsection{Toll Data Collection.}
\label{section:toll data collection}

Electronic toll collection systems use sensors and toll transponders to track vehicles. Information stored in toll records can be utilized to monitor a vehicle's movements,  making vulnerable to unauthorized tracking and user profile breaches~\cite{karim22}. 

\subsubsection{Electric Vehicle (EV) Charging.}
\label{section:ev charging}

The EV charging process involves various interactions between users and charging infrastructure. In the payment process, EV users engage in transactions with Charging Service Providers (CSP).
The chosen Charge Station (CP) generates a service order including
payment amount, charging duration, and location, which the user typically authorizes through a 
a mobile application. After the payment, the charging session begins. CPs can aggregate data from these transactions, including geographic location, charging patterns, and battery usage~\cite{Fuchs2020TrustEVTE}. Over time, this accumulated data may enable the inference of users' driving behaviors
and 
frequently visited locations~\cite{Wu2023PrivacyPreservingAT}. 
 EV charging can also be optimized to fill the overnight demand valley, reducing grid operation costs.
 However, in this process, participants need to communicate frequently, thereby
 revealing an EV owner's charging profiles~\cite{Huo2021DistributedPE}.

\subsubsection{Parking.}
\label{section:parking}
An autonomous vehicle parking system integrates service providers, parking infrastructure, users, vehicles, and authorities. It relies on secure registration and communication protocols, encrypted data handling, and user authentication mechanisms. Protecting user location and identity data is important in parking systems, as registration with central authorities for reservations can expose sensitive information and compromise privacy~\cite{locprivv,Pokhrel2021,Li2021PriParkRecPD}.

\subsection{Privacy in Vehicular Data Analysis}
\label{subsection:data analysis privacy}
In this section, we discuss use cases that target privacy aspects related to
analysis of
vehicular data.
As the SDVs continue
to advance,
organizations and institutions are increasingly interested in gathering vehicle data of various kinds for analysis purposes. Substantial amounts of
personal
data collected and processed by service providers can pose serious privacy risks.
\nm{\textcolor{black}{Research works address these concerns by proposing privacy-preserving solutions for data aggregation (\ref{section:data analysis for iov}), or architectures for privacy-preserving federated learning (\ref{section:decentralized fl}). 
Vehicular data analysis often includes learning models of vehicle or driver behavior (\ref{driver behavior}--\ref{section:predictive maintenance}). These use cases rely on sensitive data streams (e.g., driving patterns, locations, images) for training accurate models. A primary privacy concern is protecting this data from misuse and profiling. Another class of applications (\ref{section:fl-navigation}--\ref{vehicle emission}) involves learning from the vehicles' \textit{external} environment.
These applications often require collaborative sharing of sensitive data, which introduces distinct privacy challenges.
In the following, we discuss concrete use cases for vehicular data collection and analysis.
}}

\subsubsection{Data Processing in IoV}
\label{section:data analysis for iov}
Several works address the generic use case of aggregating 
 data in IoV environments. Each vehicle node provides data, partially aggregated by \textcolor{black}{Roadside Units (RSUs)} and then fully aggregated by a central server.
 The privacy of vehicle data needs to be protected from being misused by RSUs and the server~\cite{Zhou2024PrivacypreservingAV}.
Decentralized VANETs are designed to reduce 
centralization
and minimize network communication overhead by involving vehicles, RSUs, and
edge nodes to aggregate or even process data. For example, vehicles and RSUs can share real-time traffic condition information, while
edge nodes
train machine learning models and distribute 
results. In these multi-party setups, privacy has to be taken into account in all existing data flows~\cite{chenji2022}.

\subsubsection{Federated Learning (FL)}
\label{section:decentralized fl}
\label{section:decentralized fl 2}
\label{section:vehicular fog computing}

Several works address a generic scenario of training machine learning models on vehicular data using Federated Learning (FL)~\cite{bdfl2021,hu2023}. Although FL allows clients to avoid sharing raw data, a malicious server can still reveal sensitive information from the model updates.
\nm{\textcolor{black}{ Addressing the trust concerns associated with a central aggregator, Decentralized Federated Learning (DFL) has been increasingly applied in the vehicular domain. It enables   participants to
share model updates directly among themselves or with intermediate edge nodes~\cite{MartnezBeltrn2022DecentralizedFL,li2022,hu2023,bcpaperdfl}}.}

\subsubsection{\textcolor{black}{Training Driving Assistance Systems.}}
\label{driver behavior}
\label{section:use-case:training-driving-assistance}
\nm{\textcolor{black}{This use case involves training models for Advanced Driving Assistance Systems (ADAS). Particularly, lane-keeping systems are trained on driving patterns combined with image data, to predict optimal steering angles. The collected training data can expose sensitive information such as locations and driving patterns~\cite{Saleem2021SteeringAP},
\cite{Pes2020SPyCS}.}}

\subsubsection{\textcolor{black}{Detecting Misbehavior in VANETs.}}
\label{misbehavior}
\label{section:use-case:misbehavior}
\nm{\textcolor{black}{Misbehavior detection systems aim to identify malicious data sharing in vehicular networks. They analyze reported data, such as location and traffic details, sometimes combined with trust scores or feedback, to model and identify anomalies~\cite{sohan2021,sharma2021,uprety2021}.}}

\subsubsection{\textcolor{black}{Traffic Anomaly Detection.}}
\label{section:anomaly detection}

\nm{\textcolor{black}{This use case involves monitoring the behavior of surrounding vehicles, e.g., to identify unsafe or malicious driving.
An example is detecting stalking vehicles, which follow a vehicle for a long duration through various turns and speed changes. This is achieved using sensor data, such as cameras and IMU sensors, to track proximity and movements~\cite{Sun2023RemindingDO}.}}

 \subsubsection{Predictive Maintenance.}
 \label{section:predictive maintenance}

Data from vehicle sensors and historical repair records can be analyzed to model and predict potential breakdowns and schedule timely maintenance. This approach aims to minimize unexpected vehicle failures and extend vehicle lifespan~\cite{prevmain}.

\subsubsection{FL-based Navigation}
\label{section:fl-navigation}
A concrete use case of DFL in vehicular networks is collaborative learning a navigation model, studied by Kong et al.~\cite{kong2021}.
In scenarios where GPS signals are weak, such as in urban centers or tunnels, vehicles can maintain accurate localization by combining high-sampling Inertial Measurement Unit (IMU) data and low-sampling GPS data.
\textcolor{black}{This process risks exposing sensitive navigation information during the exchange of FL model updates~\cite{kong2021}.}

\subsubsection{Object Classification in CAVs.}
\label{section:object classification in cavs}

CAVs use cameras to capture images of their surroundings, which aids in tasks such as obstacle avoidance and enhancing situational awareness. However, these images may contain a vast amount of sensitive information, including faces, license plates, or locations related to the vehicle's environment~\cite{xiong2022,jinbo2020,xiong2021,xiong2019}. 

\subsubsection{Road Profile Estimation}
\label{road profile}

In this setting, multiple vehicles work together to accurately assess road conditions
and identify surface anomalies like potholes. Instead of relying on data from a single vehicle, which can be affected by
sensor limitations, the collaborative approach allows vehicles on the same road segment to share and combine their data~\cite{Gao2022PrivacyPreservingCE}.

\subsubsection{Vehicle Emission Control}
\label{vehicle emission}

Utilizing traffic light cycle data shared with other vehicles can help reduce vehicle emissions at
intersections. A reinforcement learning model processes this data to help vehicles optimize their speed for lower emissions.
Achieving this requires the collection, sharing, and analysis of privacy-sensitive vehicle data, such as speed, location, and traffic conditions~\cite{Bautista2022PrivacyAwareVE}.

\begin{table}[t!]

\centering
\caption{Overview of specific use cases with relevant privacy concerns.}
\label{table: use case}
\tiny
\begin{tabularx}{\textwidth}{p{5.5cm} p{4.5cm} c c }
\toprule
\textbf{Use Case} & \textbf{Privacy Concern} &  \textsc{Use of MPC} & \textsc{Use of HE}
  \\
\midrule

\multicolumn{4}{l}{\textbf{Location-Based Services}} \\
Points of Interest (\ref{section:POI}) & Location data & \ding{52} & \ding{55} \\
 \textcolor{black}{Navigation and Route Planning} (\ref{section:navigation}) & Location and trajectory data & \ding{52} & \ding{55} \\
 Localization (\ref{section:localization}) & Location data & \ding{52} & \ding{55} \\
Vehicular Crowdsourcing (\ref{section:spatial crowdsourcing}) & Worker's and task requester's location  &\ding{52} & \ding{52}\\
Ride-Sharing (\ref{section:ride-sharing}) & Pick-up and drop-off locations and driving patterns &\ding{52} & \ding{52}\\
Vehicle Sharing (\ref{section:vehicle sharing}) &  Booking,  transaction, and location data  &\ding{52} & \ding{55}\\

\midrule
\multicolumn{4}{l}{\textbf{Mobility Infrastructure}} \\
Toll Data Collection (\ref{section:toll data collection}) & Location data and driving patterns & \ding{55} & \ding{52}\\
 Electric Vehicle (EV) Charging (\ref{section:ev charging}) & Charging location, energy consumption patterns  &  \ding{52} & \ding{55}\\
 Parking (\ref{section:parking}) & Location and identity data & \ding{52} & \ding{52}\\
 \midrule
\multicolumn{4}{l}{\textbf{Vehicular Data Analysis}} \\
Data Processing in IoV (\ref{section:data analysis for iov}
 ) & Data leakage during training process & \ding{52} & \ding{52}\\
 Federated Learning (FL) (\ref{section:decentralized fl}) & Data leakage from model updates & \ding{52} & \ding{52}\\
Training Driving Assistants (\ref{driver behavior}) & Driving patterns, location and user's identity data & \ding{55} & \ding{55}\\
Detecting Misbehavior in VANETs (\ref{misbehavior}) & Location exposure, driving patterns  & \ding{55} & \ding{52}\\
 Traffic Anomaly Detection ~(\ref{section:anomaly detection}) & Location data and driving patterns & \ding{55} & \ding{55} \\
 Predictive Maintenance (\ref{section:predictive maintenance} ) & Location and identity data  & \ding{55} & \ding{52}\\
 FL-based Navigation (\ref{section:fl-navigation}) & Privacy disclosure during navigation updates &  \ding{55} & \ding{52}\\
 Object Classification in CAVs (\ref{section:object classification in cavs}) & Image data leakage &  \ding{52} & \ding{55}\\
Road Profile Estimation (\ref{road profile}) & Vehicle's sensitive information  & \ding{55}& \ding{55}\\

Vehicle Emission Control (\ref{vehicle emission}) & Location and speed data  &\ding{55}& \ding{55}\\
\midrule
\multicolumn{4}{l}{\textbf{Traffic Management}} \\
Message Transmission (\ref{section:message transmission}) & Location and trajectory data & \ding{55} & \ding{52}\\
Driver Profile Matching (\ref{section:feature matching}) & Personal data based on interests or destination &  \ding{52} & \ding{55}\\

Energy Storage Sharing (\ref{section:energy storage sharing}) & Energy consumption and travel patterns   & \ding{55} & \ding{55}\\

Speed Advisory (\ref{section:speed advisory}) & Vehicle speed and arrival time  &\ding{52} & \ding{55}\\
Traffic Signal Control (\ref{section:traffic signal control}) & Location data & \ding{52} & \ding{55}\\
Platooning (\ref{section:platooning}) & Location and travel route information & \ding{55} & \ding{52}\\
 Traffic Monitoring (\ref{section:traffic monitoring}) & Location data and travel patterns  & \ding{55} & \ding{52}\\
\bottomrule
\end{tabularx}
\end{table}

\subsection{Privacy in Dynamic Traffic Management and V2X Communication.}
\label{subsection:traffic management privacy}

This section discusses privacy-sensitive traffic management scenarios involving  real-time vehicular data exchange between vehicles (V2V) or between vehicles and infrastructure (V2X).
The primary privacy risks come from constantly sharing sensitive vehicle data during these 
interactions. Compared to use cases within the Mobility infrastructure (Section \ref{subsection:mobility privacy}) that involve more static exchanges, in this group, we focus on dynamic, real-time communication. 

\subsubsection{Message Transmission.}
\label{section:message transmission}
Message exchanges in VANETs enable vehicles and pedestrians to communicate with each other directly (V2V) and with infrastructure such as RSUs. 
Although increased connectivity and information flow benefit transportation systems, they also introduce privacy risks by tracking and revealing personal patterns and locations~\cite{Magaiga2018}.

\subsubsection{\nm{\textcolor{black}{Driver Profile Matching}}}
\label{section:feature matching}
 \nm{\textcolor{black}{A concrete use case in V2V exchange includes driver profile matching,}} which allows drivers to recognize and connect with others based on shared characteristics such as destinations, interests, or travel routes. For example, this process allows people to add each other as friends and share information based on similar
 interests~\cite{Wang2021ObliviousTF}.

\subsubsection{Energy Storage Sharing}
\label{section:energy storage sharing}
This use case allows multiple users to access and benefit from shared energy storage systems, either through community sharing, outsourcing to third-party energy storage operators, or peer-to-peer sharing.  This approach improves cost-effectiveness by distributing the storage capacity and associated costs among users. This process can involve the disclosure of energy consumption data, users' daily routines, or working patterns~\cite{Wang2021PrivacyPreservingES}.

\subsubsection{Speed Advisory.}
\label{section:speed advisory}
The aim of Consensus-Based Speed Advisory Systems (CSAS)  is to provide real-time, privacy-preserving speed recommendations for groups of vehicles, with a focus on reducing emissions and enhancing energy efficiency. They require collecting sensitive data, such as vehicle type, fuel consumption, and driver's arrival time while optimizing consensus speed~\cite{liu2022}.

\subsubsection{Traffic Signal Control.}
 \label{section:traffic signal control}
In traditional intelligent traffic signal control systems, users' vehicle information, such as location and speed, is transmitted to RSUs to improve service efficiency. Servers collect this data to train machine learning models, which automate the formulation of traffic signal control strategies,  optimizing road traffic management. The transmission of this vehicle information can result in privacy breaches, disclosing sensitive user details~ \cite{Ying2022PrivacySignalPT}.

\subsubsection{Platooning.}
\label{section:platooning}
Platooning is a fuel-efficient transportation method where multiple vehicles, typically trucks, follow each other in proximity on highways, 
for enhanced privacy and increased road capacity. Forming a platoon typically requires the disclosure of sensitive information, such as the real-time geographic position and intended routes of participating vehicles~\cite{Li2023RPPMAR,Zhang2022TPPRAT,Quero2023TowardsPP,Cheng2023PPRTPP}.

\subsubsection{Traffic Monitoring.}
\label{section:traffic monitoring}
Crowdsourcing-based traffic flow statistics can optimize traffic light scheduling and mitigate congestion. Gathering drivers' directional intentions via RSUs and Traffic Management Centers (TMC) may expose sensitive information ~\cite{Zhang2020VerifiableAP}.

\subsection{Summary}

In this section, we briefly reviewed a variety of privacy-sensitive use cases within the automotive domain, grouped into four subdomains, and listed in Table~\ref{table: use case}. Our review shows many cases where privacy-sensitive data is exchanged between multiple parties, such as drivers, data requesters, infrastructure providers, and service providers. This highlights the relevance of exploring privacy-preserving computing technologies, such as MPC and HE, in these multi-party environments.

\section{Applications of Privacy-Preserving Computing}
\label{section:mpc}

In this section, we examine privacy-preserving computing applications (concerning either MPC or HE) in the automotive domain proposed in the literature more closely.

\begin{table*}[t!]
\tiny
\caption{Overview of works that employ MPC and HE in the automotive domain.}\label{table:mpc}
\begin{tabularx}{\textwidth}{
>{\hsize=0.1\hsize}l 
>{\hsize=.1\hsize}l 
>{\hsize=.1\hsize}l 
>{\hsize=.0625\hsize}l 
>{\hsize=.05\hsize}l  
>{\hsize=.015\hsize}l 
>{\hsize=.012\hsize}l 
>{\hsize=.015\hsize}l}
\toprule
Reference \& Authors & 
Use Case & 
\begin{tabular}[c]{@{}c@{}}Domain\end{tabular} & 
Protocols & 
\begin{tabular}[c]{@{}c@{}}Security\\Model\end{tabular} & 
Function  & 
\begin{tabular}[c]{@{}c@{}}Integrated\\Tech.\end{tabular} &
Data\\
\midrule

\multicolumn{8}{l}{\textbf{Client-to-Client Setting}}\\
\cite{siam2018} Hussain and Koushanfar & Localization (\ref{section:localization}) & LBS & BMR (mp), GC(2p) & \textcolor{black}{\faAdjust} & Custom  & - & \textcolor{black}{\faPlay} \\
\cite{Avodji2016MeetingPI} Aïvodji et al. &  Ride-Sharing (\ref{section:ride-sharing})  &  LBS & PSI (2p) & \textcolor{black}{\faAdjust} & Matching  & - & \textcolor{black}{\faDatabase} \\
\cite{Hallgren2017} Hallgren et al. & Ride-Sharing (\ref{section:ride-sharing}) &  LBS & TPSI (2p) & \textcolor{black}{\faAdjust} & Matching  & - & \textcolor{black}{\faDatabase} \\
\cite{toppool2019} Pagnin et al. & Ride-Sharing (\ref{section:ride-sharing}) &  LBS & AHE, PSI (2p) & \textcolor{black}{\faAdjust} & Matching& - & \textcolor{black}{\faDatabase} \\
\cite{Huo2021DistributedPE} Huo et al. & EV Charging (\ref{section:ev charging})& MI & SSS (mp) & \textcolor{black}{\faAdjust} & Aggregation  & - & \textcolor{black}{\faPlay} \\
\cite{bdfl2021} Chen et al. & Federated Learning (\ref{section:decentralized fl}) & VDA & PVSS (mp) & \textcolor{black}{\faCircle} & Aggregation  & FL & \textcolor{black}{\faDatabase} \\
\cite{Wang2021ObliviousTF} Wang et al. & Driver Profile Matching (\ref{section:feature matching}) & DTM & OT, PSI (2p) &  \textcolor{black}{\faAdjust} & Matching & -  & \textcolor{black}{\faBook} \\
\cite{Magaiga2018} Magaia et al. &  Message Transmission (\ref{section:message transmission})  & DTM & AHE & \textcolor{black}{\faCircle} & Matching  & - & \textcolor{black}{\faDatabase}~\textcolor{black}{\faPlay} \\

\midrule

\multicolumn{8}{l}{\textbf{Client-to-Server Setting}}\\
\cite{Peng2024PrivacyPreservingTD} Peng et al. & Vehicular Crowdsourcing (\ref{section:spatial crowdsourcing}) &  LBS &  ASS (mp) & \textcolor{black}{\faAdjust} & Aggregation & - & \textcolor{black}{\faDatabase} \\
\cite{Kong2017} Kong et al. &  Vehicular Crowdsourcing  (\ref{section:spatial crowdsourcing})  & LBS & AHE & \textcolor{black}{\faAdjust} & Custom & - & \textcolor{black}{\faPlay} \\
\cite{Tan2018PrivateIR} Tan et al. & Points of Interest (\ref{section:POI})&  LBS & PIR (mp) & \textcolor{black}{\faAdjust} & Matching & - & \textcolor{black}{\faDatabase} \\
\cite{Zhou2020ATQ} Zhou et al. & Points of Interest (\ref{section:POI})&  LBS & OT (2p), PIR (2p) & \textcolor{black}{\faAdjust} & Custom & - & \textcolor{black}{\faDatabase} \\
\cite{Zhang2022APP} Zhang et al. & Points of Interest (\ref{section:POI}) &  LBS  & PSI (2p) &  \textcolor{black}{\faAdjust}  & Matching & DP &  \textcolor{black}{\faPlay} \\
\cite{ORide2017} Pham et al. & Ride-Sharing (\ref{section:ride-sharing}) &  LBS & SHE & \textcolor{black}{\faAdjust} & Matching  & - & \textcolor{black}{\faDatabase} \\

\cite{marie2018} Aïvodji et al. & Ride-Sharing (\ref{section:ride-sharing})&  LBS & SHE, PEQT, \textcolor{black}{SS} & \textcolor{black}{\faAdjust} & Matching & - & \textcolor{black}{\faDatabase} \\
\cite{yuan2018} He et al. &  Ride-Sharing (\ref{section:ride-sharing}) & LBS & AHE & \textcolor{black}{\faAdjust} & Matching  & - & \textcolor{black}{\faDatabase} \\
\textcolor{black}{\cite{pRidePR2019,Huang2021pRidePO}} Luo et al., \textcolor{black}{Huang et al.} & Ride-Sharing (\ref{section:ride-sharing})&  LBS & AHE, GC (2p), SHE & \textcolor{black}{\faAdjust} & Matching  & - & \textcolor{black}{\faDatabase} \\
\textcolor{black}{\cite{lpride2019} Yu et al.} & Ride-Sharing (\ref{section:ride-sharing}) &  LBS & AHE & \textcolor{black}{\faAdjust} & Matching & - & \textcolor{black}{\faDatabase} \\

\cite{yu2021} Yu et al. & Ride-Sharing (\ref{section:ride-sharing}) &  LBS & SHE & \textcolor{black}{\faAdjust} & Matching & - & \textcolor{black}{\faDatabase} \\
\cite{yuha2021} Yu et al. & Ride-Sharing (\ref{section:ride-sharing}) & LBS & AHE, GC (2p) & \textcolor{black}{\faAdjust} & Matching & - & \textcolor{black}{\faDatabase} \\
\cite{Yu2022EfficientAP} Yu et al. & Ride-Sharing (\ref{section:ride-sharing}) &  LBS & SHE & \textcolor{black}{\faAdjust} & Matching & - & \textcolor{black}{\faDatabase} \\
\textcolor{black}{\cite{Xu2022AnEA} Xu et al.} & Ride-Sharing (\ref{section:ride-sharing}) &  LBS & GM & \textcolor{black}{\faAdjust} & Matching & - & \textcolor{black}{\faPlay} \\
\cite{Zhang2023PrivacypreservingOR} Zhang et al. & Ride-Sharing (\ref{section:ride-sharing}) &  LBS & PSI (2p) & \textcolor{black}{\faCircle} & Matching & - & \textcolor{black}{\faPlay} \\
\cite{Luo2023P2RidePA} Luo et al. & Ride-Sharing (\ref{section:ride-sharing}) &  LBS & PEQT (2p) & \textcolor{black}{\faAdjust} & Matching  & - & \textcolor{black}{\faDatabase} \\
\cite{Karmakar2024QuickPoolPR} Karmakar et al. & Ride-Sharing (\ref{section:ride-sharing}) &  LBS &  FSS (3p) & \textcolor{black}{\faAdjust} & Matching & - & \textcolor{black}{\faPlay} \\
\cite{Zhou2024PrivacypreservingAV} Zhou et al. & Data Processing in IoV (\ref{section:data analysis for iov}) &  VDA & ASS (mp) & \textcolor{black}{\faCircle} & Aggregation & - & \textcolor{black}{\faPlay}   \\
\cite{liu2022} Liu et al. & Speed Advisory (\ref{section:speed advisory})& DTM & SS (mp) &\textcolor{black}{\faAdjust} & Custom & - & \textcolor{black}{\faPlay} \\
\cite{Liang2022PPRPPR} Liang   et al. & Message Transmission (\ref{section:message transmission}) & DTM & OT (2p) & \textcolor{black}{\faCircle} & Custom & - &  \textcolor{black}{\faPlay}\\
\cite{Zhang2020VerifiableAP} Zhang et al. & Traffic Monitoring (\ref{section:traffic monitoring}) & DTM & BGN & \textcolor{black}{\faAdjust} & Aggregation & DP & \textcolor{black}{\faPlay} \\

\cite{Quero2023TowardsPP} Quero et al. & Platooning (\ref{section:platooning})& DTM & CKKS & \textcolor{black}{\faAdjust} & Matching  & - & \textcolor{black}{\faPlay} \\

\cite{Zhang2022TPPRAT} Zhang et al. & Platooning (\ref{section:platooning}) & DTM & AHE & \textcolor{black}{\faCircle} & Aggregation  & - & \textcolor{black}{\faPlay} \\
\cite{Cheng2023PPRTPP} Cheng et al. & Platooning (\ref{section:platooning}) & DTM & AHE & \textcolor{black}{\faCircle} & Aggregation  & - & \textcolor{black}{\faPlay} \\
\midrule

\multicolumn{8}{l}{\textbf{Distributed Setting}}\\
\cite{junwei2021} Zhang et al. & Vehicular Crowdsourcing (\ref{section:spatial crowdsourcing}) &  LBS & AHE, OPE & \textcolor{black}{\faCircle} & Matching  & BC & \textcolor{black}{\faDatabase} \\
\cite{Xu2023PriTAECPT} Xu et al. & Vehicular Crowdsourcing (\ref{section:spatial crowdsourcing}) &  LBS & OT, PEQT (2p) & \textcolor{black}{\faAdjust} & Matching  &  - & \textcolor{black}{\faPlay} \\
\cite{Cheng2023ALP} Cheng et al. & Vehicular Crowdsourcing  (\ref{section:spatial crowdsourcing}) &  LBS  & AHE & \textcolor{black}{\faAdjust} & Matching & - & \textcolor{black}{\faPlay}  \\

\textcolor{black}{\cite{Guan2020AchievingPV} Guan et al.} & Vehicular Crowdsourcing  (\ref{section:spatial crowdsourcing}) &  LBS & SHE & \textcolor{black}{\faAdjust} & Custom & - & \textcolor{black}{\faDatabase}  \\

\cite{Yu2024EfficientPT} Yu et al. & Vehicular Crowdsourcing  (\ref{section:spatial crowdsourcing}) &  LBS & SSS (mp) & \textcolor{black}{\faAdjust} & Custom & - & \textcolor{black}{\faPlay}  \\
\textcolor{black}{\cite{Zhou2021EPNSEP} Zhou et al.} & Navigation and Route Planning  (\ref{section:navigation}) &  LBS & MPDC & \textcolor{black}{\faAdjust} & Custom & - & \textcolor{black}{\faPlay} \\

\textcolor{black}{\cite{Tiausas2023HPRoPHP} Tiausas et al.} & Navigation and Route Planning  (\ref{section:navigation}) &  LBS & PIR (mp) & \textcolor{black}{\faAdjust} & Custom & - & \textcolor{black}{\faPlay} \\

\cite{iraklis2017,iraklis2022} Symeonidis et al. & Vehicle Sharing (\ref{section:vehicle sharing}) &  LBS & SSS (mp) & \textcolor{black}{\faAdjust} & Custom  & - & \textcolor{black}{\faDatabase} \\
\cite{karim22} Karim and Rawat & Toll Data Collection (\ref{section:toll data collection})& MI & FHE & \textcolor{black}{\faAdjust} & Custom & BC & \textcolor{black}{\faDatabase}  \\
\cite{Zhang2023BSDPBS} Zhang et al. & Parking Systems  (\ref{section:parking}) & MI & AHE &  \textcolor{black}{\faAdjust} & Custom & BC & \textcolor{black}{\faDatabase} \\
\cite{Li2021PriParkRecPD} Li et al. & Parking Systems (\ref{section:parking}) & MI & AHE, PSI (2p) & \textcolor{black}{\faAdjust} & Matching  &  - & \textcolor{black}{\faPlay} \\
\cite{Amiri2019PrivacyPreservingSP} Amiri et al. & Parking Systems (\ref{section:parking}) & MI & PIR (mp) & \textcolor{black}{\faCircle} & Custom  &  BC & \textcolor{black}{\faPlay} \\
\cite{chenji2022} Chen et al. & Data Processing in IoV (\ref{section:data analysis for iov}) & VDA & FHE & \textcolor{black}{\faCircle} & Custom  & BC, DL & \textcolor{black}{\faPlay} \\
\textcolor{black}{\cite{trajectoryvanet} Liu et al.} & Data Processing in IoV (\ref{section:data analysis for iov}) & VDA & SSS (mp) & \textcolor{black}{\faAdjust} & Aggregation  & - & \textcolor{black}{\faPlay} \\

\cite{li2022} Li et al. & Federated Learning (\ref{section:vehicular fog computing}) & VDA & SSS (mp) & \textcolor{black}{\faAdjust} & Aggregation  & FL, DP & \textcolor{black}{\faPlay} \\
\cite{hu2023} Hu et al. & Federated Learning  (\ref{section:decentralized fl}) & VDA & SSS (mp), CKKS & \textcolor{black}{\faAdjust} & Aggregation  & BC, FL & \textcolor{black}{\faPlay} \\

\cite{Li2021PrivacyPreservedFL} Li et al. & Federated Learning (\ref{section:decentralized fl 2}) & VDA & DGHV & \textcolor{black}{\faCircle} & Aggregation & BC, FL & \textcolor{black}{\faDatabase} \\
\cite{kong2021} Kong et al. & FL-based Navigation (\ref{section:fl-navigation}) & VDA & SSS (mp), AHE & \textcolor{black}{\faCircle} & Aggregation  & FL, DP & \textcolor{black}{\faPlay} \\
\cite{jinbo2020,xiong2022} Xiong et al. & Object Classification in CAVs (\ref{section:object classification in cavs}) & VDA & ASS (mp) & \textcolor{black}{\faAdjust} & Custom & DL & \textcolor{black}{\faDatabase} \\

\cite{Bi2023AchievingLA} Bi et al. & Object Detection in CAVs (\ref{section:object classification in cavs}) & VDA & ASS (mp) & \textcolor{black}{\faAdjust} & Custom & DL & \textcolor{black}{\faDatabase} \\

\cite{prevmain} Kong et al. & Predictive Maintenance  (\ref{section:predictive maintenance}) & VDA  & AHE & \textcolor{black}{\faAdjust} & Aggregation & DP & \textcolor{black}{\faBook} \\
\cite{sohan2021} Gyawali et al. &  Det. Misbehavior in VANETs (\ref{misbehavior}) & VDA & AHE & \textcolor{black}{\faCircle} & Aggregation& DL & \textcolor{black}{\faPlay} \\
\cite{Wang2021PrivacyPreservingES} Wang et al. & Energy Storage Sharing  (\ref{section:energy storage sharing})& DTM & SSS (mp) & \textcolor{black}{\faCircle}  & Aggregation & BC & \textcolor{black}{\faPlay} \\
\cite{Ying2022PrivacySignalPT} Ying et al. & Traffic Signal Control (\ref{section:traffic signal control}) & DTM & ASS (2p) & \textcolor{black}{\faAdjust} & Custom & DL & \textcolor{black}{\faPlay} \\
\textcolor{black}{\cite{2025VulnerabilityMO} Adelipour et al.} & Traffic Signal Control (\ref{section:traffic signal control}) & DTM & ASS (mp) & \textcolor{black}{\faAdjust} & Custom & - & \textcolor{black}{\faPlay} \\

\midrule
\end{tabularx}
\begin{tablenotes}
		\begin{tiny}
		\item[*] 

\textbf{Application Domains: }
\textbf{LBS} --- Location-Based Services. 
\textbf{MI} --- Mobility Infrastructures. 
\textbf{VDA} --- Vehicular Data Analysis. 
\textbf{DTM} --- Dynamic Traffic Management and V2X Communications. 
        \textbf{Protocols: }
        \textbf{ASS} --- Additive Secret Sharing.
        \textbf{BGN} --- Boneh, Goh, and Nissim cryptosystem.
        \textbf{BMR} --- Beaver-Micali-Rogaway protocol.
        \textbf{DGHV}--- Dijk-Gentry-Halevi-Vaikutanathan Algorithm.
         \textbf{CKKS}---  Cheon-Kim-Kim-Song Algorithm.
        \textbf{FHE} --- Fully Homomorphic Encryption.
        \textbf{GC} --- Garbled Circuit.
        \textbf{OPE} --- Order Preserving Encryption.
        \textbf{OPRF} --- Oblivious Pseudorandom Function.
        \textbf{OT} --- Oblivious Transfer.
        \textbf{AHE } --- Additive Homomorphic Encryption (Paillier cryptosystem).
        \textbf{PEQT} --- Private Equality Test.
        \textbf{PSI} --- Private Set Intersection.
         \textbf{TPSI} --- Threshold Private Set Intersection.
        \textbf{PIR} --- Private Information Retrieval.
        \textbf{PVSS} --- Publicly Verifiable Secret Sharing.
        \textbf{SHE} --- Somewhat Homomorphic Encryption.
       \textcolor{black}{\textbf{GM} --- Goldwasser–Micali Algorithm.}
        \textbf{SSS} --- Shamir's Secret Sharing.
       \textcolor{black}{\textbf{FSS} --- Function Secret Sharing.}
      \textcolor{black}{  \textbf{MPDC} --- Multiparty Delegated Computation.}
       \textbf{Number of parties in MPC protocols: }
   \textbf{2p} --- Two-Party.
   \textbf{mp} --- Multi-Party (three or more parties).
        \textbf{Security Model: }\textcolor{black}{\faAdjust} --- semi-honest.
        \textcolor{black}{\faCircle} --- malicious. 
         \textbf{Integrated Technology: }\textcolor{black}{BC} --- Blockchain.
         \textbf{DL} --- Deep Learning.
         \textbf{DP} --- Differential Privacy.
         \textbf{FL} --- Federated Learning.
         \textbf{Evaluation: }\textcolor{black}{\faDatabase} --- Real-world data.
       \textcolor{black}{\faPlay} --- Simulated data. 
      \textcolor{black}{\faBook} --- Theoretical work.\\
		\end{tiny}
		\end{tablenotes}
\end{table*}

\subsection{Overview}
To achieve a comprehensive overview, we categorize the surveyed works into three application settings based on similar architectural setups and requirements for applying MPC or HE.
The three primary settings we focus on are: a \textit{client-to-client setting}, which involves direct data exchange between clients, such as vehicles or drivers, without intermediary infrastructure; a \textit{client-to-server} setting, where clients transfer data to one or more central server(s), for processing and/or storage; and finally, a \textit{distributed setting}, which  leverages a network of servers and edge nodes to handle computational tasks closer to data sources in a decentralized way. Following the structure of Section~\ref{section:use-cases}, we then group surveyed works within each setting by their application domains: location-based services, mobility infrastructures, vehicular data analysis, and dynamic traffic management, to show how different settings are observed in each use case group.

An overview of the surveyed works
can be found in Table~\ref{table:mpc}.
The first column lists the publications and the names of the authors. The second and third columns indicate the specific addressed use case and its application domain, respectively. The fourth column details the protocols used in the implementation of the works.
The fifth column indicates the security model considered in the publication: semi-honest and malicious. In the semi-honest security model, it is assumed that the participants follow the prescribed protocol but may be curious to derive additional information from the process. On the contrary, in the malicious security model, participants are assumed to actively attempt to undermine the protocol by modifying inputs or deviating from the protocol.

In addition, we observe that many implementations of MPC and HE serve two primary purposes in reviewed works.
First, numerous works employ MPC or HE to perform privacy-preserving aggregation, computing an aggregate (e.g., a sum) of data collected from multiple sources. 
These functions are used in various use cases, such as 
aggregating model updates in federated learning, predictive maintenance, etc. Second, several works employ MPC or HE to implement matching functionality. In these works, the technologies are used to identify pairs or subsets of data entries that meet specific criteria without revealing privacy-sensitive
information. They are widely used in contexts such as ride-sharing services and task allocation in spatial crowdsourcing. The remaining works
implement customized functionalities   
that do not fall under aggregation or matching.
Examples of custom functions include training machine learning models~\cite{chenji2022}, performing joined decision-making~\cite{Ying2022PrivacySignalPT}, and optimization~\cite{liu2022}.
We list these three categories in the sixth column.

The seventh column shows whether HE and MPC are combined with technologies such as Blockchain, Differential Privacy, Federated Learning, and Machine Learning. Finally, the last column indicates whether real-world data or simulated data was used in the papers.

\subsection{Client-to-Client Setting}
\label{section:client-to-client}
In this section, we discuss how 
MPC and HE, help to solve privacy issues in various
client-to-client
settings, where clients (e.g., vehicle drivers) directly interact with each other to share or exchange sensitive data, such as locations or travel paths. This approach eliminates the need for a server or centralized infrastructure to process, store, or transmit data, reducing the privacy risks associated with centralized systems.
In this setting, privacy risks mainly involve the potential exposure or misuse
of sensitive data by malicious clients.
To counter these risks, clients use MPC and HE to jointly process the data without revealing sensitive information to each other. Here, clients typically act as both data providers and computing parties, performing computations on their own inputs. The result of the computation can be used by all or specific parties, depending on the use case.

\begin{figure}[t!]
    \centering
    \includegraphics[trim=0cm 11.8cm 0cm 0.4cm, clip,width=0.75\textwidth]{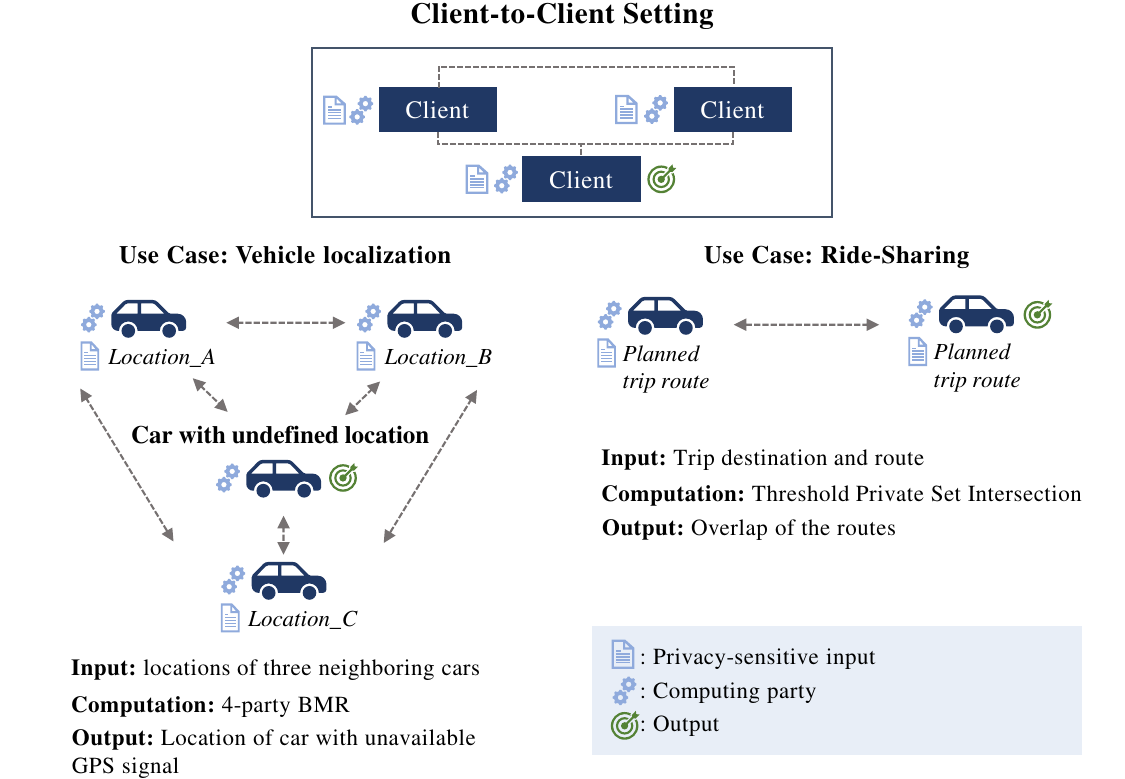}
    \caption{Generic client-to-client setting, with two specific example use cases under the same setting.  
In this setting, clients typically act as computing parties on their own inputs, with output used by all or some clients.}
    \Description{Figure showing a generic client-to-client setting illustrating three client nodes, along with two applications of this setting to the use case of vehicle localization, by using three car nodes to locate a fourth one, and to the use case of ride-sharing, where two car nodes coordinate on the overlap of their trip routes.}
    \label{fig:setting-client-to-client}
\end{figure}

Figure~\ref{fig:setting-client-to-client}
demonstrates the client-to-client setting with two specific use cases. In the first example, localization~\cite{siam2018}, a car performs secure 4-party computation with three neighboring cars to compute its location (the output) based on the locations of the neighboring cars (the inputs). The MPC allows vehicles to not reveal their actual locations to each other.
The second example shows the use of ride-sharing.
Here, two drivers compute on their planned routes using PSI protocol, allowing them to identify potential overlaps without disclosing the full routes. These examples show how different implementations of the client-to-client setting follow a similar pattern where multiple clients interact as computing parties on their own inputs.
Below, we survey existing works in detail.

\subsubsection{Privacy in Location-Based Services.}

Client-to-client interactions in LBSs involve clients sharing sensitive location data directly with each other to enable services such as localization and ride-sharing, which may compromise privacy. \nmold{
Addressing this concern in localization, Hussain and Koushanfar~\cite{siam2018} proposed a method where a lost car can compute its location with the help of three nearby cars while ensuring that the locations of all participating vehicles remain private. 
The method enables cars to communicate directly to locate a lost vehicle,  computing their mutual geometric intersections for positioning. 
}
\nmold{This work is one of the few examples of using GC in the automotive domain, presenting two alternatives:  one using a two-party GC protocol and another using a multi-party BMR protocol.
Although the BMR protocol extends GC to support multiple parties and enhances privacy, it introduces higher computational complexity. \textcolor{black}{More precisely, the GC-based protocol runs in 0.35s ($\sim$1MB of data sent), while the BMR takes 2.65s due to the use of a combined circuit (TriLoc), which integrates three instances of sub-circuits that securely compute pairwise circle intersections (Intersection) and verify whether each point lies within a third car’s range (Range), along with additional logic to select intersection points.}}

\nmold{Another group of works focuses on implementing ride-sharing scenarios, which often involve a centralized server to coordinate routes, as seen in Table~\ref{table:mpc}. 
In particular, } Aïvodji et al.~\cite{Avodji2016MeetingPI}  propose   
a PSI-based decentralized ride-sharing architecture, where a driver and a rider want to determine a pick-up and drop-off location without disclosing their origin and destination locations to each other or any third party. %
After clients encrypt their location using HE, they utilize PSI to compare them and find the common ride-sharing locations. Even though the PSI method is secure, a semi-honest adversary might infer information based on the size of the encrypted data. To address this, the authors propose to fix the size of the isochrone (a geographical boundary showing areas reachable within a certain time) for both clients to ensure indistinguishability.%

Hallgren et al.~\cite{Hallgren2017} consider the abstract model of two parties exchanging location data in ride-sharing. They introduce PrivatePool,
\nmold{supporting two ride-sharing approaches. The first,  proximity-based method, leverages HE to allow users to determine if starting and ending points are within a certain range,
without revealing locations.
The second, intersection-based approach, employs a TPSI protocol to identify overlaps in ride trajectories. 
Finally, Pagnin et al.~\cite{toppool2019} further develop this concept with TOPPool, that extends the PrivatePool~\cite{Hallgren2017} and enhances privacy-preserving ride-sharing with time-aware optimizations and the ability to handle partial schedule overlaps.  Unlike PrivatePool, TOPPool employs regular PSI to perform more efficient intersection-based matching between trips represented as sets of consecutive points and AHE for ride endpoint-based matching.
}
\textcolor{black}{Due to space constraints, we omit performance descriptions for ride-sharing papers here and in Section~\ref{section:client-to-server-lbs}, and instead provide a performance comparison in Section~\ref{performance}.}

\subsubsection{Privacy in Mobility Infrastructures.}
\nmold{Although typically mobility infrastructure services involve a centralized server, there are scenarios where data exchange occurs directly between vehicles.}
A relevant example that utilizes MPC is EV charging control, where EVs coordinate charging without relying on a centralized charging station to manage the process. To develop a privacy-preserving EV charging control using SSS, Huo and Liu~\cite{Huo2021DistributedPE} adjust the charging of all EVs so that they are all charged as required by the end of the night without surpassing their maximum charging rates. In their work, the charging profile of each EV is considered the secret and is transformed into an integer before being shared.  Each EV constructs a polynomial of degree $k$ with secret as the constant term and random coefficients, then evaluates the polynomial at predefined points and shares the results with other EVs. Using these shared evaluations, the EVs can securely reconstruct the aggregate sum of the charging profiles without revealing individual profiles. Additionally, collusion among semi-honest EVs requires at least $k$ EV's to collaborate to infer another EV's secret.

\subsubsection{Privacy in Vehicular Data Analysis.}

Decentralized Federated Learning (DFL) is increasingly used in the automotive sector during vehicle data analysis without relying on a centralized server~\cite{MartnezBeltrn2022DecentralizedFL}. V2X communication, with its vehicle mobility and limited storage capacity of nodes, benefits from DFL, as it allows direct client-to-client interactions for training collaborative models. The direct exchange of gradients between clients can potentially reveal sensitive information about training data. Addressing data leakage in autonomous vehicles during the training process, 
Chen et al.~\cite{bdfl2021} introduce a novel Byzantine-fault-tolerant decentralized FL method based on a peer-to-peer network, using a PVSS~\cite{PVSS} scheme, which enables anyone (not just the participant) to confirm the accuracy of encrypted shares. In their method, each autonomous vehicle uses PVSS to protect its data. If a share is identified as false, the responsible participant is considered malicious and will be excluded from participating in the next communication round. \textcolor{black}{Experiments show that the PVSS protocol runs in $\sim$5.2s with 512 autonomous vehicles, with secret share distribution taking  $\sim$0.97s.}

\subsubsection{Privacy in Dynamic Traffic Management and V2X Communications.}
\nmold{We observe two works leveraging privacy-preserving computing technologies within the context of vehicle-to-vehicle (V2V) communication to improve traffic management 
.}
First, to
address privacy concerns in driver profile matching, where users in vehicular social networks share information based on similar characteristics, Wang et al.~\cite{Wang2021ObliviousTF} propose solutions based on OT and PSI. The characteristics include upcoming destinations, tourist spots, the work sector, favorite sports, preferred movies, music preferences, etc. The authors use an OT protocol to design a PSI protocol with equality tests. These protocols allow two parties in a VANET to identify similar characteristics in their sets without revealing any additional information beyond the intersection.

\nmold{Second, Magaia et al.~\cite{Magaiga2018} focus on enhancing message delivery in vehicular delay-tolerant networks. The authors introduce a  routing protocol called ePRIVO, based on AHE. It solves the problem of dynamically selecting the optimal vehicle for message forwarding. 
When two vehicles meet, they use the protocol to compare their routing metrics (such es ego betweenness centrality and similarity) to determine which vehicle is a better candidate for message forwarding, without exchanging privacy-sensitive metrics directly. \textcolor{black}{Encryption and decryption take $\sim$11.03ms with 1024-bit keys;
and the use of encryption results in delivery ratio losses ranging from approximately 0.09$\%$ to 30.64$\%$ in different scenarios.}}

\subsubsection{Summary}
\nmold{In this section, we surveyed applications of MPC and HE in the client-to-client setting, 
using a diverse set of technologies, including GC-based and SS-based MPC, PSI, and HE, enabling privacy-preserving interactions. Overall, we note a limited number of such implementations, likely due to the nature of vehicular settings with large fleets, real-time communication, resource constraints, and client drop-offs. These demands pose a challenge for resource-intensive MPC and HE technologies. Consequently, we observe more solutions either involving a centralized server or adopting complex decentralized setups, which we will discuss in the following sections.}

\subsection{Client-to-Server Setting}
\label{section:client-to-server}
\label{section:server}

In this section, we discuss how
MPC and HE 
address privacy concerns in the client-to-server setting, where clients (such as vehicles or users) interact with one or more servers that aggregate and process client data. The primary privacy concern here involves the handling and potential exposure of large-scale sensitive client data by servers, which may be vulnerable to misuse or breaches. To mitigate these risks, MPC and HE enable servers to perform computations on client data without directly accessing sensitive information.

\begin{figure}[t!]
    \centering
    \includegraphics[trim=0cm 12.2cm 0cm 0.25cm, clip,width=0.75\textwidth]{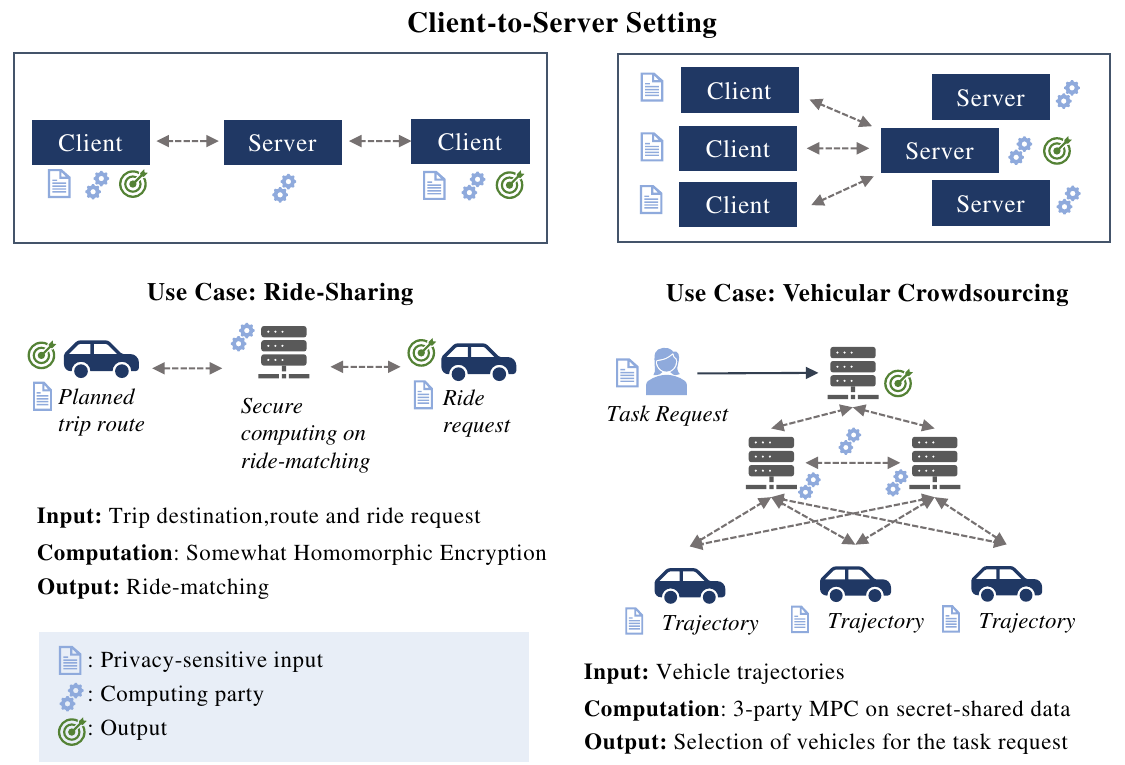}
    \caption{Generic client-to-server setting, with two specific use cases under the same setting as examples.  
In this setting, clients typically provide their private inputs to the server,
with output used by clients or server.}
    \label{fig:setting-client-to-server}
\end{figure}

We observe two primary cases within this setting. In the first case, servers typically function as service providers, performing computations on data collected from clients or providing specific services in response to client requests. Here, the server processes either encrypted data provided by clients (in the case of HE) or engages in an MPC protocol with clients, generating the desired outputs based on client data while preserving the privacy of individual inputs. 
Figure~\ref{fig:setting-client-to-server} illustrates this case on the left, with a concrete example of ride-sharing, where the server securely processes trip information from clients to generate suitable ride matches by computing on homomorphically encrypted client data. 
Note how this implementation differs from the ride-sharing use case within the client-to-client setting discussed in Section~\ref{section:client-to-client}, where clients communicated with each other directly.
The second case (right) involves an outsourced computation scenario, where multiple servers aggregate and process client data, where the result is further used by the server. This is represented in the figure (bottom right) by the vehicular crowdsourcing example, where three servers receive secret-shared route information from multiple clients and run an MPC protocol to identify vehicles suitable for a location-specific task without revealing individual vehicle locations.

These examples highlight how client-to-server implementations can leverage MPC and HE to protect privacy in applications involving sensitive data sharing and processing. Below, we examine relevant works that apply these techniques across different use cases.

\subsubsection{Privacy in Location-Based Services.}
\label{section:client-to-server-lbs}
\nmold{The client-to-server setting is naturally represented in LBSs, where clients share sensitive location data with service providers to receive location-dependent services. Service providers need to process clients' location data to deliver their services while ensuring they cannot misuse this sensitive information. We observe three distinct use cases where MPC and HE are applied: 
mobile crowdsourcing, points of interest, and ride-sharing. Below, we examine how different works address privacy concerns in each of these scenarios.}

\nmold{Within the crowdsourcing scenario, mobile crowdsensing involves users collaborating through their sensing devices to complete a shared task, in applications like traffic monitoring and road condition analysis. In vehicular networks, the key challenges lie in assessing the reliability of sensing vehicles. To address these concerns, Peng et al.~\cite{Peng2024PrivacyPreservingTD} propose a truth discovery scheme. The approach uses ASS where Sensing Users (SUs) distribute their collected sensing data to three servers in a secret-shared way. These servers jointly run a protocol to discover the most accurate information (ground truth) and to implement a quality-driven user reward mechanism. The data requester can later aggregate the shares from all servers to recover the final results. %
}
\nmold{Kong et al.~\cite{Kong2017} propose a range query scheme using AHE. 
Vehicles equipped with air pollution sensors generate encrypted data reports containing the sensed values and their location information.  
In this scheme, AHE enables the protection of the location privacy of vehicles and data requesters.
}

\nmold{
In the context of querying POI, three approaches have been developed to protect privacy while allowing vehicles to obtain service results. First, Tan et al.~\cite{Tan2018PrivateIR} propose a computationally efficient PIR-based framework for vehicular LBS. Their approach partitions a city's road network into Natural Road Segments (NRS), with each segment maintaining an association with nearby POIs within a specified distance. 
This segmentation significantly reduces the dataset size, making the PIR-based protocol more efficient. When a vehicle queries POI information, it sends a PIR request that enables service retrieval without revealing either its exact location or query interests to the server. %
Second, Zhou et al.~\cite{Zhou2020ATQ} introduce a novel POI protocol that combines multiple cryptographic techniques and is specifically designed for real road networks. The approach enables top-K POI queries through vehicle cooperation: Vehicles first obtain symmetric keys through OT, then use PIR to retrieve encrypted POI details, while network coding ensures query interest privacy during processing. Third, Zhang et al.~\cite{Zhang2022APP} propose a proximity testing scheme that uses two-party PSI to detect any location grids overlap, indicating proximity. For this scheme, Chebyshev polynomials are used to optimize computational efficiency and reduce energy consumption. %
}

As observed from Table~\ref{table:mpc}, ride-sharing applications have received significant attention in the context of LBSs, with multiple works proposing various privacy-preserving approaches. Pham et al.~\cite{ORide2017} proposed ORide, implementing SHE with optimized ciphertext packing for privacy-preserving ride-matching. \textcolor{black}{While the system allows user matching without accessing their identities or locations, Kumaraswamy et al.~\cite{Kumaraswamy2021RevisitingDA} and Murthy and Vivek~\cite{PassiveTAoride}
revealed security limitations. A semi-honest rider can infer driver locations through a location-harvesting attack  \cite{Kumaraswamy2021RevisitingDA} or a passive triangulation attack~\cite{PassiveTAoride} using the permuted distances of the riders in ORide.}
Further advancing privacy guarantees, SRide, proposed by Aïvodji et al.~\cite{marie2018}, implements ride-sharing with a two-stage approach. First, feasible matches for each rider are computed using an ASS protocol based on SHE. Then, it uses secure two-party equality testing to determine final matches.
To address the efficiency concerns, He et al.~\cite{yuan2018} proposed PRIS, which uses AHE and bilinear pairing to protect location privacy.
The system separates operations into offline (local generation of ride offers/requests) and online phases (secure matching), optimizing computational efficiency while maintaining privacy. 

\textcolor{black}{Luo et al.~\cite{pRidePR2019} propose a ride-matching protocol that 
operates in two variants: pRide and pRide$_2$. pRide encrypts rider and driver location data and distances using SHE (BGN~\cite{boneh2005}) and identifies the nearest driver via GC-based  secure comparison. pRide$_2$ uses AHE (Paillier) for distance computation, GC for secure
comparisons, while applying data packing and graph partitioning to reduce computation and communication costs.  Huang et al.~\cite{Huang2021pRidePO} present an SHE (FV~\cite{Fan2012SomewhatPF})-based improved version of both the pRide and pRide$_2$ schemes, 
combining secure distance computation with a ride request prediction model for optimized driver matching. However, Murthy and Vivek~\cite{Murthy2022DriverLH} revealed that  pRide$_2$ scheme
\cite{Huang2021pRidePO} is vulnerable to driver location inference attacks, where a semi-honest rider uses decrypted distance values that were homomorphically blinded to recover the underlying distances and infer the locations of at least 80$\%$ of the drivers responding to a single ride request.}

\textcolor{black}{
 Yu et al.~\cite{lpride2019} proposed lpRide which enables efficient shortest road distance computation over encrypted rider and taxi locations using a lightweight encryption scheme based on the modified Paillier cryptosystem~\cite{modifiedpa}. In addition, it securely compares two distances over the corresponding blinded ciphertexts to find the closest taxi. 
 However, Vivek~\cite{Vivek2021AttacksOA} showed that in lpRide, the modified Paillier cryptosystem is vulnerable to a key recovery attack, allowing any semi-honest rider or driver to extract the secret keys of other users and learn the location of riders.}

Expanding the scope to group scenarios, Yu et al.~\cite{yu2021} developed PGRide. The system uses SHE with ciphertext packing to compute aggregate distances between multiple riders and potential drivers in encrypted form. While supporting efficient group matching through separated offline (key generation) and online (secure computation) operations, the system cannot optimize the actual pickup route for multiple riders.%
Focusing on dynamic scheduling, Yu et al.~\cite{yuha2021} introduced PSRide. Unlike previous approaches requiring preset locations, PSRide allows real-time matching and schedule modifications even with active rides. The system uses AHE with cipher packing for computation and GCs for secure schedule feasibility checks.
While this enables flexible ride-sharing, their use of upper-bound travel time estimates can lead to suboptimal matches. %
Another approach is proposed by Yu et al.~\cite{Yu2022EfficientAP}, EPRide, allowing riders to submit encrypted ride requests to a server, which matches them with nearby taxis that regularly update their encrypted locations using SHE. The crypto server generates cryptographic keys and collaborates with the server to execute the secure comparison protocol based on HE to match riders with the nearest available taxi.%

\textcolor{black}{
Leveraging the XOR-homomorphic property of the Goldwasser–Micali (GM) encryption algorithm, Xu et al.~\cite{Xu2022AnEA} proposed TAROT, route-matching scheme for ride-sharing services. It claims to enable  secure equality testing and route similarity computation between encrypted location points accurately, avoiding plaintext exposure during the matching process. However, subsequent passive attacks by Vargheese and Vivek~\cite{Vargheese2023AttackOT} demonstrate that TAROT's GM-based equality determination algorithm leaks the Hamming weight of XORed encrypted location vectors during route similarity computation. By exploiting this leakage, colluding semi-honest users can infer other users' sensitive location data through two distinct passive attacks. In the first attack, adversaries select specific location points to infer the target’s data. In the second attack, the location points of
the colluding adversary are arbitrarily placed.}

Taking a different approach, Zhang et al.~\cite{Zhang2023PrivacypreservingOR} proposed a PSI-based approach. Instead of working with exact locations, their system represents user positions as sets of nearby POIs and then uses PSI to determine matching potential. This allows the platform to facilitate matches without accessing exact user locations, sharing driver information when sufficient geographical overlap exists.

Addressing scalability challenges, Luo et al.~\cite{Luo2023P2RidePA} introduced P$^2$Ride. Instead of using computationally expensive GCs, the system reduces matching to non-interactive PEQT using an overlapping partition system. This approach significantly reduces computational and communication overhead, making it more practical for large-scale deployment than GC-based solutions. %

Karmakar et al.~\cite{Karmakar2024QuickPoolPR} proposed QuickPool, introducing two complementary approaches for simultaneous privacy-preserving ride-matching. The first uses pseudorandom functions for route intersection matching, while the second employs \textcolor{black}{Function Secret Sharing (FSS)~\cite{fss}} to match users based on the proximity of trip endpoints. The system evaluates match compatibility through threshold-based distance comparisons while maintaining location privacy through computation.

\subsubsection{Privacy in Vehicular Data Analysis.}

In IoV, aggregating sensitive data from vehicles to centralized servers for analysis requires privacy-preserving approaches to prevent the exposure of information during data exchanges. To aggregate vehicle perception data for analysis, Zhou et al.~\cite{Zhou2024PrivacypreservingAV} propose a
data aggregation scheme PPVDA. This scheme employs homomorphic MAC and SS to achieve lightweight, verifiable data aggregation, supporting multidimensional data inner product computation. The system operates by dividing each piece of sensed data from Vehicular Nodes (VNs) into multiple additive shares. These shares are then distributed to different RSUs. Each RSU combines its received shares and creates a partial proof, which is then sent to the central server for the reconstruction of the full data and computation of the final aggregation result. 
We observe similar SS-based approaches for vehicular data analysis in distributed setups in Section~\ref{distributed VDA}.

\subsubsection{Privacy in Dynamic Traffic Management and V2X Communications.}

In a client-to-server setting, dynamic traffic management in vehicular networks utilizes the continuous exchange of data between vehicles and servers to optimize tasks such as emission control, energy management, and traffic management. Below, we see how different  techniques are tailored to this setting.

Optimizing emissions and energy use often requires vehicles to share speed and emission data with servers. Liu et al. \cite{liu2022} introduce MPC-CSAS, a SS-based solution for recommending common speeds to a group of vehicles. In the conventional approach, vehicles $A$ and $B$ send their speed-emission mapping values directly to a base station, which aggregates these values and recommends the optimal speed.  $A$ and $B$ split their speed-emission mappings into shares, keeping some locally and sharing others. They aggregate their local and shared data and send the results to the base station, which calculates the emissions for each speed and recommends the optimal speed. \textcolor{black}{MPC-CSAS achieves real-time performance by computing the optimal speed in a single iteration using SS, with total communication less than 3KB for 20 vehicles and less than 5ms runtime.}

Route planning often involves pre-sharing routes with RSUs to speed up authentication. Liang et al.~\cite{Liang2022PPRPPR} propose OT-based route planning scheme,  where a vehicle securely obtains information about RSUs along its planned route with the help of a Certificate Authority (CA), without the CA knowing which specific RSUs the vehicle has chosen. The vehicle uses this pre-shared RSU information to authenticate with RSUs as it enters its coverage. 

Monitoring traffic flow at intersections requires driver data, introducing concerns about the privacy of their movements. Zhang et al.~\cite{Zhang2020VerifiableAP} propose VPTS, a crowdsourcing-based traffic monitoring scheme that ensures the privacy-preserving collection of traffic flow statistics at road intersections. This scheme utilizes HE and DP to secure traffic data.  The process begins with initializing an instance of the BGN cryptosystem~\cite{bgn}, where a trusted authority sets up public and private keys, selects a hash function, and communicates with the server for encrypted traffic direction data handling. Drivers report traffic conditions to an RSU that encrypts and aggregates the data before sending it to the server for processing. The server decrypts this data to predict and manage future traffic flow, which is used to control infrastructure, such as traffic light scheduling. \textcolor{black}{VPTS requires $\sim$26ms per driver for encryption, commitment, and signing, and 1.565s on the RSU for 200 drivers, with a communication overhead of 0.071KB per driver and 0.051KB on RSU-side.}

Three approaches address privacy concerns in platoon formation and management. 
To enable clients to find and join nearby platoons without compromising their location privacy, Quero et al.~\cite{Quero2023TowardsPP} utilize the BFV~\cite{Brakerski2012FullyHE} and \textcolor{black}{CKKS~\cite{CKKS}}encryption schemes, which are FHE methods capable of performing multiple additions and a finite number of multiplications on encrypted data. Clients send encrypted platoon requests to the server, indicating their desired platoon location. The server responds with encrypted platoon identifiers, allowing clients to privately select and contact platoons. \textcolor{black}{Their experiments show that each plaintext-ciphertext multiplication takes 1.99ms and total client-server interaction is under 6ms. A single processor core can serve up to 500 clients per second, making the system scalable to 100,000 daily users with a 10--20 core server.
}

 To address the challenge of selecting reliable platoon leaders, Zhang et al.~\cite{Zhang2022TPPRAT} propose a trust-based platoon recommendation scheme called TPPR, which helps potential user vehicles avoid selecting malicious head vehicles. The TPPR uses AHE to ensure secure communication between the lead vehicle and other vehicles when joining a platoon. Once the trip ends, both the lead vehicle and the joining vehicle send their driving reports, such as handshake proof and trust value, to RSUs. The RSU verifies the joining vehicle's legitimacy and calculates the lead vehicle's reliability rating using trust score and feedback. The service provider evaluates the joining vehicle's performance and shares it with the trusted authority for forecasting future actions based on historical behavior. \textcolor{black}{TPPR aggregates ciphertexts for 100 users in 5.6ms and generates handshakes on vehicles in 18ms.}

Addressing another concern in platooning, Cheng et al.~\cite{Cheng2023PPRTPP} use HE to develop a recommendation system for vehicular platoons aimed at accurately calculating feedback about the lead vehicle's performance. Each vehicle encrypts its feedback score using additive homomorphic properties, ensuring that the platoon head vehicle's reputation can be computed on the aggregated encrypted data while maintaining individual score confidentiality. In the reputation score evaluation phase, AHE is applied to calculate the distances between encrypted feedback scores and to aggregate these encrypted values,  allowing the determination of reputation scores while preserving the individual feedback data. Finally, trusted authorities and servers use the output for a recommendation system for vehicular platoons. \textcolor{black}{Evaluation results show that AHE decryption takes 1.356ms (4 exponential operations) and encryption along with multiplication and hash operations run in 1.1673ms.}

\nmold{
\subsubsection{Summary}
In this section, we surveyed applications of MPC and HE in the client-to-server setting. The solutions address privacy concerns across a diverse range of services, from location-based applications like ride-sharing and POI queries to traffic management services like platoon formation. These implementations are typically designed to scale with multiple clients, either computing on encrypted data from multiple clients (HE) or processing client data in secret-shared form (MPC). 
In the next section, we see how these techniques develop further in  distributed settings.}

\subsection{Distributed Setting}
\label{section:distributed}
In this section, we discuss the distributed setting, which extends beyond the traditional client-to-server model introduced in Section~\ref{section:server}. In this setting, computational tasks are offloaded to a network of multiple servers, edge nodes, fog nodes, or other distributed resources.
Edge and fog computing processes data at or near the data generation source, minimizing latency without the need to transmit data to distant servers. A layered structure supports handling large-scale data-intensive applications in vehicular networks with thousands of clients.

\begin{figure}[t!]

    \centering
    \includegraphics[trim=0cm 8.6cm 0cm 0.2cm, clip,width=0.8\textwidth]{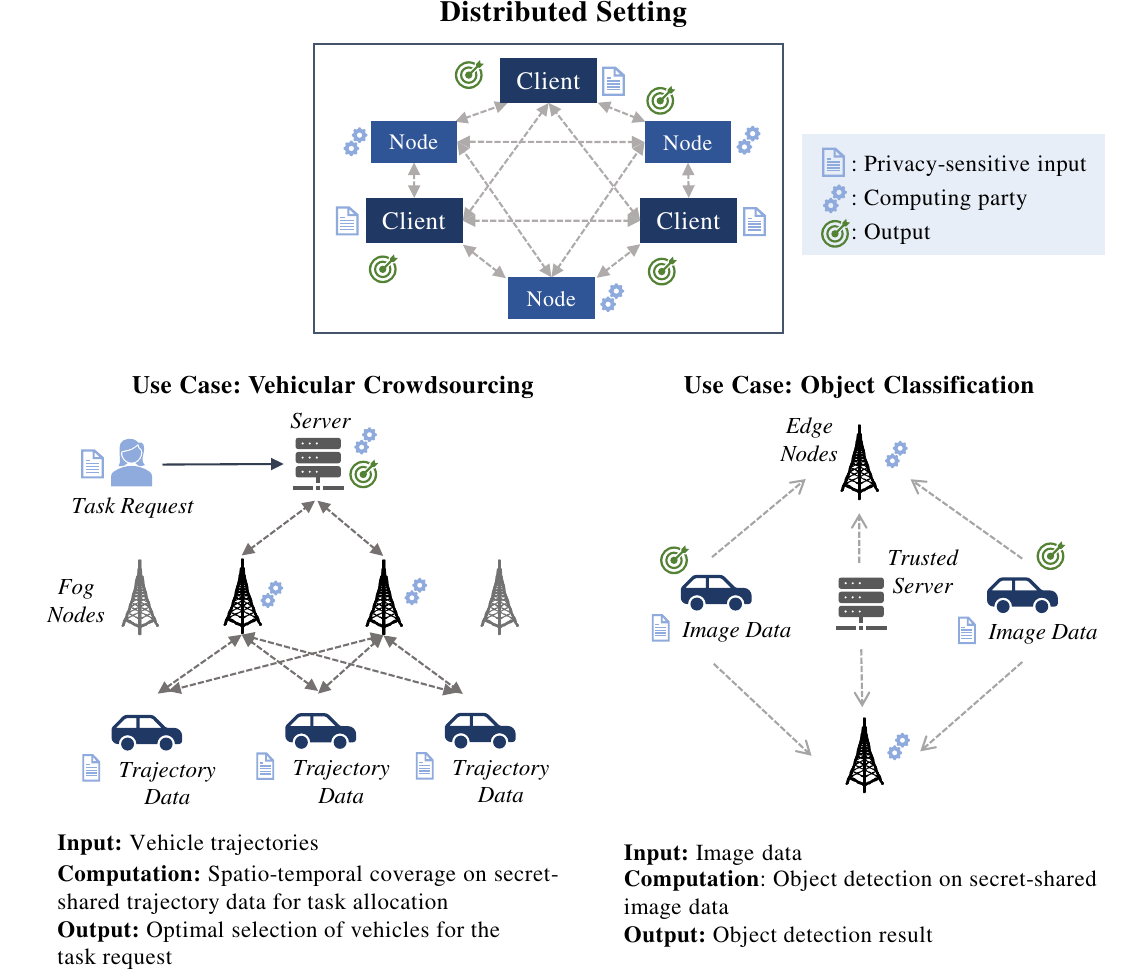}
    \caption{Generic distributed setting, with two specific use cases under the same setting as examples.  
In this setting, clients typically provide  their inputs to intermediate computing edge/fog nodes, with output used by all or some clients, or aggregated by a server.}
    \label{fig:setting-distributed}
\end{figure}

We illustrate two use cases within this distributed setting in Figure~\ref{fig:setting-distributed}. In the first example, vehicular crowdsourcing, vehicles share trajectory data to two non-colluding fog nodes. These nodes can run an MPC protocol with a server, to select suitable vehicles for a location-based task.
Compared to the vehicular crowdsourcing example in the client-to-server setting (see  Figure~\ref{fig:setting-client-to-server}), this implementation can \textcolor{black}{scale to} large number of computing nodes.
The second example (bottom right) involves object classification, where two non-colluding edge nodes receive image data from vehicles, and run computer vision inference in a 2-party MPC protocol, allowing for privacy-preserving identification of objects in the images.
Both examples illustrate how MPC and HE can support privacy in applications involving resource-intensive computations across a decentralized network. Below, we review existing research that applies these techniques to various distributed
use cases.

\subsubsection{Privacy in Location-Based Sevices.}
\nmold{
To protect location privacy in vehicle-based spatial crowdsourcing in IoV, Zhang et al.~\cite{junwei2021} propose a decentralized PriSC framework. Their scheme involves requesters (such as vehicle services and traffic management) encrypting sensitive location policies using AHE, 
while vehicle workers submit their encrypted locations through Order-Preserving Encryption (OPE)~\cite{ope} to a blockchain and verify their eligibility for the tasks. Workers also provide location proofs, enabling requesters to verify authenticity while preserving privacy, all recorded on the blockchain.  \textcolor{black}{PriSC incurs $\sim$1.7s computation per worker, $\sim$1.93s for encrypting 10 location policies with Paillier-1024, and $\sim$27.5ms for OPE-based location record generation with a $2^{40}$ plaintext space.}
Another approach is the PriTAEC scheme developed by Xu et al.~\cite{Xu2023PriTAECPT}, which utilizes OT and PEQT for securing task assignments. In this work, requesters and drivers submit their location data encoded through Hilbert curves and Bloom filters to edge nodes for range queries. In the proposed optimized model, edge nodes execute OT protocols with requesters during an offline phase, while drivers assist edge nodes to finalize the task. The edge computing framework helps reduce communication latency while preserving the location privacy. \textcolor{black}{Specifically, PriTAEC achieves task assignment within $\sim$53ms, where OT accounts for $\sim$12ms and PEQT for $\sim$40ms.}

Cheng et al.~\cite{Cheng2023ALP} address the task of reputation management in vehicle crowdsensing and propose a PPRM scheme. It is similar to the work by Peng et al.~\cite{Peng2024PrivacyPreservingTD} (discussed in Section~\ref{section:client-to-server-lbs}) but works in a different setting. To protect the privacy of sensing data and validate its authenticity, Cheng et al. use the Paillier algorithm for encryption and apply a comparison algorithm for Paillier ciphertext for data verification. Then, the cloud server transmits reputation feedback reports to the reputation center to efficiently update the sensing vehicles' reputation values.

\textcolor{black}{
Guan et al.~\cite{Guan2020AchievingPV} introduced a task allocation scheme for content dissemination in vehicular networks that uses SHE. In their approach, two non-colluding servers collaboratively select a certain number of vehicles to cover a near-optimal city area based on encrypted vehicle trajectory data. Although the scheme accounts for the high mobility of vehicles, processing large volumes of encrypted location data results in significant computational and communication overhead.}
Yu et al. ~\cite{Yu2024EfficientPT} address task allocation in fog computing using a grid-based region encoding, where users' locations or trajectories are encrypted into binary arrays called region codes, the scheme applies bit-wise XOR-based secret sharing to split and transmit the codes to two fog servers. The cloud server then collaborates with both fog servers to allocate tasks without revealing user locations.

\textcolor{black}{To address location privacy concerns in traffic navigation, Zhou et al.\cite{Zhou2021EPNSEP} propose a lightweight cryptographic primitive, Multiparty Delegated Computation (MPDC). It allows two non-colluding servers to perform secure addition, multiplication, and comparison over location data encrypted under different keys, providing similar privacy guarantees of MPC and FHE-based approaches with lower communication and computation overhead. Allowing a client to privately retrieve precomputed route segments without revealing which segment is being accessed, Tiausas et al.~\cite{Tiausas2023HPRoPHP} introduces HPRoP,  a PIR-based route planning system. More precisely, their approach provides privacy-preserving location and trajectory data processing by adding dummy queries alongside the real route planning request within a hierarchical road network structure.}

Another LBS addressed in this setting is vehicle sharing, where users can share their vehicles via digital keys, known as Access Tokens (AT).} HERMES, proposed by Symeonidis et al.~\cite{iraklis2022}, is a scalable and privacy-enhancing vehicular access system that extends SePCAR~\cite{iraklis2017} by utilizing MPC to manage vehicle ATs across non-colluding servers. In HERMES, vehicle keys and booking details are kept private, as each server holds only a share of the secret data. The system optimizes MPC protocols by using 
AES-CBC-MAC for Boolean circuits and HtMAC for arithmetic circuits, minimizing communication rounds and computational overhead. \textcolor{black}{This system allows for rapid AT generation in $\sim$30.30ms, managing  up to $546$ operation per second, while providing the scalability needed for real-world applications, such as rental companies overseeing 1000 vehicles.}

\subsubsection{Privacy in Mobility Infrastructures.}

As can be observed in Table~\ref{table:mpc}, most privacy-preserving solutions for infrastructure-based vehicle services align with distributed settings, involving interactions between vehicles, RSUs, servers, and blockchain networks. These approaches are particularly relevant for preserving privacy as vehicle data moves through various infrastructure layers, such as toll systems, parking facilities, and traffic management platforms.

One relevant example is TollsOnly, a privacy risk reduction model proposed by Karim and Rawat~\cite{karim22}. The model addresses the need for secure and controlled sharing of electronic toll transponder data with smart city infrastructure planners, to help manage traffic congestion while preserving driver's location privacy. TollsOnly employs FHE to perform computations on encrypted toll data. In addition, a blockchain-based mechanism grants users control over how and when their encrypted data is shared with authorized entities.

Parking services in a distributed setting involve interactions between vehicles, parking providers, and decentralized units such as RSUs to securely manage and process parking data while preserving driver privacy. Zhang et al.~\cite{Zhang2023BSDPBS} propose a blockchain-based smart parking scheme called BSDP to protect the privacy of Vehicular Sensor Network (VSN) participants and provide reliable data aggregation. The scheme employs the Paillier Cryptosystem with Threshold Decryption (PCTD) to securely encrypt and aggregate location and driving speed data from various VSN participants. Two adjacent RSUs collaborate to perform privacy-preserving data aggregation. Secure Hidden Vector Encryption (SHVE) is employed to handle encrypted location queries. \textcolor{black}{BSDP aggregates data in under 150ms when 100 vehicles and parking requests in 2.5ms per driver.}

Another blockchain-based approach for parking was proposed by Amiri et al.~\cite{Amiri2019PrivacyPreservingSP}. They leverage PIR to enable drivers to privately retrieve parking offers from a consortium blockchain, using Reed-Solomon codes to generate coded queries.
\textcolor{black}{Their evaluations show that  PIR incurs low communication overhead with $\sim$3.5KB and $\sim$1ms computation for parking reservation, making it practical for real-world deployment.} However, the complexity of PIR schemes generally increases with the number of nodes, which could impact scalability in very large networks.

With a similar concern, to protect the location privacy of drivers during the parking space detection and matching process, Li et al.~\cite{Li2021PriParkRecPD} introduced PriParkRec. It utilizes OPRF and PSI to enable drivers to match available parking spaces provided by a semi-honest parking service provider without revealing their exact location. Anonymous credentials allow users to authenticate without disclosing their identities, while AHE ensures confidentiality during aggregated data operations.

\subsubsection{Privacy in Vehicular Data Analysis.}
\label{distributed VDA}

\nmold{Many recent work address scalable vehicular data analysis through a distributed paradigm, largely to support privacy-preserving machine learning.
In particular,} Chen et al.~\cite{chenji2022} propose a Decentralized Privacy-preserving Deep Learning (DPDL) model for VANETs that aims to reduce network congestion and provide low-latency services. This decentralized approach shifts data processing from central cloud servers to Edge Computing (EC) nodes. 
The transportation data is encrypted using FHE before being input into local DPDL models on each EC node for training. \textcolor{black}{Evaluation results indicate low communication latency, rising from 287.2ms to 823.6ms as the number of vehicles scales from 25 to 150.}

\textcolor{black}{Liu et al.~\cite{trajectoryvanet} addressed real-time lane-changing trajectory prediction in VANETs based on SSS. Vehicles collect driving data and distribute it in secret-shared form to multiple RSUs, which collaboratively train an Adaboost~\cite{adaboost} algorithm. Secure sub-protocols allow RSUs to jointly compute model updates, error rates, and trajectory predictions.
}

\nmold{Li et al. \cite{li2022} focus on addressing the limitations related to resource overhead in cloud-assisted fog computing, that} involves multiple interactions between the cloud service center and the fog nodes, which can cause delays and increase resource overhead. They employ SSS to enable user gradients to split into multiple secret shares and distribute them among fog nodes. Specifically, the $(T, N)$ threshold property prevents collusion among up to $T-1$ fog nodes and supports up to $N-T$ fog nodes, where $N$ represents the number of fog nodes, and $T$ stands for the threshold value. Their work distinguishes itself by enabling computation across an unlimited number of nodes $N$, offering greater scalability than other MPC methods that are typically limited to 2--4 nodes.

Hu et al.~\cite{hu2023} propose DSSFL, a DFL-based data-sharing scheme for the IoV, integrating SSS and HE.
In this approach, vehicles fragment their local model parameters using SSS, distributing these encrypted fragments to multiple RSUs. The use of HE (specifically, the CKKS scheme) allows RSUs to perform secure aggregation on these encrypted fragments.
The results are then sent back to vehicles, which use Lagrange interpolation to recover the global model securely. Although HE adds computational overhead, 
 the model still achieves a high accuracy of approximately $86\%$ due to efficient data handling and secure aggregation methods. \textcolor{black}{DSSFL uses SSS with a polynomial degree of 5, requiring at least 6 out of 10 RSUs for model reconstruction. Each vehicle sends 10 encrypted fragments per round.
 Model training converges in $\sim$9 minutes over 20 training rounds.}

Li et al.~\cite{Li2021PrivacyPreservedFL} present a FL framework for autonomous vehicles based on HE and ZKPs to protect model updates from both semi-honest servers and potentially malicious vehicles. 
HE enables vehicles to share encrypted model updates securely without disclosing raw data, while ZKP  ensures anonymous identity verification. This dual approach improves model accuracy and reduces training loss, but it faces latency challenges in high-mobility environments that could benefit from further optimization for real-time performance.
\nmold{A common limitation in privacy-preserving FL schemes is the challenge of handling new users and user dropouts~\cite{bonawitzfl,verify}.} Kong et al.~\cite{kong2021} propose a privacy-preserving aggregation scheme for FL in vehicular fog computing. Their approach combines SSS with a homomorphic threshold encryption scheme to ensure that client data remains private during model aggregation. SSS enables the system to tolerate user dropouts by establishing a minimum threshold for the number of user shares required for aggregation.

Beyond general 
solutions for privacy-preserving machine learning and FL, several works focus on concrete applications for smart and autonomous vehicles.
Xiong et al.~\cite{jinbo2020} propose an edge-assisted framework for privacy-preserving object classification.
In this ASS-based approach, a vehicle splits an image into shares for two non-colluding edge servers, which
cooperatively
process them with a deep learning model.
The same authors in~\cite{xiong2022} introduce a refined lightweight ASS-based scheme with enhanced security using chaotic map encryption and proposing a multi-party extension to tolerate offline servers.
 \textcolor{black}{Bi et al.~\cite{Bi2023AchievingLA} extend the approach to full object detection with their P2OD framework, which implements a secure equivalent of Faster R-CNN to privately compute both object features and their bounding boxes.}
\nm{\textcolor{black}{These works show that it is possible to offload intensive computer vision tasks from vehicles to edge servers in a privacy-preserving manner.}}

Kong et al.~\cite{prevmain} propose a privacy-preserving scheme based on HE and DP for continuous data collection in vehicular fog to implement predictive maintenance, aiming to detect the anomalies of vehicles and offer early warnings in ITSs.
In this scheme, the Paillier cryptosystem is used to encrypt individual sensory data pieces, facilitating the secure aggregation of multiple data reports directly at the fog nodes before sending these encrypted data to the cloud server. DP is applied by adding noise to each aggregated result, making it difficult to infer individual data points.

Gyawali et al.~\cite{sohan2021} proposes a privacy-preserving misbehavior detection system in VANETS. In this work, vehicles evaluate messages from neighboring vehicles and send weighted, encrypted feedback scores to the Local Authority (LA), utilizing a modified ElGamal cryptosystem. The LA aggregates these encrypted scores without accessing the individual values and then forwards the aggregated result to the Trusted Authority (TA). The TA decrypts this result to update the vehicle reputation scores. The authors emphasize that in contrast to the Paillier cryptosystem, which causes considerable delays, the lightweight and efficient ElGamal-based encryption, with additive homomorphic properties, is better suited for misbehavior detection systems. Additionally, implementing batch verification in a bilinear system further reduces delays  \textcolor{black}{to 8.23ms per encrypted feedback at the vehicle-side and 7.61ms for decryption and verification}.

\subsubsection{Privacy in Dynamic Traffic Management and V2X Communications.}

Energy storage sharing involves challenges in maintaining the privacy of clients' energy demands.
To address these issues, Wang et al.~\cite{Wang2021PrivacyPreservingES} propose a solution combining blockchain and SSS, enabling secure service scheduling without revealing individual users' demands. Users share their individual energy demands in a secret-shared way and generate ZKPs to verify the consistency of their commitments before calculating the energy storage service schedule and agreeing on payments. They then submit these payments to the ledger with proofs, execute the service, and request settlement, with the operator verifying transactions through signed receipts on the ledger.

Ying et al.~\cite{Ying2022PrivacySignalPT} propose PrivacySignal to address the vulnerability in traffic control systems where the transmission of vehicle data, such as location and speed, potentially leads to privacy breaches. In this system, vehicles divide their location and speed data into secret shares that are processed by RSUs.
To demonstrate the feasibility of PrivacySignal in the multi-party setting, the authors assert that its sub-protocols, such as secure addition, multiplication, and comparison, are MPC-
compatible,
Indeed, %
indicating the system's potential for scaling to more RSUs in practical~implementations. \textcolor{black}{PrivacySignal incurs a runtime overhead ranging from 0.003s to 0.672s and a communication overhead between 4.9KB and 45.8KB across its secure sub-protocols, by employing ASS protocols, making it suitable for real-time applications.}
 \textcolor{black}{Recently, Adelipour et al.~\cite{2025VulnerabilityMO} proposed another approach to traffic signal control systems to enable secure green time durations, i.e., the time intervals during which specific traffic lights remain green, in urban networks. To achieve this, they use SSS to distribute real-time traffic data across multiple shares, allowing semi-honest servers to jointly compute green time signals without accessing the raw data. 
}

\nmold{
\subsubsection{Summary}
In this section, we reviewed applications of HE and MPC in distributed settings. These implementations typically leverage multiple computing nodes to process data closer to the source. The solutions aim to support large dynamic vehicle fleets and address diverse use cases, from scalable location-based services to real-time data analysis. We observe increasing adoption of hybrid approaches combining multiple privacy-preserving technologies. While distributed architectures offer improved scalability compared to traditional client-server models, they introduce additional
complexity and require careful privacy analysis.
 }
 
\section{Takeaways and Discussion}
\label{section:discussion}

In this section, we review the applicability of MPC and HE in the automotive domain, based on the analysis in the previous Section~\ref{section:mpc}, and highlight several challenges in applying these technologies.

\subsection{Primitives}

\nmold{We observe that both MPC and HE are widely applied to secure computing scenarios in the automotive domain, and are useful across a variety of use cases. MPC and HE are sometimes employed interchangeably within the same use cases, offering similar, though not identical, privacy guarantees and resource requirements. For instance, ride-sharing and vehicular crowdsourcing scenarios show applications of both HE and MPC-based solutions.}

 \nmold{In particular, AHE and SHE are frequently adopted across ride-sharing, platooning, and crowdsourcing applications, utilizing privacy-preserving arithmetic operations. While FHE theoretically supports more complex operations, it is currently less suited for multi-client 
 collaborative %
 scenarios, as it typically requires data to be encrypted under the same key.
 Multi-key FHE schemes provide a promising alternative by enabling computations on data encrypted by different parties with their individual keys~\cite{Mukherjee2016TwoRM}. However, many such schemes are limited to working over single-bit ciphertexts and may not be suitable for encrypting large datasets~\cite{Yuan2022AnEO}.
 A notable challenge for AHE and SHE-based solutions is the increased size of encrypted messages, which leads to higher computational and bandwidth demands. Nonetheless, several reviewed optimized approaches like hybrid PSI and SHE schemes \cite{Hallgren2017}, and \textcolor{black}{ ciphertext-packing techniques~\cite{packedciphertext}} help reduce the payload sizes, reducing the computational and communication overhead when handling large-scale systems.}

MPC approaches also demonstrate their utility in the automotive domain, particularly those based on Secret Sharing (SS). SS-based non-linear computation requires more communication rounds compared to HE and GC-based approaches. However, they are advantageous for linear computations, having relatively low costs during the input-independent computation in the online phase~\cite{Zhang2021PrivacyPreservingDL}.
These approaches prove particularly effective in outsourced settings, such as Federated Learning~\cite{kong2021,Li2021PrivacyPreservedFL} and vehicular crowdsourcing~\cite{Yu2024EfficientPT}, where clients can upload their privacy-sensitive data to external servers for computation, enabling clients to remain offline between iterations and reducing computational overhead. In non-outsourced scenarios such as EV charging~\cite{Huo2021DistributedPE,Wang2021PrivacyPreservingES}, speed advisory~\cite{liu2022}, ride-sharing~\cite{Karmakar2024QuickPoolPR} and vehicle sharing~\cite{iraklis2017,iraklis2022}, secret sharing can be employed to protect sensitive information, while clients need to be online during data exchange process.

In contrast, there are only a few GC-based MPC approaches to privacy issues in the automotive domain. For instance, we can see several GC-based solutions for vehicle localization and ride-sharing scenarios. 
In the localization use case~\cite{siam2018}, we can see the comparison-based nonlinear functions can be efficiently implemented in a two-party setting utilizing GC. However, considering the large number of clients in ride-sharing, GC protocols incur high communication and computation overhead. For instance, Luo et al.~\cite{pRidePR2019} used GC to lightweight the distance comparison computation between rider and driver; however, it is not practical for real-world ride-sharing applications.
 
We observe that several papers utilize PSI
in distinct automotive scenarios
with matching as main function: %
parking~\cite{Li2021PriParkRecPD}, ride-sharing\cite{Avodji2016MeetingPI,Hallgren2017,toppool2019,Zhang2023PrivacypreservingOR,Karmakar2024QuickPoolPR}, profile matching~\cite{Wang2021ObliviousTF}, and POIs~\cite{Zhang2022APP}.
The intersection of multiple sets can be calculated iteratively by performing pairwise intersections.
Extending the two-party PSI protocol to a multi-party setting in certain  automotive use cases may not be straightforward.
However, there are several works on Multi-Party PSI (MPSI)
\cite{Bay2022PracticalMP,Zhang2019EfficientMP}. 
Although MPSI comes with higher communication overhead and complexity, it reduces the limitations of PSI among a larger group of participants~\cite{Escalera2023PrivateSI}. Despite the limitations of both techniques, considering the trade-offs, MPSI might be a new research direction for automotive scenarios such as ride-sharing or spatial crowdsourcing.

Our study shows that both MPC and HE are effective for implementing privacy-preserving aggregation of sensitive vehicular data. In particular, integrating MPC with FL enables model training on client data without exposing individual model parameters. In an FL setup, the communication of model updates can be done via MPC, which enhances the privacy of the system. Similarly, HE can also be a suitable choice for privacy-preserving aggregation in respective scenarios. HE efficiently supports the arithmetic addition, which aligns well with the frequent requirement in vehicular data analysis and FL to compute aggregated averages based on the client's data.

Our analysis reveals limited adoption of MPC and HE in client-to-client settings, such as in V2V/V2X contexts, primarily due to the requirements for dynamic, real-time communication. Recent developments in efficient two-party MPC protocols~\cite{Lu2024MaliciouslySM, Friedman20242PCMPCET} offer promising solutions to enhance the applicability of these use cases. In contrast, we observe an increasing number of works addressing scalable and distributed settings. These complex scenarios often require not only privacy-preserving computation but also verification of client inputs, achieved through  PVSS~\cite{Stadler1996PubliclyVS} as demonstrated in~\cite{bdfl2021}, or through MPC protocols with authenticated inputs~\cite{Dutta2022ComputeBV}. For server-side vehicular data analysis applications, additional privacy measures for protecting computation outputs become relevant, such as combining MPC and HE with differential privacy techniques~\cite{Sebert2022ProtectingDF,Wang2021PrivacyPreservingES}.

\subsection{Performance}
\label{performance}

\textcolor{black}{
Given the diversity of system settings, protocols, and evaluation metrics, a direct performance comparison across all surveyed papers is not straightforward.  The surveyed papers span a wide spectrum, from highly performant, lightweight systems to computation or communication-heavy systems, which makes it challenging to draw generalized performance evaluations.}

\textcolor{black}{
Direct comparisons are possible for specific applications such as ride-sharing. For example, in the client-to-server setting, the SHE-based ORide~\cite{ORide2017} approach shows high computation overhead from 
0.2s to $\sim$114s
for different algorithms on the server side. This overhead was later reduced to 37.13s in PRIS~\cite{yuan2018} using AHE and bilinear pairing. SRide~\cite{marie2018} introduces a more efficient two-stage approach with 0.519s for match computation (SHE-based SS) and 0.005s for equality testing, requiring 62KB and 31KB of communication for riders and drivers, respectively, with 1000 drivers. More recent works further improve efficiency: pRide2~\cite{pRidePR2019} achieves 0.0051s rider-side and 9.1s server-side runtime with only 256 bytes of communication
per user;
EPRide~\cite{Yu2022EfficientAP} reaches $\sim$0.0006s client-side runtime and 8.67MB server-side communication for 2000 taxis; PGRide~\cite{yu2021} supports group matching with $\sim$1.25s server runtime and less than 3.2 MB communication for 2000 drivers; and PSRide~\cite{yuha2021} enables dynamic scheduling with $\sim$1.5s server runtime and 2.3 MB communication for 6000 taxis.  
P$^2$Ride  ~\cite{Luo2023P2RidePA} replaces GC with non-interactive PEQT, reducing rider-side computation from 58.6s to 0.47s and driver-side from 60.1s to 4.7s, with 784 bytes and 9.8KB communication, respectively. 
In contrast, client-to-client protocols show worse performance. The PrivatePool~\cite{Hallgren2017} exhibits high runtime with trajectory size (0.022s for 32 segments to 96.78s for 1024), whereas TOPPool~\cite{toppool2019} reduces this to less than 0.31s.
Aïvodji et al.~\cite{marie2018} report runtimes between 0.48–0.67s for PSI-based matching.}
\textcolor{black}{Overall, client-to-server solutions offer better scalability and lower per-client latency, while client-to-client approaches are better suited for small-scale setups.}

\nm{\textcolor{black}{
Although the ride-sharing shows a clear trend towards efficiency, applications that involve large-scale or data-intensive computations still exhibit significant performance overheads.
For example, end-to-end object classification by Xiong et al.~\cite{xiong2022} requires over 20s and 327~MB of communication, while  
the object detection task by Bi et al.~\cite{Bi2023AchievingLA} takes 190s and over 6~GB for a single detection. Similarly, the HPRoP algorithm for route planning requires $\sim$23.55s to compute a complete route, a notable latency despite being an improvement over prior work.
}}

\nm{\textcolor{black}{Overall, our analysis shows that MPC and HE are feasible in
 scenarios
that do not require continuous real-time updates and can tolerate moderate latency. In addition, recent protocol optimizations and hybrid approaches have reduced overheads and improved scalability, particularly in client–to–server and distributed settings. However, MPC and HE likely remain infeasible in use cases with large numbers of clients, frequent real-time interactions, and complex system design.}}

\subsection{Datasets}
 We observe that several
 surveyed works
 evaluate their solutions on real-world datasets while others use simulated data only
 (see Table~\ref{table:mpc}). We also observe that only a few works perform an evaluation on relatively large datasets containing hundreds of thousands of records or more~\cite{ORide2017,junwei2021,Hallgren2017,Luo2023P2RidePA,Li2021PrivacyPreservedFL,Quero2023TowardsPP}.
 \nm{\textcolor{black}{We believe that only the evaluation on real-world
 datasets (recently surveyed in \cite{Bari2025DatasetsIV, Liu2024ASO})
can validate the applicability of the
 solutions, while the large-scale evaluation can prepare the solutions for deployment in large national fleets.
The release of new large-scale
datasets, and unified}}
evaluation
benchmarks for HE and MPC based on them, would provide great value for privacy research in automotive applications, as authors could evaluate their works in realistic settings. 

\subsection{Security Model}
We find that most MPC and HE works in our paper assume a semi-honest security model, where participating parties honestly follow a protocol. However, in practical applications, real-world threats often come from malicious attackers. Therefore, it is important to study the performance of protocols that offer protection against actively malicious adversaries. Several works (see Table~\ref{table:mpc})
demonstrate the applicability of malicious security models in the automotive domain. Protocols designed under the assumption of semi-honest adversaries can often be modified to protect against malicious adversaries. However, this modification significantly increases cost, often to levels impractical for real-world applications.
Overall, the extension of MPC to larger-scale applications remains a challenge, particularly under a fully malicious security model~\cite{Evans2019API}.

\section{Future Directions}
\label{section: future}

Although extensive research has already addressed many challenges related to privacy issues in the automotive domain, there are still several key research problems that need further exploration. We highlight several key areas below to inspire future directions.

  \subsubsection*{Unaddressed Use Cases.}
  Our analysis in Sections~\ref{section:use-cases}-\ref{section:mpc} reveals several privacy-sensitive automotive use cases that remain unaddressed by either MPC or HE, 
  \nm{\textcolor{black}{including traffic anomaly detection, road profile estimation, vehicle emission control, etc.}} Addressing these use cases is a clear opportunity for future work. Furthermore, we observe that several use cases leverage HE but not MPC, such as predictive maintenance, misbehavior detection, platooning, etc. Similarly, some use cases employ MPC but lack HE-based solutions, e.g., querying points of interest, localization, EV charging, etc. Given that MPC and HE
  can be applied interchangeably in many scenarios, exploring alternative implementations of existing solutions using the other technology, and comparing their respective requirements and guarantees, could offer valuable insights for automotive applications.%

  \subsubsection*{Baseline Implementations.}
The surveyed 
research works typically propose comprehensive solutions addressing multiple requirements for specific use cases. For instance, numerous ride-sharing solutions are proposed, each with different complexity levels and privacy guarantees. However, the field often lacks baseline implementations of fundamental MPC and HE protocols for common automotive scenarios. Proof-of-concept implementations, evaluated on sample data, would enable systematic evaluation of various approaches. 
Such implementation
would help researchers assess how different requirements (e.g., moving from semi-honest to malicious security models) affect system performance, and determine the applicability of technologies in corresponding use cases.%

  \subsubsection*{Real-World Datasets and Benchmarks.}
    Existing works mostly evaluate their approaches to simulated vehicular data.  We need publicly available real-world datasets to improve the accuracy and applicability of privacy-preserving solutions in real-world automotive scenarios.
    Similarly, 
    one of the future directions can be focusing on establishing standard benchmarks with unified datasets and metrics, allowing for the comparison of performance and scalability across privacy-preserving solutions for similar automotive use cases and settings.

      \subsubsection*{Security Model.}

 \textcolor{black}{The security model in privacy-preserving approaches needs to be strengthened to ensure robust solutions by moving from semi-honest to malicious security in practical applications. Recent advances in MPC demonstrate that malicious security can be achieved efficiently using techniques such as authenticated secret sharing (e.g., SPDZ-style IT-MACs) and optimized preprocessing protocols (e.g., MASCOT) \cite{Feng2022ConcretelyES}.
 Applying and benchmarking maliciously secure MPC protocols in privacy-sensitive vehicular use cases remains an open and promising direction.}

    \subsubsection*{Dynamic Data.}  Most existing approaches omit client dropouts and focus on static vehicular data processing. The development of flexible, dynamic datasets and solutions that can cope with dynamic users might be a future research direction that needs to be addressed. 

    \subsubsection*{Broader Exploration of Use Cases and Privacy-Preserving Technologies.} In Section~\ref{section:use-cases}, our work specifically reviewed automotive use cases that were explicitly identified as privacy-sensitive or implemented privacy-preserving solutions to analyze the applicability of HE and MPC in those. Future research could expand this scope to systematically analyze a broader range of automotive use cases, identify \textit{new} privacy-sensitive use cases, and study suitable solutions. Finally, while our work concentrated on MPC and HE, similar systematic analysis could be conducted for other PETs, such as DP, anonymization techniques, zero-knowledge proofs, or a combination of those.

\subsubsection*{\nm{\textcolor{black}{Alignment with Privacy Regulations.}}}
\nm{\textcolor{black}{
A promising future direction includes aligning
solutions with privacy regulations such as GDPR and CCPA. We find that only eight works surveyed in Section~\ref{section:mpc} discuss the relation to such frameworks:
some suggest their solutions help minimize data disclosure to ensure  compliance \cite{Hallgren2017, toppool2019}; others highlight that regulations complicate use case adoption without such solutions \cite{Xu2022AnEA, liu2022}, or serve as a general motivation for applying PETs in considered scenarios \cite{Zhang2020VerifiableAP, karim22, Wang2021PrivacyPreservingES, 2025VulnerabilityMO}. Future work can include a detailed legal assessment of the proposed schemes. Recent analysis shows MPC's classification as a GDPR-compliant anonymization technique depends on the deployment setting, such as the legal relationship between the parties computing on the secret shares~\cite{legalBecker2025MultiPartyCI}. Future research can evaluate which automotive settings (e.g., specific client-to-client or distributed configurations) best align with such regulations. Furthermore, research could explore how complementary technologies, such as applying differential privacy to computation outputs, can enhance both the technical privacy guarantees and the solution's legal standing.
}}

\subsubsection*{\nm{\textcolor{black}{Application to Existing Automotive Standards.}}}
\textcolor{black}{
Most surveyed papers frame their solutions in the V2X or CAVs context; however, they rarely align with established automotive standards such as ISO/SAE 21434, AUTOSAR, or
IEEE 802.11p.
Some works reference these standards and incorporate related simulation parameters such as message size, latency, or communication range \cite{siam2018, Liang2022PPRPPR,Zhang2022TPPRAT,Cheng2023PPRTPP,chenji2022,Li2021PrivacyPreservedFL}.
Others mention standards such as IEEE 802.11p or AUTOSAR only at a conceptual level,
without integrating them into the system architecture or evaluation \cite{Magaiga2018,Kong2017,Zhou2024PrivacypreservingAV,sohan2021}. Future research direction should align 
proposed solutions with these regulations and evaluate them on automotive hardware (e.g., ECUs, embedded SoCs) under realistic
network conditions.
}

\subsubsection*{\textcolor{black}{Formal Analysis.}}
\nm{\textcolor{black}{We observe that the surveyed works
offer  different levels of security analysis. Some works provide formal, simulation-based or game-based proofs (e.g., \cite{Hallgren2017, Karmakar2024QuickPoolPR}). Other works (e.g.,\cite{siam2018, marie2018}) argue for their security 
based on the security of their underlying cryptographic primitives without providing a formal proof for the complete system.
Vulnerabilities (e.g., ~\cite{Vivek2021AttacksOA, Kumaraswamy2021RevisitingDA, PassiveTAoride}) found in some schemes
(see Section~\ref{section:client-to-server-lbs})
further demonstrate the need for a rigorous security analysis.
A key direction for future research is therefore to perform comprehensive security analysis for proposed solutions.
For use cases with multiple privacy-preserving solutions, such as ride-sharing or vehicular crowdsourcing, it would be beneficial to perform a \textit{comparative} analysis of the guarantees offered by different approaches.
Strengthening the formal analysis will improve trust and facilitate the adoption of PETs in the automotive domain.
}}

 \subsubsection*{\textcolor{black}{Recent Advancements in PETs}}

\nm{\textcolor{black}{Finally, we point out several developments in the fields of PETs, outside the automotive domain, that can be transferrable to vehicular applications.}}

\nm{\textcolor{black}{Significant work has been done on improving MPC and HE with hybrid protocols for deep learning applications. 
Specifically, combining different cryptographic primitives, such as HE for linear operations and SS or GCs for non-linear functions, can significantly reduce computation and communication overhead compared to single-primitive solutions \cite{Zhang2021PrivacyPreservingDL, Gamiz2024Challenges}. The vehicular data analysis papers we surveyed mainly rely on a single primitive. Future work can explore adapting these hybrid approaches for vehicular use cases, such as trajectory prediction, object classification, or predictive maintenance.}}
\textcolor{black}{Furthermore, the choice of MPC protocol should align with automotive network characteristics. As discussed in recent works~\cite{Feng2022ConcretelyES,Ng2023SoKC},
SS-based approaches are better suited for low-latency settings such as edge-based local area network (LAN). In contrast, GC protocols and 
non-interactive HE-based protocols are often more appropriate for wide area networks (WANs).}

\textcolor{black}{Beyond protocol design, the field is advancing with the development of usable compilers and frameworks (such as CrypTen, EzPC, MP-SPDZ) that abstract cryptographic complexity and automatically translate high-level code and DL interfaces into optimized MPC protocols~\cite{Ng2023SoKC,mpspdz}. These advancements can simplify and accelerate adoption in vehicular applications.}

\textcolor{black}{Finally, research in secure aggregation for FL has evolved
to include integrity and verification mechanisms, and ensure robustness against malicious participants~\cite{SoKSA}. 
In addition, complementary research explores secure FL architectures in distributed settings such as IoT networks~\cite{WhenFL,SecuringFL}, software defined networks~\cite{mitfed} and wireless sensor networks~\cite{fl3rev}. These works may offer transferable insights to collaborative vehicular applications such as FL-based navigation, misbehavior detection, object classification, and predictive maintenance.}

\section{Conclusion}
\label{section:conclusion}

In this paper, we offer a thorough analysis of current MPC and HE applications in the automotive domain. First, we identified and categorized a set of privacy-sensitive use cases relevant to modern automotive architectures and setups.
The privacy use cases they examine mainly focus on privacy in the contexts of location-based services, mobility infrastructure, vehicular data analysis, and dynamic traffic management. Second, we studied existing works applying MPC and HE to the selected privacy-related use cases in detail. Based on our comprehensive analysis, existing MPC and HE applications in privacy-sensitive automotive scenarios offer promising directions for research and development toward privacy-preserving, deployable, and scalable computing approaches in the automotive domain. Finally, we highlight areas for future research in this field.

\bibliographystyle{ACM-Reference-Format}
\bibliography{main}


\begin{thebibliography}{150}


\ifx \showCODEN    \undefined \def \showCODEN     #1{\unskip}     \fi
\ifx \showDOI      \undefined \def \showDOI       #1{#1}\fi
\ifx \showISBNx    \undefined \def \showISBNx     #1{\unskip}     \fi
\ifx \showISBNxiii \undefined \def \showISBNxiii  #1{\unskip}     \fi
\ifx \showISSN     \undefined \def \showISSN      #1{\unskip}     \fi
\ifx \showLCCN     \undefined \def \showLCCN      #1{\unskip}     \fi
\ifx \shownote     \undefined \def \shownote      #1{#1}          \fi
\ifx \showarticletitle \undefined \def \showarticletitle #1{#1}   \fi
\ifx \showURL      \undefined \def \showURL       {\relax}        \fi
\providecommand\bibfield[2]{#2}
\providecommand\bibinfo[2]{#2}
\providecommand\natexlab[1]{#1}
\providecommand\showeprint[2][]{arXiv:#2}

\bibitem[Acar et~al\mbox{.}(2018)]%
        {acar2018survey}
\bibfield{author}{\bibinfo{person}{Abbas Acar}, \bibinfo{person}{Hidayet Aksu}, \bibinfo{person}{A~Selcuk Uluagac}, {and} \bibinfo{person}{Mauro Conti}.} \bibinfo{year}{2018}\natexlab{}.
\newblock \showarticletitle{A survey on homomorphic encryption schemes: Theory and implementation}.
\newblock \bibinfo{journal}{\emph{ACM Computing Surveys (Csur)}} \bibinfo{volume}{51}, \bibinfo{number}{4} (\bibinfo{year}{2018}), \bibinfo{pages}{1--35}.
\newblock


\bibitem[Adelipour et~al\mbox{.}(2025)]%
        {2025VulnerabilityMO}
\bibfield{author}{\bibinfo{person}{Saeed Adelipour}, \bibinfo{person}{Enayatollah Amiri~Darreh Razgahi}, {and} \bibinfo{person}{Mohammad Haeri}.} \bibinfo{year}{2025}\natexlab{}.
\newblock \showarticletitle{Vulnerability Mitigation of Urban Traffic Control Against Cyberattacks Using Secure Multi-Party Computation}.
\newblock \bibinfo{journal}{\emph{IEEE Transactions on Intelligent Transportation Systems}}  \bibinfo{volume}{26} (\bibinfo{year}{2025}), \bibinfo{pages}{4568--4578}.
\newblock


\bibitem[Agrawal et~al\mbox{.}(2004)]%
        {ope}
\bibfield{author}{\bibinfo{person}{Rakesh Agrawal}, \bibinfo{person}{Jerry Kiernan}, \bibinfo{person}{Ramakrishnan Srikant}, {and} \bibinfo{person}{Yirong Xu}.} \bibinfo{year}{2004}\natexlab{}.
\newblock \showarticletitle{Order preserving encryption for numeric data}. In \bibinfo{booktitle}{\emph{Proceedings of the 2004 ACM SIGMOD International Conference on Management of Data}} (Paris, France) \emph{(\bibinfo{series}{SIGMOD '04})}. \bibinfo{publisher}{Association for Computing Machinery}, \bibinfo{address}{New York, NY, USA}, \bibinfo{pages}{563–574}.
\newblock
\showISBNx{1581138598}


\bibitem[A\"{\i}vodji et~al\mbox{.}(2016)]%
        {Avodji2016MeetingPI}
\bibfield{author}{\bibinfo{person}{Ulrich~Matchi A\"{\i}vodji}, \bibinfo{person}{Sébastien Gambs}, \bibinfo{person}{Marie-José Huguet}, {and} \bibinfo{person}{Marc-Olivier Killijian}.} \bibinfo{year}{2016}\natexlab{}.
\newblock \showarticletitle{Meeting points in ridesharing: A privacy-preserving approach}.
\newblock \bibinfo{journal}{\emph{Transportation Research Part C: Emerging Technologies}}  \bibinfo{volume}{72} (\bibinfo{year}{2016}), \bibinfo{pages}{239--253}.
\newblock
\showISSN{0968-090X}


\bibitem[A\"{\i}vodji et~al\mbox{.}(2018)]%
        {marie2018}
\bibfield{author}{\bibinfo{person}{Ulrich~Matchi A\"{\i}vodji}, \bibinfo{person}{K\'{e}vin Huguenin}, \bibinfo{person}{Marie-Jos\'{e} Huguet}, {and} \bibinfo{person}{Marc-Olivier Killijian}.} \bibinfo{year}{2018}\natexlab{}.
\newblock \showarticletitle{SRide: A Privacy-Preserving Ridesharing System}. In \bibinfo{booktitle}{\emph{Proceedings of the 11th ACM Conference on Security \& Privacy in Wireless and Mobile Networks}} (Stockholm, Sweden) \emph{(\bibinfo{series}{WiSec '18})}. \bibinfo{publisher}{Association for Computing Machinery}, \bibinfo{address}{New York, NY, USA}, \bibinfo{pages}{40–50}.
\newblock
\showISBNx{9781450357319}


\bibitem[Amiri et~al\mbox{.}(2019)]%
        {Amiri2019PrivacyPreservingSP}
\bibfield{author}{\bibinfo{person}{Wesam~Al Amiri}, \bibinfo{person}{Mohamed Baza}, \bibinfo{person}{Karim~A. Banawan}, \bibinfo{person}{Mohamed Mahmoud}, \bibinfo{person}{Waleed~S. Alasmary}, {and} \bibinfo{person}{Kemal Akkaya}.} \bibinfo{year}{2019}\natexlab{}.
\newblock \showarticletitle{Privacy-Preserving Smart Parking System Using Blockchain and Private Information Retrieval}.
\newblock \bibinfo{journal}{\emph{2019 International Conference on Smart Applications, Communications and Networking (SmartNets)}} (\bibinfo{year}{2019}), \bibinfo{pages}{1--6}.
\newblock


\bibitem[Aziz et~al\mbox{.}(2019)]%
        {aziz2019privacy}
\bibfield{author}{\bibinfo{person}{Md~Momin~Al Aziz}, \bibinfo{person}{Md~Nazmus Sadat}, \bibinfo{person}{Dima Alhadidi}, \bibinfo{person}{Shuang Wang}, \bibinfo{person}{Xiaoqian Jiang}, \bibinfo{person}{Cheryl~L Brown}, {and} \bibinfo{person}{Noman Mohammed}.} \bibinfo{year}{2019}\natexlab{}.
\newblock \showarticletitle{Privacy-preserving techniques of genomic data—a survey}.
\newblock \bibinfo{journal}{\emph{Briefings in bioinformatics}} \bibinfo{volume}{20}, \bibinfo{number}{3} (\bibinfo{year}{2019}), \bibinfo{pages}{887--895}.
\newblock


\bibitem[Bari et~al\mbox{.}(2025)]%
        {Bari2025DatasetsIV}
\bibfield{author}{\bibinfo{person}{Bifta~Sama Bari}, \bibinfo{person}{Deepak Puthal}, {and} \bibinfo{person}{Kumar Yelamarthi}.} \bibinfo{year}{2025}\natexlab{}.
\newblock \showarticletitle{Datasets in Vehicular Communication Systems: A Review of Current Trends and Future Prospects}.
\newblock \bibinfo{journal}{\emph{SN Comput. Sci.}}  \bibinfo{volume}{6} (\bibinfo{year}{2025}), \bibinfo{pages}{210}.
\newblock


\bibitem[Bautista et~al\mbox{.}(2022)]%
        {Bautista2022PrivacyAwareVE}
\bibfield{author}{\bibinfo{person}{Pablo Andr{\'e}s~Barbecho Bautista}, \bibinfo{person}{Luis~Felipe Urquiza-Aguiar}, {and} \bibinfo{person}{M{\'o}nica Aguilar-Igartua}.} \bibinfo{year}{2022}\natexlab{}.
\newblock \showarticletitle{Privacy-Aware Vehicle Emissions Control System for Traffic Light Intersections}.
\newblock \bibinfo{journal}{\emph{Proceedings of the 19th ACM International Symposium on Performance Evaluation of Wireless Ad Hoc, Sensor, \& Ubiquitous Networks}} (\bibinfo{year}{2022}).
\newblock


\bibitem[Bay et~al\mbox{.}(2022)]%
        {Bay2022PracticalMP}
\bibfield{author}{\bibinfo{person}{Aslı Bay}, \bibinfo{person}{Zekeriya Erkin}, \bibinfo{person}{Jaap-Henk Hoepman}, \bibinfo{person}{Simona Samardjiska}, {and} \bibinfo{person}{Jelle Vos}.} \bibinfo{year}{2022}\natexlab{}.
\newblock \showarticletitle{Practical Multi-Party Private Set Intersection Protocols}.
\newblock \bibinfo{journal}{\emph{IEEE Transactions on Information Forensics and Security}}  \bibinfo{volume}{17} (\bibinfo{year}{2022}), \bibinfo{pages}{1--15}.
\newblock


\bibitem[Beaver et~al\mbox{.}(1990)]%
        {bmr}
\bibfield{author}{\bibinfo{person}{D. Beaver}, \bibinfo{person}{S. Micali}, {and} \bibinfo{person}{P. Rogaway}.} \bibinfo{year}{1990}\natexlab{}.
\newblock \showarticletitle{The round complexity of secure protocols}. In \bibinfo{booktitle}{\emph{Proceedings of the Twenty-Second Annual ACM Symposium on Theory of Computing}} (Baltimore, MD, USA) \emph{(\bibinfo{series}{STOC '90})}. \bibinfo{publisher}{Association for Computing Machinery}, \bibinfo{address}{New York, NY, USA}, \bibinfo{pages}{503–513}.
\newblock
\showISBNx{0897913612}


\bibitem[Becker et~al\mbox{.}(2025)]%
        {legalBecker2025MultiPartyCI}
\bibfield{author}{\bibinfo{person}{Sebastian Becker}, \bibinfo{person}{Christoph B{\"o}sch}, \bibinfo{person}{Benjamin Hettwer}, \bibinfo{person}{Thomas Hoeren}, \bibinfo{person}{Merlin Rombach}, \bibinfo{person}{Sven Trieflinger}, {and} \bibinfo{person}{Hossein Yalame}.} \bibinfo{year}{2025}\natexlab{}.
\newblock \showarticletitle{Multi-Party Computation in Corporate Data Processing: Legal and Technical Insights}.
\newblock \bibinfo{journal}{\emph{IACR Cryptol. ePrint Arch.}}  \bibinfo{volume}{2025} (\bibinfo{year}{2025}), \bibinfo{pages}{463}.
\newblock
\urldef\tempurl%
\url{https://api.semanticscholar.org/CorpusID:276962245}
\showURL{%
\tempurl}


\bibitem[Beltr{\'a}n et~al\mbox{.}(2022)]%
        {MartnezBeltrn2022DecentralizedFL}
\bibfield{author}{\bibinfo{person}{Enrique Tom{\'a}s~Mart{\'i}nez Beltr{\'a}n}, \bibinfo{person}{Mario~Quiles P{\'e}rez}, \bibinfo{person}{Pedro Miguel~S'anchez S'anchez}, \bibinfo{person}{Sergio~L'opez Bernal}, \bibinfo{person}{G{\'e}r{\^o}me Bovet}, \bibinfo{person}{Manuel~Gil P{\'e}rez}, \bibinfo{person}{Gregorio~Mart'inez P'erez}, {and} \bibinfo{person}{Alberto~Huertas Celdr'an}.} \bibinfo{year}{2022}\natexlab{}.
\newblock \showarticletitle{Decentralized Federated Learning: Fundamentals, State of the Art, Frameworks, Trends, and Challenges}.
\newblock \bibinfo{journal}{\emph{IEEE Communications Surveys \& Tutorials}}  \bibinfo{volume}{25} (\bibinfo{year}{2022}), \bibinfo{pages}{2983--3013}.
\newblock


\bibitem[Bi et~al\mbox{.}(2023)]%
        {Bi2023AchievingLA}
\bibfield{author}{\bibinfo{person}{Renwan Bi}, \bibinfo{person}{Jinbo Xiong}, \bibinfo{person}{Youliang Tian}, \bibinfo{person}{Qi Li}, {and} \bibinfo{person}{Kim‐Kwang~Raymond Choo}.} \bibinfo{year}{2023}\natexlab{}.
\newblock \showarticletitle{Achieving Lightweight and Privacy-Preserving Object Detection for Connected Autonomous Vehicles}.
\newblock \bibinfo{journal}{\emph{IEEE Internet of Things Journal}}  \bibinfo{volume}{10} (\bibinfo{year}{2023}), \bibinfo{pages}{2314--2329}.
\newblock


\bibitem[Blakley(1979)]%
        {Blakley1899SafeguardingCK}
\bibfield{author}{\bibinfo{person}{G.~R. Blakley}.} \bibinfo{year}{1979}\natexlab{}.
\newblock \showarticletitle{Safeguarding cryptographic keys}. In \bibinfo{booktitle}{\emph{1979 International Workshop on Managing Requirements Knowledge (MARK)}}. \bibinfo{pages}{313--318}.
\newblock


\bibitem[Bonawitz et~al\mbox{.}(2016)]%
        {bonawitzfl}
\bibfield{author}{\bibinfo{person}{Keith Bonawitz}, \bibinfo{person}{Vladimir Ivanov}, \bibinfo{person}{Ben Kreuter}, \bibinfo{person}{Antonio Marcedone}, \bibinfo{person}{H.~Brendan McMahan}, \bibinfo{person}{Sarvar Patel}, \bibinfo{person}{Daniel Ramage}, \bibinfo{person}{Aaron Segal}, {and} \bibinfo{person}{Karn Seth}.} \bibinfo{year}{2016}\natexlab{}.
\newblock \bibinfo{title}{Practical Secure Aggregation for Federated Learning on User-Held Data}.
\newblock
\newblock


\bibitem[Boneh et~al\mbox{.}(2005)]%
        {boneh2005}
\bibfield{author}{\bibinfo{person}{Dan Boneh}, \bibinfo{person}{Eu-Jin Goh}, {and} \bibinfo{person}{Kobbi Nissim}.} \bibinfo{year}{2005}\natexlab{}.
\newblock \showarticletitle{Evaluating 2-DNF Formulas on Ciphertexts}. In \bibinfo{booktitle}{\emph{Proceedings of the Second International Conference on Theory of Cryptography}} (Cambridge, MA) \emph{(\bibinfo{series}{TCC'05})}. \bibinfo{publisher}{Springer-Verlag}, \bibinfo{address}{Berlin, Heidelberg}, \bibinfo{pages}{325–341}.
\newblock
\showISBNx{3540245731}


\bibitem[Boyle et~al\mbox{.}(2020)]%
        {fss}
\bibfield{author}{\bibinfo{person}{Elette Boyle}, \bibinfo{person}{Nishanth Chandran}, \bibinfo{person}{Niv Gilboa}, \bibinfo{person}{Divya Gupta}, \bibinfo{person}{Yuval Ishai}, \bibinfo{person}{Nishant Kumar}, {and} \bibinfo{person}{Mayank Rathee}.} \bibinfo{year}{2020}\natexlab{}.
\newblock \showarticletitle{Function Secret Sharing for Mixed-Mode and Fixed-Point Secure Computation}.
\newblock \bibinfo{journal}{\emph{IACR Cryptol. ePrint Arch.}}  \bibinfo{volume}{2020} (\bibinfo{year}{2020}), \bibinfo{pages}{1392}.
\newblock


\bibitem[Brakerski(2012)]%
        {Brakerski2012FullyHE}
\bibfield{author}{\bibinfo{person}{Zvika Brakerski}.} \bibinfo{year}{2012}\natexlab{}.
\newblock \showarticletitle{Fully Homomorphic Encryption without Modulus Switching from Classical GapSVP}. In \bibinfo{booktitle}{\emph{Advances in Cryptology -- CRYPTO 2012}}, \bibfield{editor}{\bibinfo{person}{Reihaneh Safavi-Naini} {and} \bibinfo{person}{Ran Canetti}} (Eds.). \bibinfo{publisher}{Springer Berlin Heidelberg}, \bibinfo{address}{Berlin, Heidelberg}, \bibinfo{pages}{868--886}.
\newblock
\showISBNx{978-3-642-32009-5}


\bibitem[Brakerski et~al\mbox{.}(2013)]%
        {packedciphertext}
\bibfield{author}{\bibinfo{person}{Zvika Brakerski}, \bibinfo{person}{Craig Gentry}, {and} \bibinfo{person}{Shai Halevi}.} \bibinfo{year}{2013}\natexlab{}.
\newblock \showarticletitle{Packed Ciphertexts in LWE-Based Homomorphic Encryption}. In \bibinfo{booktitle}{\emph{International Conference on Theory and Practice of Public Key Cryptography}}.
\newblock


\bibitem[cai Zhou et~al\mbox{.}(2024)]%
        {Zhou2024PrivacypreservingAV}
\bibfield{author}{\bibinfo{person}{Fu cai Zhou}, \bibinfo{person}{Qiyu Wu}, \bibinfo{person}{Pengfei Wu}, \bibinfo{person}{Jian Xu}, {and} \bibinfo{person}{Da Feng}.} \bibinfo{year}{2024}\natexlab{}.
\newblock \showarticletitle{Privacy-preserving and verifiable data aggregation for Internet of Vehicles}.
\newblock \bibinfo{journal}{\emph{Comput. Commun.}}  \bibinfo{volume}{218} (\bibinfo{year}{2024}), \bibinfo{pages}{198--208}.
\newblock


\bibitem[Chen et~al\mbox{.}(2022)]%
        {chenji2022}
\bibfield{author}{\bibinfo{person}{Jianguo Chen}, \bibinfo{person}{Kenli Li}, {and} \bibinfo{person}{Philip~S. Yu}.} \bibinfo{year}{2022}\natexlab{}.
\newblock \showarticletitle{Privacy-Preserving Deep Learning Model for Decentralized VANETs Using Fully Homomorphic Encryption and Blockchain}.
\newblock \bibinfo{journal}{\emph{IEEE Transactions on Intelligent Transportation Systems}} \bibinfo{volume}{23}, \bibinfo{number}{8} (\bibinfo{year}{2022}), \bibinfo{pages}{11633--11642}.
\newblock


\bibitem[Chen et~al\mbox{.}(2021)]%
        {bdfl2021}
\bibfield{author}{\bibinfo{person}{Jin-Hua Chen}, \bibinfo{person}{Min-Rong Chen}, \bibinfo{person}{Guo-Qiang Zeng}, {and} \bibinfo{person}{Jia-Si Weng}.} \bibinfo{year}{2021}\natexlab{}.
\newblock \showarticletitle{BDFL: A Byzantine-Fault-Tolerance Decentralized Federated Learning Method for Autonomous Vehicle}.
\newblock \bibinfo{journal}{\emph{IEEE Transactions on Vehicular Technology}} \bibinfo{volume}{70}, \bibinfo{number}{9} (\bibinfo{year}{2021}), \bibinfo{pages}{8639--8652}.
\newblock


\bibitem[Cheng et~al\mbox{.}(2023b)]%
        {Cheng2023PPRTPP}
\bibfield{author}{\bibinfo{person}{Hongyuan Cheng}, \bibinfo{person}{Xianchao Zhang}, \bibinfo{person}{Jingkang Yang}, {and} \bibinfo{person}{Yining Liu}.} \bibinfo{year}{2023}\natexlab{b}.
\newblock \showarticletitle{PPRT: Privacy Preserving and Reliable Trust-Aware Platoon Recommendation Scheme in IoV}.
\newblock \bibinfo{journal}{\emph{IEEE Systems Journal}} \bibinfo{volume}{17}, \bibinfo{number}{3} (\bibinfo{year}{2023}), \bibinfo{pages}{4922--4933}.
\newblock


\bibitem[Cheng et~al\mbox{.}(2023a)]%
        {Cheng2023ALP}
\bibfield{author}{\bibinfo{person}{Yudan Cheng}, \bibinfo{person}{Jianfeng Ma}, \bibinfo{person}{Zhiquan Liu}, \bibinfo{person}{Yongdong Wu}, \bibinfo{person}{Kaimin Wei}, {and} \bibinfo{person}{Caiqin Dong}.} \bibinfo{year}{2023}\natexlab{a}.
\newblock \showarticletitle{A Lightweight Privacy Preservation Scheme With Efficient Reputation Management for Mobile Crowdsensing in Vehicular Networks}.
\newblock \bibinfo{journal}{\emph{IEEE Transactions on Dependable and Secure Computing}}  \bibinfo{volume}{20} (\bibinfo{year}{2023}), \bibinfo{pages}{1771--1788}.
\newblock


\bibitem[Cheon et~al\mbox{.}(2017)]%
        {CKKS}
\bibfield{author}{\bibinfo{person}{Jung~Hee Cheon}, \bibinfo{person}{Andrey Kim}, \bibinfo{person}{Miran Kim}, {and} \bibinfo{person}{Yongsoo Song}.} \bibinfo{year}{2017}\natexlab{}.
\newblock \showarticletitle{Homomorphic Encryption for Arithmetic of Approximate Numbers}. In \bibinfo{booktitle}{\emph{International Conference on the Theory and Application of Cryptology and Information Security}}.
\newblock


\bibitem[Chor et~al\mbox{.}(1995)]%
        {Chor1995PrivateIR}
\bibfield{author}{\bibinfo{person}{Benny Chor}, \bibinfo{person}{Eyal Kushilevitz}, \bibinfo{person}{Oded Goldreich}, {and} \bibinfo{person}{Madhu Sudan}.} \bibinfo{year}{1995}\natexlab{}.
\newblock \showarticletitle{Private information retrieval}.
\newblock \bibinfo{journal}{\emph{Proceedings of IEEE 36th Annual Foundations of Computer Science}} (\bibinfo{year}{1995}), \bibinfo{pages}{41--50}.
\newblock


\bibitem[Dutta et~al\mbox{.}(2022)]%
        {Dutta2022ComputeBV}
\bibfield{author}{\bibinfo{person}{Moumita Dutta}, \bibinfo{person}{Chaya Ganesh}, \bibinfo{person}{Sikhar Patranabis}, {and} \bibinfo{person}{Nitin Singh}.} \bibinfo{year}{2022}\natexlab{}.
\newblock \showarticletitle{Compute, but Verify: Efficient Multiparty Computation over Authenticated Inputs}.
\newblock \bibinfo{journal}{\emph{IACR Cryptol. ePrint Arch.}}  \bibinfo{volume}{2022} (\bibinfo{year}{2022}), \bibinfo{pages}{1648}.
\newblock


\bibitem[Evans et~al\mbox{.}(2018)]%
        {Evans2019API}
\bibfield{author}{\bibinfo{person}{David Evans}, \bibinfo{person}{Vladimir Kolesnikov}, {and} \bibinfo{person}{Mike Rosulek}.} \bibinfo{year}{2018}\natexlab{}.
\newblock \showarticletitle{A Pragmatic Introduction to Secure Multi-Party Computation}.
\newblock \bibinfo{journal}{\emph{Foundations and Trends in Privacy and Security}} \bibinfo{volume}{2}, \bibinfo{number}{2-3} (\bibinfo{year}{2018}), \bibinfo{pages}{70--246}.
\newblock
\showISSN{2474-1558}


\bibitem[Fan and Vercauteren(2012)]%
        {Fan2012SomewhatPF}
\bibfield{author}{\bibinfo{person}{Junfeng Fan} {and} \bibinfo{person}{Frederik Vercauteren}.} \bibinfo{year}{2012}\natexlab{}.
\newblock \bibinfo{title}{Somewhat Practical Fully Homomorphic Encryption}.
\newblock \bibinfo{howpublished}{Cryptology ePrint Archive, Paper 2012/144}.
\newblock


\bibitem[Feng and Yang(2022)]%
        {Feng2022ConcretelyES}
\bibfield{author}{\bibinfo{person}{Dengguo Feng} {and} \bibinfo{person}{Kang Yang}.} \bibinfo{year}{2022}\natexlab{}.
\newblock \showarticletitle{Concretely efficient secure multi-party computation protocols: survey and more}.
\newblock \bibinfo{journal}{\emph{Secur. Saf.}}  \bibinfo{volume}{1} (\bibinfo{year}{2022}), \bibinfo{pages}{2021001}.
\newblock


\bibitem[Freedman et~al\mbox{.}(2004)]%
        {Freedman2004EfficientPM}
\bibfield{author}{\bibinfo{person}{Michael~J. Freedman}, \bibinfo{person}{Kobbi Nissim}, {and} \bibinfo{person}{Benny Pinkas}.} \bibinfo{year}{2004}\natexlab{}.
\newblock \showarticletitle{Efficient Private Matching and Set Intersection}. In \bibinfo{booktitle}{\emph{International Conference on the Theory and Application of Cryptographic Techniques}}.
\newblock


\bibitem[Friedman et~al\mbox{.}(2024)]%
        {Friedman20242PCMPCET}
\bibfield{author}{\bibinfo{person}{Offir Friedman}, \bibinfo{person}{Avichai Marmor}, \bibinfo{person}{Dolev Mutzari}, \bibinfo{person}{Omer Sadika}, \bibinfo{person}{Yehonatan~C. Scaly}, \bibinfo{person}{Yuval Spiizer}, {and} \bibinfo{person}{Avishay Yanai}.} \bibinfo{year}{2024}\natexlab{}.
\newblock \showarticletitle{2PC-MPC: Emulating Two Party ECDSA in Large-Scale MPC}.
\newblock \bibinfo{journal}{\emph{IACR Cryptol. ePrint Arch.}}  \bibinfo{volume}{2024} (\bibinfo{year}{2024}), \bibinfo{pages}{253}.
\newblock


\bibitem[Fuchs et~al\mbox{.}(2020)]%
        {Fuchs2020TrustEVTE}
\bibfield{author}{\bibinfo{person}{Andreas Fuchs}, \bibinfo{person}{Dustin Kern}, \bibinfo{person}{Christoph Krau{\ss}}, {and} \bibinfo{person}{Maria Zhdanova}.} \bibinfo{year}{2020}\natexlab{}.
\newblock \showarticletitle{TrustEV: trustworthy electric vehicle charging and billing}.
\newblock \bibinfo{journal}{\emph{Proceedings of the 35th Annual ACM Symposium on Applied Computing}} (\bibinfo{year}{2020}).
\newblock


\bibitem[Gamiz et~al\mbox{.}(2024)]%
        {Gamiz2024Challenges}
\bibfield{author}{\bibinfo{person}{Idoia Gamiz}, \bibinfo{person}{Cristina Regueiro}, \bibinfo{person}{Oscar Lage}, \bibinfo{person}{Eduardo Jacob}, {and} \bibinfo{person}{Jasone Astorga}.} \bibinfo{year}{2024}\natexlab{}.
\newblock \showarticletitle{Challenges and future research directions in secure multi-party computation for resource-constrained devices and large-scale computations}.
\newblock \bibinfo{journal}{\emph{Int. J. Inf. Sec.}}  \bibinfo{volume}{24} (\bibinfo{year}{2024}), \bibinfo{pages}{27}.
\newblock


\bibitem[Gao et~al\mbox{.}(2022)]%
        {Gao2022PrivacyPreservingCE}
\bibfield{author}{\bibinfo{person}{Huan Gao}, \bibinfo{person}{Zhaojian Li}, {and} \bibinfo{person}{Yongqiang Wang}.} \bibinfo{year}{2022}\natexlab{}.
\newblock \showarticletitle{Privacy-Preserving Collaborative Estimation for Networked Vehicles With Application to Collaborative Road Profile Estimation}.
\newblock \bibinfo{journal}{\emph{IEEE Transactions on Intelligent Transportation Systems}}  \bibinfo{volume}{23} (\bibinfo{year}{2022}), \bibinfo{pages}{17301--17311}.
\newblock


\bibitem[Gentry et~al\mbox{.}(2010)]%
        {bgn}
\bibfield{author}{\bibinfo{person}{Craig Gentry}, \bibinfo{person}{Shai Halevi}, {and} \bibinfo{person}{Vinod Vaikuntanathan}.} \bibinfo{year}{2010}\natexlab{}.
\newblock \showarticletitle{A Simple BGN-Type Cryptosystem from LWE}.
\newblock \bibinfo{journal}{\emph{IACR Cryptol. ePrint Arch.}}  \bibinfo{volume}{2010} (\bibinfo{year}{2010}), \bibinfo{pages}{182}.
\newblock


\bibitem[Goldreich(1998)]%
        {goldreich1998secure}
\bibfield{author}{\bibinfo{person}{Oded Goldreich}.} \bibinfo{year}{1998}\natexlab{}.
\newblock \showarticletitle{Secure multi-party computation}.
\newblock \bibinfo{journal}{\emph{Manuscript. Preliminary version}} \bibinfo{volume}{78}, \bibinfo{number}{110} (\bibinfo{year}{1998}), \bibinfo{pages}{1--108}.
\newblock


\bibitem[Guan et~al\mbox{.}(2020)]%
        {Guan2020AchievingPV}
\bibfield{author}{\bibinfo{person}{Yunguo Guan}, \bibinfo{person}{Rongxing Lu}, \bibinfo{person}{Yandong Zheng}, \bibinfo{person}{Jun Shao}, {and} \bibinfo{person}{Guiyi Wei}.} \bibinfo{year}{2020}\natexlab{}.
\newblock \showarticletitle{Achieving Privacy-Preserving Vehicle Selection for Effective Content Dissemination in Smart Cities}.
\newblock \bibinfo{journal}{\emph{GLOBECOM 2020 - 2020 IEEE Global Communications Conference}} (\bibinfo{year}{2020}), \bibinfo{pages}{1--6}.
\newblock


\bibitem[Gyawali et~al\mbox{.}(2021)]%
        {sohan2021}
\bibfield{author}{\bibinfo{person}{Sohan Gyawali}, \bibinfo{person}{Yi Qian}, {and} \bibinfo{person}{Rose~Qingyang Hu}.} \bibinfo{year}{2021}\natexlab{}.
\newblock \showarticletitle{A Privacy-Preserving Misbehavior Detection System in Vehicular Communication Networks}.
\newblock \bibinfo{journal}{\emph{IEEE Transactions on Vehicular Technology}} \bibinfo{volume}{70}, \bibinfo{number}{6} (\bibinfo{year}{2021}), \bibinfo{pages}{6147--6158}.
\newblock


\bibitem[Hallgren et~al\mbox{.}(2017)]%
        {Hallgren2017}
\bibfield{author}{\bibinfo{person}{Per Hallgren}, \bibinfo{person}{Claudio Orlandi}, {and} \bibinfo{person}{Andrei Sabelfeld}.} \bibinfo{year}{2017}\natexlab{}.
\newblock \showarticletitle{PrivatePool: Privacy-Preserving Ridesharing}. In \bibinfo{booktitle}{\emph{2017 IEEE 30th Computer Security Foundations Symposium (CSF)}}. \bibinfo{publisher}{IEEE}, \bibinfo{address}{Santa Barbara, CA, USA}, \bibinfo{pages}{276--291}.
\newblock


\bibitem[Hataba et~al\mbox{.}(2022)]%
        {Hataba2022SecurityAP}
\bibfield{author}{\bibinfo{person}{Muhammad Hataba}, \bibinfo{person}{Ahmed Sherif}, \bibinfo{person}{Mohamed Mahmoud}, \bibinfo{person}{Mohamed Abdallah}, {and} \bibinfo{person}{Waleed Alasmary}.} \bibinfo{year}{2022}\natexlab{}.
\newblock \showarticletitle{Security and Privacy Issues in Autonomous Vehicles: A Layer-Based Survey}.
\newblock \bibinfo{journal}{\emph{IEEE Open Journal of the Communications Society}}  \bibinfo{volume}{3} (\bibinfo{year}{2022}), \bibinfo{pages}{811--829}.
\newblock


\bibitem[He et~al\mbox{.}(2018)]%
        {yuan2018}
\bibfield{author}{\bibinfo{person}{Yuanyuan He}, \bibinfo{person}{Jianbing Ni}, \bibinfo{person}{Xinyu Wang}, \bibinfo{person}{Ben Niu}, \bibinfo{person}{Fenghua Li}, {and} \bibinfo{person}{Xuemin Shen}.} \bibinfo{year}{2018}\natexlab{}.
\newblock \showarticletitle{Privacy-Preserving Partner Selection for Ride-Sharing Services}.
\newblock \bibinfo{journal}{\emph{IEEE Transactions on Vehicular Technology}} \bibinfo{volume}{67}, \bibinfo{number}{7} (\bibinfo{year}{2018}), \bibinfo{pages}{5994--6005}.
\newblock


\bibitem[Houda et~al\mbox{.}(2022)]%
        {WhenFL}
\bibfield{author}{\bibinfo{person}{Zakaria Abou~El Houda}, \bibinfo{person}{Bouziane Brik}, \bibinfo{person}{Adlen Ksentini}, \bibinfo{person}{Lyes Khoukhi}, {and} \bibinfo{person}{Mohsen Guizani}.} \bibinfo{year}{2022}\natexlab{}.
\newblock \showarticletitle{When Federated Learning Meets Game Theory: A Cooperative Framework to Secure IIoT Applications on Edge Computing}.
\newblock \bibinfo{journal}{\emph{IEEE Transactions on Industrial Informatics}}  \bibinfo{volume}{18} (\bibinfo{year}{2022}), \bibinfo{pages}{7988--7997}.
\newblock


\bibitem[Houda et~al\mbox{.}(2023a)]%
        {mitfed}
\bibfield{author}{\bibinfo{person}{Zakaria Abou~El Houda}, \bibinfo{person}{Abdelhakim~Senhaji Hafid}, {and} \bibinfo{person}{Lyes Khoukhi}.} \bibinfo{year}{2023}\natexlab{a}.
\newblock \showarticletitle{MiTFed: A Privacy Preserving Collaborative Network Attack Mitigation Framework Based on Federated Learning Using SDN and Blockchain}.
\newblock \bibinfo{journal}{\emph{IEEE Transactions on Network Science and Engineering}}  \bibinfo{volume}{10} (\bibinfo{year}{2023}), \bibinfo{pages}{1985--2001}.
\newblock


\bibitem[Houda et~al\mbox{.}(2024a)]%
        {fl3rev}
\bibfield{author}{\bibinfo{person}{Zakaria Abou~El Houda}, \bibinfo{person}{Hajar Moudoud}, {and} \bibinfo{person}{Bouziane Brik}.} \bibinfo{year}{2024}\natexlab{a}.
\newblock \showarticletitle{Federated Deep Reinforcement Learning for Efficient Jamming Attack Mitigation in O-RAN}.
\newblock \bibinfo{journal}{\emph{IEEE Transactions on Vehicular Technology}}  \bibinfo{volume}{73} (\bibinfo{year}{2024}), \bibinfo{pages}{9334--9343}.
\newblock


\bibitem[Houda et~al\mbox{.}(2023b)]%
        {SecuringFL}
\bibfield{author}{\bibinfo{person}{Zakaria Abou~El Houda}, \bibinfo{person}{Hajar Moudoud}, \bibinfo{person}{Bouziane Brik}, {and} \bibinfo{person}{Lyes Khoukhi}.} \bibinfo{year}{2023}\natexlab{b}.
\newblock \showarticletitle{Securing Federated Learning through Blockchain and Explainable AI for Robust Intrusion Detection in IoT Networks}.
\newblock \bibinfo{journal}{\emph{IEEE INFOCOM 2023 - IEEE Conference on Computer Communications Workshops (INFOCOM WKSHPS)}} (\bibinfo{year}{2023}), \bibinfo{pages}{1--6}.
\newblock


\bibitem[Houda et~al\mbox{.}(2024b)]%
        {bcpaperdfl}
\bibfield{author}{\bibinfo{person}{Zakaria Abou~El Houda}, \bibinfo{person}{Hajar Moudoud}, \bibinfo{person}{Bouziane Brik}, {and} \bibinfo{person}{Lyes Khoukhi}.} \bibinfo{year}{2024}\natexlab{b}.
\newblock \showarticletitle{Blockchain-Enabled Federated Learning for Enhanced Collaborative Intrusion Detection in Vehicular Edge Computing}.
\newblock \bibinfo{journal}{\emph{IEEE Transactions on Intelligent Transportation Systems}}  \bibinfo{volume}{25} (\bibinfo{year}{2024}), \bibinfo{pages}{7661--7672}.
\newblock


\bibitem[Hu et~al\mbox{.}(2023)]%
        {hu2023}
\bibfield{author}{\bibinfo{person}{Xiaoya Hu}, \bibinfo{person}{Ruiqin Li}, \bibinfo{person}{Licheng Wang}, \bibinfo{person}{Yuqiao Ning}, {and} \bibinfo{person}{Kaoru Ota}.} \bibinfo{year}{2023}\natexlab{}.
\newblock \showarticletitle{A Data Sharing Scheme Based on Federated Learning in IoV}.
\newblock \bibinfo{journal}{\emph{IEEE Transactions on Vehicular Technology}} \bibinfo{volume}{72}, \bibinfo{number}{9} (\bibinfo{year}{2023}), \bibinfo{pages}{11644--11656}.
\newblock


\bibitem[hua Liang et~al\mbox{.}(2022)]%
        {Liang2022PPRPPR}
\bibfield{author}{\bibinfo{person}{Yan hua Liang}, \bibinfo{person}{Yining Liu}, {and} \bibinfo{person}{Brij~Bhooshan Gupta}.} \bibinfo{year}{2022}\natexlab{}.
\newblock \showarticletitle{PPRP: Preserving-Privacy Route Planning Scheme in VANETs}.
\newblock \bibinfo{journal}{\emph{ACM Transactions on Internet Technology}}  \bibinfo{volume}{22} (\bibinfo{year}{2022}), \bibinfo{pages}{1 -- 18}.
\newblock


\bibitem[Huang et~al\mbox{.}(2021)]%
        {Huang2021pRidePO}
\bibfield{author}{\bibinfo{person}{Junxin Huang}, \bibinfo{person}{Yuchuan Luo}, \bibinfo{person}{Shaojing Fu}, \bibinfo{person}{Ming Xu}, {and} \bibinfo{person}{Bowen Hu}.} \bibinfo{year}{2021}\natexlab{}.
\newblock \showarticletitle{pRide: Privacy-Preserving Online Ride Hailing Matching System With Prediction}.
\newblock \bibinfo{journal}{\emph{IEEE Transactions on Vehicular Technology}}  \bibinfo{volume}{70} (\bibinfo{year}{2021}), \bibinfo{pages}{7413--7425}.
\newblock


\bibitem[Huo and Liu(2022)]%
        {Huo2021DistributedPE}
\bibfield{author}{\bibinfo{person}{Xiang Huo} {and} \bibinfo{person}{Mingxi Liu}.} \bibinfo{year}{2022}\natexlab{}.
\newblock \showarticletitle{Distributed privacy-preserving electric vehicle charging control based on secret sharing}.
\newblock \bibinfo{journal}{\emph{Electric Power Systems Research}}  \bibinfo{volume}{211} (\bibinfo{year}{2022}), \bibinfo{pages}{108357}.
\newblock
\showISSN{0378-7796}


\bibitem[Hussain and Koushanfar(2018)]%
        {siam2018}
\bibfield{author}{\bibinfo{person}{Siam~Umar Hussain} {and} \bibinfo{person}{Farinaz Koushanfar}.} \bibinfo{year}{2018}\natexlab{}.
\newblock \showarticletitle{P3: Privacy Preserving Positioning for Smart Automotive Systems}.
\newblock \bibinfo{journal}{\emph{ACM Trans. Des. Autom. Electron. Syst.}} \bibinfo{volume}{23}, \bibinfo{number}{6}, Article \bibinfo{articleno}{79} (\bibinfo{date}{Nov} \bibinfo{year}{2018}), \bibinfo{numpages}{19}~pages.
\newblock
\showISSN{1084-4309}


\bibitem[Jiang et~al\mbox{.}(2021)]%
        {Jiang2021LocationPM}
\bibfield{author}{\bibinfo{person}{Hongbo Jiang}, \bibinfo{person}{Jie Li}, \bibinfo{person}{Ping Zhao}, \bibinfo{person}{Fanzi Zeng}, \bibinfo{person}{Zhu Xiao}, {and} \bibinfo{person}{Arun Iyengar}.} \bibinfo{year}{2021}\natexlab{}.
\newblock \showarticletitle{Location Privacy-preserving Mechanisms in Location-based Services}.
\newblock \bibinfo{journal}{\emph{ACM Computing Surveys (CSUR)}}  \bibinfo{volume}{54} (\bibinfo{year}{2021}), \bibinfo{pages}{1 -- 36}.
\newblock


\bibitem[Kargl et~al\mbox{.}(2013)]%
        {dpits}
\bibfield{author}{\bibinfo{person}{Frank Kargl}, \bibinfo{person}{Arik Friedman}, {and} \bibinfo{person}{Roksana Boreli}.} \bibinfo{year}{2013}\natexlab{}.
\newblock \showarticletitle{Differential privacy in intelligent transportation systems}. In \bibinfo{booktitle}{\emph{Wireless Network Security}}.
\newblock


\bibitem[Karim and Rawat(2022)]%
        {karim22}
\bibfield{author}{\bibinfo{person}{Hassan Karim} {and} \bibinfo{person}{Danda~B. Rawat}.} \bibinfo{year}{2022}\natexlab{}.
\newblock \showarticletitle{TollsOnly Please—Homomorphic Encryption for Toll Transponder Privacy in Internet of Vehicles}.
\newblock \bibinfo{journal}{\emph{IEEE Internet of Things Journal}} \bibinfo{volume}{9}, \bibinfo{number}{4} (\bibinfo{year}{2022}), \bibinfo{pages}{2627--2636}.
\newblock


\bibitem[Karmakar et~al\mbox{.}(2024)]%
        {Karmakar2024QuickPoolPR}
\bibfield{author}{\bibinfo{person}{Banashri Karmakar}, \bibinfo{person}{Shyam Murthy}, \bibinfo{person}{Arpita Patra}, {and} \bibinfo{person}{Protik Paul}.} \bibinfo{year}{2024}\natexlab{}.
\newblock \showarticletitle{QuickPool: Privacy-Preserving Ride-Sharing Service}.
\newblock \bibinfo{journal}{\emph{IACR Cryptol. ePrint Arch.}}  \bibinfo{volume}{2024} (\bibinfo{year}{2024}), \bibinfo{pages}{1109}.
\newblock


\bibitem[Keller(2020)]%
        {mpspdz}
\bibfield{author}{\bibinfo{person}{Marcel Keller}.} \bibinfo{year}{2020}\natexlab{}.
\newblock \showarticletitle{{MP-SPDZ}: A Versatile Framework for Multi-Party Computation}. In \bibinfo{booktitle}{\emph{Proceedings of the 2020 ACM SIGSAC Conference on Computer and Communications Security}}.
\newblock


\bibitem[Kong et~al\mbox{.}(2017)]%
        {Kong2017}
\bibfield{author}{\bibinfo{person}{Qinglei Kong}, \bibinfo{person}{Rongxing Lu}, \bibinfo{person}{Maode Ma}, {and} \bibinfo{person}{Haiyong Bao}.} \bibinfo{year}{2017}\natexlab{}.
\newblock \showarticletitle{Achieve Location Privacy-Preserving Range Query in Vehicular Sensing}.
\newblock \bibinfo{journal}{\emph{Sensors}} \bibinfo{volume}{17}, \bibinfo{number}{8} (\bibinfo{year}{2017}), \bibinfo{numpages}{16}~pages.
\newblock
\showISSN{1424-8220}
\urldef\tempurl%
\url{https://www.mdpi.com/1424-8220/17/8/1829}
\showURL{%
\tempurl}


\bibitem[Kong et~al\mbox{.}(2021a)]%
        {prevmain}
\bibfield{author}{\bibinfo{person}{Qinglei Kong}, \bibinfo{person}{Rongxing Lu}, \bibinfo{person}{Feng Yin}, {and} \bibinfo{person}{Shuguang Cui}.} \bibinfo{year}{2021}\natexlab{a}.
\newblock \showarticletitle{Privacy-Preserving Continuous Data Collection for Predictive Maintenance in Vehicular Fog-Cloud}.
\newblock \bibinfo{journal}{\emph{IEEE Transactions on Intelligent Transportation Systems}} \bibinfo{volume}{22}, \bibinfo{number}{8} (\bibinfo{year}{2021}), \bibinfo{pages}{5060--5070}.
\newblock


\bibitem[Kong et~al\mbox{.}(2021b)]%
        {kong2021}
\bibfield{author}{\bibinfo{person}{Qinglei Kong}, \bibinfo{person}{Feng Yin}, \bibinfo{person}{Rongxing Lu}, \bibinfo{person}{Beibei Li}, \bibinfo{person}{Xiaohong Wang}, \bibinfo{person}{Shuguang Cui}, {and} \bibinfo{person}{Ping Zhang}.} \bibinfo{year}{2021}\natexlab{b}.
\newblock \showarticletitle{Privacy-Preserving Aggregation for Federated Learning-Based Navigation in Vehicular Fog}.
\newblock \bibinfo{journal}{\emph{IEEE Transactions on Industrial Informatics}} \bibinfo{volume}{17}, \bibinfo{number}{12} (\bibinfo{year}{2021}), \bibinfo{pages}{8453--8463}.
\newblock


\bibitem[Kumaraswamy et~al\mbox{.}(2021)]%
        {Kumaraswamy2021RevisitingDA}
\bibfield{author}{\bibinfo{person}{Deepak Kumaraswamy}, \bibinfo{person}{Shyam Murthy}, {and} \bibinfo{person}{Srinivas Vivek}.} \bibinfo{year}{2021}\natexlab{}.
\newblock \showarticletitle{Revisiting Driver Anonymity in ORide}.
\newblock \bibinfo{journal}{\emph{ArXiv}}  \bibinfo{volume}{abs/2101.06419} (\bibinfo{year}{2021}).
\newblock


\bibitem[Li et~al\mbox{.}(2023a)]%
        {Li2023RPPMAR}
\bibfield{author}{\bibinfo{person}{Runchuan Li}, \bibinfo{person}{Zhiquan Liu}, \bibinfo{person}{Yong Ma}, \bibinfo{person}{Yunni Xia}, \bibinfo{person}{Yudan Cheng}, \bibinfo{person}{Lin Wan}, {and} \bibinfo{person}{Jianfeng Ma}.} \bibinfo{year}{2023}\natexlab{a}.
\newblock \showarticletitle{RPPM: A Reputation-Based and Privacy-Preserving Platoon Management Scheme in Vehicular Networks}.
\newblock \bibinfo{journal}{\emph{IEEE Transactions on Intelligent Transportation Systems}} (\bibinfo{year}{2023}), \bibinfo{pages}{1--14}.
\newblock


\bibitem[Li et~al\mbox{.}(2023b)]%
        {Li2023ASO}
\bibfield{author}{\bibinfo{person}{Xiaoguo Li}, \bibinfo{person}{Bowen Zhao}, \bibinfo{person}{Guomin Yang}, \bibinfo{person}{Tao Xiang}, \bibinfo{person}{Jian Weng}, {and} \bibinfo{person}{Robert~H. Deng}.} \bibinfo{year}{2023}\natexlab{b}.
\newblock \showarticletitle{A Survey of Secure Computation Using Trusted Execution Environments}.
\newblock  (\bibinfo{year}{2023}).
\newblock
\showeprint[arxiv]{2302.12150}~[cs.CR]


\bibitem[Li et~al\mbox{.}(2022)]%
        {li2022}
\bibfield{author}{\bibinfo{person}{Yiran Li}, \bibinfo{person}{Hongwei Li}, \bibinfo{person}{Guowen Xu}, \bibinfo{person}{Tao Xiang}, {and} \bibinfo{person}{Rongxing Lu}.} \bibinfo{year}{2022}\natexlab{}.
\newblock \showarticletitle{Practical Privacy-Preserving Federated Learning in Vehicular Fog Computing}.
\newblock \bibinfo{journal}{\emph{IEEE Transactions on Vehicular Technology}} \bibinfo{volume}{71}, \bibinfo{number}{5} (\bibinfo{year}{2022}), \bibinfo{pages}{4692--4705}.
\newblock


\bibitem[Li et~al\mbox{.}(2021b)]%
        {Li2021PrivacyPreservedFL}
\bibfield{author}{\bibinfo{person}{Yijing Li}, \bibinfo{person}{Xiaofeng Tao}, \bibinfo{person}{Xuefei Zhang}, \bibinfo{person}{Junjie Liu}, {and} \bibinfo{person}{Jin Xu}.} \bibinfo{year}{2021}\natexlab{b}.
\newblock \showarticletitle{Privacy-Preserved Federated Learning for Autonomous Driving}.
\newblock \bibinfo{journal}{\emph{IEEE Transactions on Intelligent Transportation Systems}}  \bibinfo{volume}{23} (\bibinfo{year}{2021}), \bibinfo{pages}{8423--8434}.
\newblock


\bibitem[Li et~al\mbox{.}(2021a)]%
        {Li2021PriParkRecPD}
\bibfield{author}{\bibinfo{person}{Zengpeng Li}, \bibinfo{person}{Mamoun Alazab}, \bibinfo{person}{Sahil Garg}, {and} \bibinfo{person}{M.~Shamim Hossain}.} \bibinfo{year}{2021}\natexlab{a}.
\newblock \showarticletitle{PriParkRec: Privacy-Preserving Decentralized Parking Recommendation Service}.
\newblock \bibinfo{journal}{\emph{IEEE Transactions on Vehicular Technology}} \bibinfo{volume}{70}, \bibinfo{number}{5} (\bibinfo{year}{2021}), \bibinfo{pages}{4037--4050}.
\newblock


\bibitem[Lin and Tzeng(2005)]%
        {Lin2005AnES}
\bibfield{author}{\bibinfo{person}{Hsiao-Ying Lin} {and} \bibinfo{person}{Wen-Guey Tzeng}.} \bibinfo{year}{2005}\natexlab{}.
\newblock \showarticletitle{An Efficient Solution to the Millionaires' Problem Based on Homomorphic Encryption}.
\newblock \bibinfo{journal}{\emph{IACR Cryptol. ePrint Arch.}}  \bibinfo{volume}{2005} (\bibinfo{year}{2005}), \bibinfo{pages}{43}.
\newblock


\bibitem[Lindell and Pinkas(2011)]%
        {Lindell2011SecureTC}
\bibfield{author}{\bibinfo{person}{Yehuda Lindell} {and} \bibinfo{person}{Benny Pinkas}.} \bibinfo{year}{2011}\natexlab{}.
\newblock \showarticletitle{Secure Two-Party Computation via Cut-and-Choose Oblivious Transfer}.
\newblock \bibinfo{journal}{\emph{Journal of Cryptology}}  \bibinfo{volume}{25} (\bibinfo{year}{2011}), \bibinfo{pages}{680--722}.
\newblock


\bibitem[Liu et~al\mbox{.}(2024a)]%
        {trajectoryvanet}
\bibfield{author}{\bibinfo{person}{Dengzhi Liu}, \bibinfo{person}{Geng Yu}, \bibinfo{person}{Yongdong Ding}, \bibinfo{person}{Zhaoman Zhong}, {and} \bibinfo{person}{Chen Wang}.} \bibinfo{year}{2024}\natexlab{a}.
\newblock \showarticletitle{Privacy Preserving Multi-Party Computation With Secret Sharing for Trajectory Prediction in VANETs}.
\newblock \bibinfo{journal}{\emph{IEEE Transactions on Vehicular Technology}}  \bibinfo{volume}{73} (\bibinfo{year}{2024}), \bibinfo{pages}{18666--18677}.
\newblock


\bibitem[Liu et~al\mbox{.}(2022)]%
        {liu2022}
\bibfield{author}{\bibinfo{person}{Mingming Liu}, \bibinfo{person}{Long Cheng}, \bibinfo{person}{Yingqi Gu}, \bibinfo{person}{Ying Wang}, \bibinfo{person}{Qingzhi Liu}, {and} \bibinfo{person}{Noel~E. O’Connor}.} \bibinfo{year}{2022}\natexlab{}.
\newblock \showarticletitle{MPC-CSAS: Multi-Party Computation for Real-Time Privacy-Preserving Speed Advisory Systems}.
\newblock \bibinfo{journal}{\emph{IEEE Transactions on Intelligent Transportation Systems}} \bibinfo{volume}{23}, \bibinfo{number}{6} (\bibinfo{year}{2022}), \bibinfo{pages}{5887--5893}.
\newblock


\bibitem[Liu et~al\mbox{.}(2024b)]%
        {Liu2024ASO}
\bibfield{author}{\bibinfo{person}{Mingyu Liu}, \bibinfo{person}{Ekim Yurtsever}, \bibinfo{person}{Jonathan Fossaert}, \bibinfo{person}{Xingcheng Zhou}, \bibinfo{person}{Walter Zimmer}, \bibinfo{person}{Yuning Cui}, \bibinfo{person}{Bare~Luka Žagar}, {and} \bibinfo{person}{Alois Knoll}.} \bibinfo{year}{2024}\natexlab{b}.
\newblock \showarticletitle{A Survey on Autonomous Driving Datasets: Statistics, Annotation Quality, and a Future Outlook}.
\newblock \bibinfo{journal}{\emph{IEEE Transactions on Intelligent Vehicles}}  \bibinfo{volume}{9} (\bibinfo{year}{2024}), \bibinfo{pages}{7138--7164}.
\newblock


\bibitem[Lu et~al\mbox{.}(2024)]%
        {Lu2024MaliciouslySM}
\bibfield{author}{\bibinfo{person}{Yibiao Lu}, \bibinfo{person}{Bingsheng Zhang}, {and} \bibinfo{person}{Kui Ren}.} \bibinfo{year}{2024}\natexlab{}.
\newblock \showarticletitle{Maliciously Secure MPC From Semi-Honest 2PC in the Server-Aided Model}.
\newblock \bibinfo{journal}{\emph{IEEE Transactions on Dependable and Secure Computing}}  \bibinfo{volume}{21} (\bibinfo{year}{2024}), \bibinfo{pages}{3109--3125}.
\newblock


\bibitem[Luo et~al\mbox{.}(2023)]%
        {Luo2023P2RidePA}
\bibfield{author}{\bibinfo{person}{Yuchuan Luo}, \bibinfo{person}{Shaojing Fu}, \bibinfo{person}{Xiaohua Jia}, \bibinfo{person}{Ming Xu}, {and} \bibinfo{person}{Yingwen Chen}.} \bibinfo{year}{2023}\natexlab{}.
\newblock \showarticletitle{P2Ride: Practical and Privacy-Preserving Ride-Matching Scheme for Ridesharing}.
\newblock \bibinfo{journal}{\emph{IEEE Transactions on Intelligent Transportation Systems}} \bibinfo{volume}{24}, \bibinfo{number}{3} (\bibinfo{year}{2023}), \bibinfo{pages}{3584--3593}.
\newblock


\bibitem[Luo et~al\mbox{.}(2019)]%
        {pRidePR2019}
\bibfield{author}{\bibinfo{person}{Yuchuan Luo}, \bibinfo{person}{Xiaohua Jia}, \bibinfo{person}{Shaojing Fu}, {and} \bibinfo{person}{Ming Xu}.} \bibinfo{year}{2019}\natexlab{}.
\newblock \showarticletitle{pRide: Privacy-Preserving Ride Matching Over Road Networks for Online Ride-Hailing Service}.
\newblock \bibinfo{journal}{\emph{IEEE Transactions on Information Forensics and Security}} \bibinfo{volume}{14}, \bibinfo{number}{7} (\bibinfo{year}{2019}), \bibinfo{pages}{1791--1802}.
\newblock


\bibitem[Magaia et~al\mbox{.}(2018)]%
        {Magaiga2018}
\bibfield{author}{\bibinfo{person}{Naercio Magaia}, \bibinfo{person}{Carlos Borrego}, \bibinfo{person}{Paulo~Rogério Pereira}, {and} \bibinfo{person}{Miguel Correia}.} \bibinfo{year}{2018}\natexlab{}.
\newblock \showarticletitle{ePRIVO: An Enhanced PRIvacy-preserVing Opportunistic Routing Protocol for Vehicular Delay-Tolerant Networks}.
\newblock \bibinfo{journal}{\emph{IEEE Transactions on Vehicular Technology}} \bibinfo{volume}{67}, \bibinfo{number}{11} (\bibinfo{year}{2018}), \bibinfo{pages}{11154--11168}.
\newblock


\bibitem[Mansouri et~al\mbox{.}(2023)]%
        {SoKSA}
\bibfield{author}{\bibinfo{person}{Mohamad Mansouri}, \bibinfo{person}{Melek {\"O}nen}, \bibinfo{person}{Wafa~Ben Jaballah}, {and} \bibinfo{person}{Mauro Conti}.} \bibinfo{year}{2023}\natexlab{}.
\newblock \showarticletitle{SoK: Secure Aggregation Based on Cryptographic Schemes for Federated Learning}.
\newblock \bibinfo{journal}{\emph{Proc. Priv. Enhancing Technol.}}  \bibinfo{volume}{2023} (\bibinfo{year}{2023}), \bibinfo{pages}{140--157}.
\newblock


\bibitem[Morales et~al\mbox{.}(2023)]%
        {Escalera2023PrivateSI}
\bibfield{author}{\bibinfo{person}{Daniel Morales}, \bibinfo{person}{Isaac Agudo}, {and} \bibinfo{person}{Javier Lopez}.} \bibinfo{year}{2023}\natexlab{}.
\newblock \showarticletitle{Private set intersection: A systematic literature review}.
\newblock \bibinfo{journal}{\emph{Computer Science Review}}  \bibinfo{volume}{49} (\bibinfo{year}{2023}), \bibinfo{pages}{100567}.
\newblock
\showISSN{1574-0137}


\bibitem[Mukherjee and Wichs(2016)]%
        {Mukherjee2016TwoRM}
\bibfield{author}{\bibinfo{person}{Pratyay Mukherjee} {and} \bibinfo{person}{Daniel Wichs}.} \bibinfo{year}{2016}\natexlab{}.
\newblock \showarticletitle{Two Round Multiparty Computation via Multi-key FHE}. In \bibinfo{booktitle}{\emph{Advances in Cryptology -- EUROCRYPT 2016}}, \bibfield{editor}{\bibinfo{person}{Marc Fischlin} {and} \bibinfo{person}{Jean-S{\'e}bastien Coron}} (Eds.). \bibinfo{publisher}{Springer Berlin Heidelberg}, \bibinfo{address}{Berlin, Heidelberg}, \bibinfo{pages}{735--763}.
\newblock
\showISBNx{978-3-662-49896-5}


\bibitem[Mundhe et~al\mbox{.}(2021)]%
        {Mundhe2021ACS}
\bibfield{author}{\bibinfo{person}{Pravin Mundhe}, \bibinfo{person}{Shekhar Verma}, {and} \bibinfo{person}{S. Venkatesan}.} \bibinfo{year}{2021}\natexlab{}.
\newblock \showarticletitle{A comprehensive survey on authentication and privacy-preserving schemes in VANETs}.
\newblock \bibinfo{journal}{\emph{Computer Science Review}}  \bibinfo{volume}{41} (\bibinfo{year}{2021}), \bibinfo{pages}{100411}.
\newblock
\showISSN{1574-0137}


\bibitem[Murthy and Vivek(2022a)]%
        {Murthy2022DriverLH}
\bibfield{author}{\bibinfo{person}{Shyam Murthy} {and} \bibinfo{person}{Srinivas Vivek}.} \bibinfo{year}{2022}\natexlab{a}.
\newblock \showarticletitle{Driver Locations Harvesting Attack on pRide}. In \bibinfo{booktitle}{\emph{International Conference on Network and System Security}}.
\newblock


\bibitem[Murthy and Vivek(2022b)]%
        {PassiveTAoride}
\bibfield{author}{\bibinfo{person}{Shyam Murthy} {and} \bibinfo{person}{Srinivas Vivek}.} \bibinfo{year}{2022}\natexlab{b}.
\newblock \showarticletitle{Passive Triangulation Attack on ORide}.
\newblock \bibinfo{journal}{\emph{ArXiv}}  \bibinfo{volume}{abs/2208.12216} (\bibinfo{year}{2022}).
\newblock


\bibitem[Nabeel et~al\mbox{.}(2013)]%
        {modifiedpa}
\bibfield{author}{\bibinfo{person}{Mohamed Nabeel}, \bibinfo{person}{Stefan Appel}, \bibinfo{person}{Elisa Bertino}, {and} \bibinfo{person}{Alejandro~P. Buchmann}.} \bibinfo{year}{2013}\natexlab{}.
\newblock \showarticletitle{Privacy Preserving Context Aware Publish Subscribe Systems}. In \bibinfo{booktitle}{\emph{International Conference on Network and System Security}}.
\newblock


\bibitem[Nelson and Olovsson(2017)]%
        {IntroducingDP}
\bibfield{author}{\bibinfo{person}{Boel Nelson} {and} \bibinfo{person}{Tomas Olovsson}.} \bibinfo{year}{2017}\natexlab{}.
\newblock \showarticletitle{Introducing Differential Privacy to the Automotive Domain: Opportunities and Challenges}.
\newblock \bibinfo{journal}{\emph{2017 IEEE 86th Vehicular Technology Conference (VTC-Fall)}} (\bibinfo{year}{2017}), \bibinfo{pages}{1--7}.
\newblock


\bibitem[Ng and Chow(2023)]%
        {Ng2023SoKC}
\bibfield{author}{\bibinfo{person}{Lucien K.~L. Ng} {and} \bibinfo{person}{Sherman S.~M. Chow}.} \bibinfo{year}{2023}\natexlab{}.
\newblock \showarticletitle{SoK: Cryptographic Neural-Network Computation}.
\newblock \bibinfo{journal}{\emph{2023 IEEE Symposium on Security and Privacy (SP)}} (\bibinfo{year}{2023}), \bibinfo{pages}{497--514}.
\newblock


\bibitem[Pagnin et~al\mbox{.}(2019)]%
        {toppool2019}
\bibfield{author}{\bibinfo{person}{Elena Pagnin}, \bibinfo{person}{Gunnar Gunnarsson}, \bibinfo{person}{Pedram Talebi}, \bibinfo{person}{Claudio Orlandi}, {and} \bibinfo{person}{Andrei Sabelfeld}.} \bibinfo{year}{2019}\natexlab{}.
\newblock \showarticletitle{TOPPool: Time-aware Optimized Privacy-Preserving Ridesharing}.
\newblock \bibinfo{journal}{\emph{Proceedings on Privacy Enhancing Technologies}}  \bibinfo{volume}{2019} (\bibinfo{year}{2019}), \bibinfo{pages}{93--111}.
\newblock


\bibitem[Paillier(1999)]%
        {Paillier1999PublicKeyCB}
\bibfield{author}{\bibinfo{person}{Pascal Paillier}.} \bibinfo{year}{1999}\natexlab{}.
\newblock \showarticletitle{Public-Key Cryptosystems Based on Composite Degree Residuosity Classes}. In \bibinfo{booktitle}{\emph{International Conference on the Theory and Application of Cryptographic Techniques}}.
\newblock


\bibitem[Peng et~al\mbox{.}(2024)]%
        {Peng2024PrivacyPreservingTD}
\bibfield{author}{\bibinfo{person}{Tao Peng}, \bibinfo{person}{Wentao Zhong}, \bibinfo{person}{Guojun Wang}, \bibinfo{person}{Entao Luo}, \bibinfo{person}{Shui Yu}, \bibinfo{person}{Yining Liu}, \bibinfo{person}{Yi Yang}, {and} \bibinfo{person}{Xuyun Zhang}.} \bibinfo{year}{2024}\natexlab{}.
\newblock \showarticletitle{Privacy-Preserving Truth Discovery Based on Secure Multi-Party Computation in Vehicle-Based Mobile Crowdsensing}.
\newblock \bibinfo{journal}{\emph{IEEE Transactions on Intelligent Transportation Systems}}  \bibinfo{volume}{25} (\bibinfo{year}{2024}), \bibinfo{pages}{7767--7779}.
\newblock


\bibitem[Pes{\'e} et~al\mbox{.}(2020)]%
        {Pes2020SPyCS}
\bibfield{author}{\bibinfo{person}{Mert~D. Pes{\'e}}, \bibinfo{person}{Xiaoying Pu}, {and} \bibinfo{person}{Kang~G. Shin}.} \bibinfo{year}{2020}\natexlab{}.
\newblock \showarticletitle{SPy: Car Steering Reveals Your Trip Route!}
\newblock \bibinfo{journal}{\emph{Proceedings on Privacy Enhancing Technologies}}  \bibinfo{volume}{2020} (\bibinfo{year}{2020}), \bibinfo{pages}{155 -- 174}.
\newblock


\bibitem[Pham et~al\mbox{.}(2017)]%
        {ORide2017}
\bibfield{author}{\bibinfo{person}{Anh Pham}, \bibinfo{person}{Italo Dacosta}, \bibinfo{person}{Guillaume Endignoux}, \bibinfo{person}{Juan Ramon~Troncoso Pastoriza}, \bibinfo{person}{Kevin Huguenin}, {and} \bibinfo{person}{Jean-Pierre Hubaux}.} \bibinfo{year}{2017}\natexlab{}.
\newblock \showarticletitle{{ORide}: A {Privacy-Preserving} yet Accountable {Ride-Hailing} Service}. In \bibinfo{booktitle}{\emph{26th USENIX Security Symposium (USENIX Security 17)}}. \bibinfo{publisher}{USENIX Association}, \bibinfo{address}{Vancouver, BC}, \bibinfo{pages}{1235--1252}.
\newblock
\showISBNx{978-1-931971-40-9}


\bibitem[Pinkas et~al\mbox{.}(2014)]%
        {Pinkas2014FasterPS}
\bibfield{author}{\bibinfo{person}{Benny Pinkas}, \bibinfo{person}{T. Schneider}, {and} \bibinfo{person}{Michael Zohner}.} \bibinfo{year}{2014}\natexlab{}.
\newblock \showarticletitle{Faster Private Set Intersection Based on OT Extension}. In \bibinfo{booktitle}{\emph{USENIX Security Symposium}}.
\newblock


\bibitem[Pokhrel et~al\mbox{.}(2021)]%
        {Pokhrel2021}
\bibfield{author}{\bibinfo{person}{Shiva~Raj Pokhrel}, \bibinfo{person}{Youyang Qu}, \bibinfo{person}{Surya Nepal}, {and} \bibinfo{person}{Surjit Singh}.} \bibinfo{year}{2021}\natexlab{}.
\newblock \showarticletitle{Privacy-Aware Autonomous Valet Parking: Towards Experience Driven Approach}.
\newblock \bibinfo{journal}{\emph{IEEE Transactions on Intelligent Transportation Systems}} \bibinfo{volume}{22}, \bibinfo{number}{8} (\bibinfo{year}{2021}), \bibinfo{pages}{5352--5363}.
\newblock


\bibitem[Qiu et~al\mbox{.}(2022)]%
        {chenxi2022}
\bibfield{author}{\bibinfo{person}{Chenxi Qiu}, \bibinfo{person}{Anna Squicciarini}, \bibinfo{person}{Ce Pang}, \bibinfo{person}{Ning Wang}, {and} \bibinfo{person}{Ben Wu}.} \bibinfo{year}{2022}\natexlab{}.
\newblock \showarticletitle{Location Privacy Protection in Vehicle-Based Spatial Crowdsourcing via Geo-Indistinguishability}.
\newblock \bibinfo{journal}{\emph{IEEE Transactions on Mobile Computing}} \bibinfo{volume}{21}, \bibinfo{number}{7} (\bibinfo{year}{2022}), \bibinfo{pages}{2436--2450}.
\newblock


\bibitem[Quero et~al\mbox{.}(2023)]%
        {Quero2023TowardsPP}
\bibfield{author}{\bibinfo{person}{Nicolas Quero}, \bibinfo{person}{Aymen Boudguiga}, \bibinfo{person}{Renaud Sirdey}, {and} \bibinfo{person}{Nadir Karam}.} \bibinfo{year}{2023}\natexlab{}.
\newblock \showarticletitle{Towards Privacy-Preserving Platooning Services by means of Homomorphic Encryption}. In \bibinfo{booktitle}{\emph{Symposium on Vehicle Security and Privacy (VehicleSec) 2023}}. \bibinfo{publisher}{Internet Society}, \bibinfo{address}{San Diego, CA, USA}, \bibinfo{numpages}{7}~pages.
\newblock


\bibitem[Rivest et~al\mbox{.}(1978)]%
        {Rivest1978AMF}
\bibfield{author}{\bibinfo{person}{Ronald~L. Rivest}, \bibinfo{person}{Adi Shamir}, {and} \bibinfo{person}{Leonard~M. Adleman}.} \bibinfo{year}{1978}\natexlab{}.
\newblock \showarticletitle{A method for obtaining digital signatures and public-key cryptosystems}.
\newblock \bibinfo{journal}{\emph{Commun. ACM}}  \bibinfo{volume}{26} (\bibinfo{year}{1978}), \bibinfo{pages}{96--99}.
\newblock


\bibitem[Saleem et~al\mbox{.}(2021)]%
        {Saleem2021SteeringAP}
\bibfield{author}{\bibinfo{person}{Hajira Saleem}, \bibinfo{person}{Faisal Riaz}, \bibinfo{person}{Leonardo Mostarda}, \bibinfo{person}{Muaz~A. Niazi}, \bibinfo{person}{Ammar Rafiq}, {and} \bibinfo{person}{Saqib Saeed}.} \bibinfo{year}{2021}\natexlab{}.
\newblock \showarticletitle{Steering Angle Prediction Techniques for Autonomous Ground Vehicles: A Review}.
\newblock \bibinfo{journal}{\emph{IEEE Access}}  \bibinfo{volume}{9} (\bibinfo{year}{2021}), \bibinfo{pages}{78567--78585}.
\newblock


\bibitem[Schoenmakers(1999)]%
        {PVSS}
\bibfield{author}{\bibinfo{person}{Berry Schoenmakers}.} \bibinfo{year}{1999}\natexlab{}.
\newblock \showarticletitle{A Simple Publicly Verifiable Secret Sharing Scheme and Its Application to Electronic Voting}. In \bibinfo{booktitle}{\emph{Advances in Cryptology --- CRYPTO' 99}}, \bibfield{editor}{\bibinfo{person}{Michael Wiener}} (Ed.). \bibinfo{publisher}{Springer Berlin Heidelberg}, \bibinfo{address}{Berlin, Heidelberg}, \bibinfo{pages}{148--164}.
\newblock
\showISBNx{978-3-540-48405-9}


\bibitem[Shamir(1979)]%
        {Shamir1979HowTS}
\bibfield{author}{\bibinfo{person}{Adi Shamir}.} \bibinfo{year}{1979}\natexlab{}.
\newblock \showarticletitle{How to share a secret}.
\newblock \bibinfo{journal}{\emph{Commun. ACM}} \bibinfo{volume}{22}, \bibinfo{number}{11} (\bibinfo{date}{Nov} \bibinfo{year}{1979}), \bibinfo{pages}{612–613}.
\newblock
\showISSN{0001-0782}


\bibitem[Sharma and Liu(2021)]%
        {sharma2021}
\bibfield{author}{\bibinfo{person}{Prinkle Sharma} {and} \bibinfo{person}{Hong Liu}.} \bibinfo{year}{2021}\natexlab{}.
\newblock \showarticletitle{A Machine-Learning-Based Data-Centric Misbehavior Detection Model for Internet of Vehicles}.
\newblock \bibinfo{journal}{\emph{IEEE Internet of Things Journal}} \bibinfo{volume}{8}, \bibinfo{number}{6} (\bibinfo{year}{2021}), \bibinfo{pages}{4991--4999}.
\newblock


\bibitem[Stadler(1996)]%
        {Stadler1996PubliclyVS}
\bibfield{author}{\bibinfo{person}{Markus Stadler}.} \bibinfo{year}{1996}\natexlab{}.
\newblock \showarticletitle{Publicly Verifiable Secret Sharing}.
\newblock


\bibitem[Sun and Srinivsan(2023)]%
        {Sun2023RemindingDO}
\bibfield{author}{\bibinfo{person}{Wei Sun} {and} \bibinfo{person}{Kannan Srinivsan}.} \bibinfo{year}{2023}\natexlab{}.
\newblock \showarticletitle{Reminding Drivers of the Stalking Vehicles on the Road}.
\newblock \bibinfo{journal}{\emph{Proceedings Inaugural International Symposium on Vehicle Security \& Privacy}} (\bibinfo{year}{2023}).
\newblock


\bibitem[Sun et~al\mbox{.}(2020)]%
        {sun2020}
\bibfield{author}{\bibinfo{person}{Xiaoqiang Sun}, \bibinfo{person}{F.~Richard Yu}, \bibinfo{person}{Peng Zhang}, \bibinfo{person}{Weixin Xie}, {and} \bibinfo{person}{Xiang Peng}.} \bibinfo{year}{2020}\natexlab{}.
\newblock \showarticletitle{A Survey on Secure Computation Based on Homomorphic Encryption in Vehicular Ad Hoc Networks}.
\newblock \bibinfo{journal}{\emph{Sensors}} \bibinfo{volume}{20}, \bibinfo{number}{15} (\bibinfo{year}{2020}), \bibinfo{numpages}{31}~pages.
\newblock
\showISSN{1424-8220}


\bibitem[Sutradhar et~al\mbox{.}(2024)]%
        {Sutradhar2024ASO}
\bibfield{author}{\bibinfo{person}{Kartick Sutradhar}, \bibinfo{person}{Beena~G Pillai}, \bibinfo{person}{Ruhul Amin}, {and} \bibinfo{person}{Dayanand~Lal Narayan}.} \bibinfo{year}{2024}\natexlab{}.
\newblock \showarticletitle{A survey on privacy-preserving authentication protocols for secure vehicular communication}.
\newblock \bibinfo{journal}{\emph{Comput. Commun.}}  \bibinfo{volume}{219} (\bibinfo{year}{2024}), \bibinfo{pages}{1--18}.
\newblock


\bibitem[Symeonidis et~al\mbox{.}(2017)]%
        {iraklis2017}
\bibfield{author}{\bibinfo{person}{Iraklis Symeonidis}, \bibinfo{person}{Abdelrahaman Aly}, \bibinfo{person}{Mustafa~Asan Mustafa}, \bibinfo{person}{Bart Mennink}, \bibinfo{person}{Siemen Dhooghe}, {and} \bibinfo{person}{Bart Preneel}.} \bibinfo{year}{2017}\natexlab{}.
\newblock \showarticletitle{SePCAR: A Secure and Privacy-Enhancing Protocol for Car Access Provision}. In \bibinfo{booktitle}{\emph{Computer Security -- ESORICS 2017}}, \bibfield{editor}{\bibinfo{person}{Simon~N. Foley}, \bibinfo{person}{Dieter Gollmann}, {and} \bibinfo{person}{Einar Snekkenes}} (Eds.). \bibinfo{publisher}{Springer International Publishing}, \bibinfo{address}{Cham}, \bibinfo{pages}{475--493}.
\newblock
\showISBNx{978-3-319-66399-9}


\bibitem[Symeonidis et~al\mbox{.}(2022)]%
        {iraklis2022}
\bibfield{author}{\bibinfo{person}{Iraklis Symeonidis}, \bibinfo{person}{Dragos Rotaru}, \bibinfo{person}{Mustafa~A. Mustafa}, \bibinfo{person}{Bart Mennink}, \bibinfo{person}{Bart Preneel}, {and} \bibinfo{person}{Panos Papadimitratos}.} \bibinfo{year}{2022}\natexlab{}.
\newblock \showarticletitle{HERMES: Scalable, Secure, and Privacy-Enhancing Vehicular Sharing-Access System}.
\newblock \bibinfo{journal}{\emph{IEEE Internet of Things Journal}} \bibinfo{volume}{9}, \bibinfo{number}{1} (\bibinfo{year}{2022}), \bibinfo{pages}{129--151}.
\newblock


\bibitem[Sébert et~al\mbox{.}(2022)]%
        {Sebert2022ProtectingDF}
\bibfield{author}{\bibinfo{person}{Arnaud~Grivet Sébert}, \bibinfo{person}{Renaud Sirdey}, \bibinfo{person}{Oana Stan}, {and} \bibinfo{person}{Cédric Gouy-Pailler}.} \bibinfo{year}{2022}\natexlab{}.
\newblock \bibinfo{title}{Protecting Data from all Parties: Combining FHE and DP in Federated Learning}.
\newblock
\newblock
\showeprint[arxiv]{2205.04330}~[cs.CR]


\bibitem[Tan et~al\mbox{.}(2018)]%
        {Tan2018PrivateIR}
\bibfield{author}{\bibinfo{person}{Zheng Tan}, \bibinfo{person}{Cheng Wang}, \bibinfo{person}{Mengchu Zhou}, {and} \bibinfo{person}{Luomeng Zhang}.} \bibinfo{year}{2018}\natexlab{}.
\newblock \showarticletitle{Private information retrieval in vehicular location-based services}.
\newblock \bibinfo{journal}{\emph{2018 IEEE 4th World Forum on Internet of Things (WF-IoT)}} (\bibinfo{year}{2018}), \bibinfo{pages}{56--61}.
\newblock


\bibitem[Tiausas et~al\mbox{.}(2023)]%
        {Tiausas2023HPRoPHP}
\bibfield{author}{\bibinfo{person}{Francis~Jerome Tiausas}, \bibinfo{person}{K. Yasumoto}, \bibinfo{person}{Jose~Paolo Talusan}, \bibinfo{person}{Hayato Yamana}, \bibinfo{person}{Hirozumi Yamaguchi}, \bibinfo{person}{Shameek Bhattacharjee}, \bibinfo{person}{Abhishek Dubey}, {and} \bibinfo{person}{Sajal~K. Das}.} \bibinfo{year}{2023}\natexlab{}.
\newblock \showarticletitle{HPRoP: Hierarchical Privacy-preserving Route Planning for Smart Cities}.
\newblock \bibinfo{journal}{\emph{ACM Transactions on Cyber-Physical Systems}}  \bibinfo{volume}{7} (\bibinfo{year}{2023}), \bibinfo{pages}{1 -- 25}.
\newblock


\bibitem[Uprety et~al\mbox{.}(2021)]%
        {uprety2021}
\bibfield{author}{\bibinfo{person}{Aashma Uprety}, \bibinfo{person}{Danda~B. Rawat}, {and} \bibinfo{person}{Jiang Li}.} \bibinfo{year}{2021}\natexlab{}.
\newblock \showarticletitle{Privacy Preserving Misbehavior Detection in IoV Using Federated Machine Learning}. In \bibinfo{booktitle}{\emph{2021 IEEE 18th Annual Consumer Communications \& Networking Conference (CCNC)}}. \bibinfo{publisher}{IEEE}, \bibinfo{address}{Las Vegas, NV, USA}, \bibinfo{pages}{1--6}.
\newblock


\bibitem[Vargheese and Vivek(2023)]%
        {Vargheese2023AttackOT}
\bibfield{author}{\bibinfo{person}{Meghana Vargheese} {and} \bibinfo{person}{Srinivas Vivek}.} \bibinfo{year}{2023}\natexlab{}.
\newblock \showarticletitle{Attack on the Privacy-Preserving Carpooling Service TAROT}. In \bibinfo{booktitle}{\emph{International Conferences on Information Science and System}}.
\newblock


\bibitem[{Visual Capitalist Team}(2023)]%
        {visualcapitalist2023}
\bibfield{author}{\bibinfo{person}{{Visual Capitalist Team}}.} \bibinfo{year}{2023}\natexlab{}.
\newblock \bibinfo{title}{Network Overload}.
\newblock
\newblock
\urldef\tempurl%
\url{https://www.visualcapitalist.com/network-overload/}
\showURL{%
\tempurl}
\newblock
\shownote{Accessed: Aug 15, 2023}.


\bibitem[Vivek(2021)]%
        {Vivek2021AttacksOA}
\bibfield{author}{\bibinfo{person}{Srinivas Vivek}.} \bibinfo{year}{2021}\natexlab{}.
\newblock \showarticletitle{Attacks on a Privacy-Preserving Publish-Subscribe System and a Ride-Hailing Service}. In \bibinfo{booktitle}{\emph{IMA Conference on Cryptography and Coding}}.
\newblock


\bibitem[Wang et~al\mbox{.}(2021a)]%
        {Wang2021PrivacyPreservingES}
\bibfield{author}{\bibinfo{person}{Nan Wang}, \bibinfo{person}{Sid Chi-Kin Chau}, {and} \bibinfo{person}{Yue Zhou}.} \bibinfo{year}{2021}\natexlab{a}.
\newblock \showarticletitle{Privacy-Preserving Energy Storage Sharing with Blockchain}.
\newblock \bibinfo{journal}{\emph{Proceedings of the Twelfth ACM International Conference on Future Energy Systems}} (\bibinfo{year}{2021}).
\newblock


\bibitem[Wang et~al\mbox{.}(2021b)]%
        {Wang2021ObliviousTF}
\bibfield{author}{\bibinfo{person}{Xianmin Wang}, \bibinfo{person}{Xiaohui Kuang}, \bibinfo{person}{Jin Li}, \bibinfo{person}{Jing Li}, \bibinfo{person}{Xiaofeng Chen}, {and} \bibinfo{person}{Zheli Liu}.} \bibinfo{year}{2021}\natexlab{b}.
\newblock \showarticletitle{Oblivious Transfer for Privacy-Preserving in VANET’s Feature Matching}.
\newblock \bibinfo{journal}{\emph{IEEE Transactions on Intelligent Transportation Systems}}  \bibinfo{volume}{22} (\bibinfo{year}{2021}), \bibinfo{pages}{4359--4366}.
\newblock


\bibitem[Woo et~al\mbox{.}(2018)]%
        {woo2018localization}
\bibfield{author}{\bibinfo{person}{Ami Woo}, \bibinfo{person}{Baris Fidan}, {and} \bibinfo{person}{William~W. Melek}.} \bibinfo{year}{2018}\natexlab{}.
\newblock \showarticletitle{Localization for Autonomous Driving}.
\newblock In \bibinfo{booktitle}{\emph{Handbook of Position Location: Theory, Practice, and Advances} (\bibinfo{edition}{2nd} ed.)}, \bibfield{editor}{\bibinfo{person}{Seyed A.~(Reza) Zekavat} {and} \bibinfo{person}{R.~Michael Buehrer}} (Eds.). \bibinfo{publisher}{John Wiley \& Sons, Ltd}, \bibinfo{address}{Hoboken, NJ, USA}, Chapter~29, \bibinfo{pages}{1051--1087}.
\newblock
\showISBNx{9781119434610}


\bibitem[Wu et~al\mbox{.}(2021)]%
        {surveyblockchain}
\bibfield{author}{\bibinfo{person}{Songqi Wu}, \bibinfo{person}{Jin Li}, \bibinfo{person}{Fenghui Duan}, \bibinfo{person}{Yueming Lu}, \bibinfo{person}{Xu Zhang}, {and} \bibinfo{person}{Jiefu Gan}.} \bibinfo{year}{2021}\natexlab{}.
\newblock \showarticletitle{The Survey on the development of Secure Multi-Party Computing in the blockchain}. In \bibinfo{booktitle}{\emph{2021 IEEE Sixth International Conference on Data Science in Cyberspace (DSC)}}. \bibinfo{pages}{1--7}.
\newblock


\bibitem[Wu et~al\mbox{.}(2023)]%
        {Wu2023PrivacyPreservingAT}
\bibfield{author}{\bibinfo{person}{Yan Wu}, \bibinfo{person}{Can Zhang}, {and} \bibinfo{person}{Liehuang Zhu}.} \bibinfo{year}{2023}\natexlab{}.
\newblock \showarticletitle{Privacy-Preserving and Traceable Blockchain-Based Charging Payment Scheme for Electric Vehicles}.
\newblock \bibinfo{journal}{\emph{IEEE Internet of Things Journal}}  \bibinfo{volume}{10} (\bibinfo{year}{2023}), \bibinfo{pages}{21254--21265}.
\newblock


\bibitem[Xiong et~al\mbox{.}(2022)]%
        {xiong2022}
\bibfield{author}{\bibinfo{person}{Jinbo Xiong}, \bibinfo{person}{Renwan Bi}, \bibinfo{person}{Youliang Tian}, \bibinfo{person}{Ximeng Liu}, {and} \bibinfo{person}{Dapeng Wu}.} \bibinfo{year}{2022}\natexlab{}.
\newblock \showarticletitle{Toward Lightweight, Privacy-Preserving Cooperative Object Classification for Connected Autonomous Vehicles}.
\newblock \bibinfo{journal}{\emph{IEEE Internet of Things Journal}} \bibinfo{volume}{9}, \bibinfo{number}{4} (\bibinfo{year}{2022}), \bibinfo{pages}{2787--2801}.
\newblock


\bibitem[Xiong et~al\mbox{.}(2020)]%
        {jinbo2020}
\bibfield{author}{\bibinfo{person}{Jinbo Xiong}, \bibinfo{person}{Renwan Bi}, \bibinfo{person}{Mingfeng Zhao}, \bibinfo{person}{Jingda Guo}, {and} \bibinfo{person}{Qing Yang}.} \bibinfo{year}{2020}\natexlab{}.
\newblock \showarticletitle{Edge-Assisted Privacy-Preserving Raw Data Sharing Framework for Connected Autonomous Vehicles}.
\newblock \bibinfo{journal}{\emph{IEEE Wireless Communications}} \bibinfo{volume}{27}, \bibinfo{number}{3} (\bibinfo{year}{2020}), \bibinfo{pages}{24--30}.
\newblock


\bibitem[Xiong et~al\mbox{.}(2021)]%
        {xiong2021}
\bibfield{author}{\bibinfo{person}{Zuobin Xiong}, \bibinfo{person}{Zhipeng Cai}, \bibinfo{person}{Qilong Han}, \bibinfo{person}{Arwa Alrawais}, {and} \bibinfo{person}{Wei Li}.} \bibinfo{year}{2021}\natexlab{}.
\newblock \showarticletitle{ADGAN: Protect Your Location Privacy in Camera Data of Auto-Driving Vehicles}.
\newblock \bibinfo{journal}{\emph{IEEE Transactions on Industrial Informatics}} \bibinfo{volume}{17}, \bibinfo{number}{9} (\bibinfo{year}{2021}), \bibinfo{pages}{6200--6210}.
\newblock


\bibitem[Xiong et~al\mbox{.}(2019)]%
        {xiong2019}
\bibfield{author}{\bibinfo{person}{Zuobin Xiong}, \bibinfo{person}{Wei Li}, \bibinfo{person}{Qilong Han}, {and} \bibinfo{person}{Zhipeng Cai}.} \bibinfo{year}{2019}\natexlab{}.
\newblock \showarticletitle{Privacy-Preserving Auto-Driving: A GAN-Based Approach to Protect Vehicular Camera Data}. In \bibinfo{booktitle}{\emph{2019 IEEE International Conference on Data Mining (ICDM)}}. \bibinfo{publisher}{IEEE}, \bibinfo{address}{Beijing, China}, \bibinfo{pages}{668--677}.
\newblock


\bibitem[Xu et~al\mbox{.}(2020)]%
        {verify}
\bibfield{author}{\bibinfo{person}{Guowen Xu}, \bibinfo{person}{Hongwei Li}, \bibinfo{person}{Sen Liu}, \bibinfo{person}{Kan Yang}, {and} \bibinfo{person}{Xiaodong Lin}.} \bibinfo{year}{2020}\natexlab{}.
\newblock \showarticletitle{VerifyNet: Secure and Verifiable Federated Learning}.
\newblock \bibinfo{journal}{\emph{IEEE Transactions on Information Forensics and Security}}  \bibinfo{volume}{15} (\bibinfo{year}{2020}), \bibinfo{pages}{911--926}.
\newblock


\bibitem[Xu et~al\mbox{.}(2022)]%
        {Xu2022AnEA}
\bibfield{author}{\bibinfo{person}{Qi Xu}, \bibinfo{person}{Hui Zhu}, \bibinfo{person}{Yandong Zheng}, \bibinfo{person}{Jiaqi Zhao}, \bibinfo{person}{Rongxing Lu}, {and} \bibinfo{person}{Hui Li}.} \bibinfo{year}{2022}\natexlab{}.
\newblock \showarticletitle{An Efficient and Privacy-Preserving Route Matching Scheme for Carpooling Services}.
\newblock \bibinfo{journal}{\emph{IEEE Internet of Things Journal}}  \bibinfo{volume}{9} (\bibinfo{year}{2022}), \bibinfo{pages}{19890--19902}.
\newblock


\bibitem[Xu et~al\mbox{.}(2023)]%
        {Xu2023PriTAECPT}
\bibfield{author}{\bibinfo{person}{Zihui Xu}, \bibinfo{person}{Lei Wu}, \bibinfo{person}{Chengyi Qin}, \bibinfo{person}{Su Li}, \bibinfo{person}{Songnian Zhang}, {and} \bibinfo{person}{Rongxing Lu}.} \bibinfo{year}{2023}\natexlab{}.
\newblock \showarticletitle{PriTAEC: Privacy-Preserving Task Assignment Based on Oblivious Transfer and Edge Computing in VANET}.
\newblock \bibinfo{journal}{\emph{IEEE Transactions on Vehicular Technology}} \bibinfo{volume}{72}, \bibinfo{number}{4} (\bibinfo{year}{2023}), \bibinfo{pages}{4996--5009}.
\newblock


\bibitem[Yang et~al\mbox{.}(2019)]%
        {Yang2019ACS}
\bibfield{author}{\bibinfo{person}{Yang Yang}, \bibinfo{person}{Xindi Huang}, \bibinfo{person}{Ximeng Liu}, \bibinfo{person}{Hongju Cheng}, \bibinfo{person}{Jian Weng}, \bibinfo{person}{Xiangyang Luo}, {and} \bibinfo{person}{Victor Chang}.} \bibinfo{year}{2019}\natexlab{}.
\newblock \showarticletitle{A Comprehensive Survey on Secure Outsourced Computation and Its Applications}.
\newblock \bibinfo{journal}{\emph{IEEE Access}}  \bibinfo{volume}{7} (\bibinfo{year}{2019}), \bibinfo{pages}{159426--159465}.
\newblock


\bibitem[Yao(1982)]%
        {yao1982}
\bibfield{author}{\bibinfo{person}{Andrew~C. Yao}.} \bibinfo{year}{1982}\natexlab{}.
\newblock \showarticletitle{Protocols for secure computations}. In \bibinfo{booktitle}{\emph{23rd Annual Symposium on Foundations of Computer Science (SFCS 1982)}}. \bibinfo{publisher}{IEEE}, \bibinfo{address}{Chicago, IL, USA}, \bibinfo{pages}{160--164}.
\newblock


\bibitem[Yao(1986)]%
        {Yao1986HowTG}
\bibfield{author}{\bibinfo{person}{Andrew Chi-Chih Yao}.} \bibinfo{year}{1986}\natexlab{}.
\newblock \showarticletitle{How to generate and exchange secrets}. In \bibinfo{booktitle}{\emph{27th Annual Symposium on Foundations of Computer Science (SFCS 1986)}}. \bibinfo{publisher}{IEEE}, \bibinfo{address}{Toronto, ON, Canada}, \bibinfo{pages}{162--167}.
\newblock


\bibitem[Ying et~al\mbox{.}(2022)]%
        {Ying2022PrivacySignalPT}
\bibfield{author}{\bibinfo{person}{Zuobin Ying}, \bibinfo{person}{Shuanglong Cao}, \bibinfo{person}{Ximeng Liu}, \bibinfo{person}{Zhuo Ma}, \bibinfo{person}{Jianfeng Ma}, {and} \bibinfo{person}{Robert~H. Deng}.} \bibinfo{year}{2022}\natexlab{}.
\newblock \showarticletitle{PrivacySignal: Privacy-Preserving Traffic Signal Control for Intelligent Transportation System}.
\newblock \bibinfo{journal}{\emph{IEEE Transactions on Intelligent Transportation Systems}} \bibinfo{volume}{23}, \bibinfo{number}{9} (\bibinfo{year}{2022}), \bibinfo{pages}{16290--16303}.
\newblock


\bibitem[Yoshizawa et~al\mbox{.}(2022)]%
        {Yoshizawa2022ASO}
\bibfield{author}{\bibinfo{person}{Takahito Yoshizawa}, \bibinfo{person}{Dave Singel{\'e}e}, \bibinfo{person}{Jan~Tobias M{\"u}hlberg}, \bibinfo{person}{St{\'e}phane Delbruel}, \bibinfo{person}{Amirhosein Taherkordi}, \bibinfo{person}{Danny Hughes}, {and} \bibinfo{person}{Bart Preneel}.} \bibinfo{year}{2022}\natexlab{}.
\newblock \showarticletitle{A Survey of Security and Privacy Issues in V2X Communication Systems}.
\newblock \bibinfo{journal}{\emph{Comput. Surveys}}  \bibinfo{volume}{55} (\bibinfo{year}{2022}), \bibinfo{pages}{1 -- 36}.
\newblock


\bibitem[Yu et~al\mbox{.}(2022)]%
        {Yu2022EfficientAP}
\bibfield{author}{\bibinfo{person}{Haining Yu}, \bibinfo{person}{Xiaohua Jia}, \bibinfo{person}{Hongli Zhang}, {and} \bibinfo{person}{Jiangang Shu}.} \bibinfo{year}{2022}\natexlab{}.
\newblock \showarticletitle{Efficient and Privacy-Preserving Ride Matching Using Exact Road Distance in Online Ride Hailing Services}.
\newblock \bibinfo{journal}{\emph{IEEE Transactions on Services Computing}}  \bibinfo{volume}{15} (\bibinfo{year}{2022}), \bibinfo{pages}{1841--1854}.
\newblock


\bibitem[Yu et~al\mbox{.}(2021a)]%
        {yuha2021}
\bibfield{author}{\bibinfo{person}{Haining Yu}, \bibinfo{person}{Xiaohua Jia}, \bibinfo{person}{Hongli Zhang}, \bibinfo{person}{Xiangzhan Yu}, {and} \bibinfo{person}{Jiangang Shu}.} \bibinfo{year}{2021}\natexlab{a}.
\newblock \showarticletitle{PSRide: Privacy-Preserving Shared Ride Matching for Online Ride Hailing Systems}.
\newblock \bibinfo{journal}{\emph{IEEE Transactions on Dependable and Secure Computing}} \bibinfo{volume}{18}, \bibinfo{number}{3} (\bibinfo{year}{2021}), \bibinfo{pages}{1425--1440}.
\newblock


\bibitem[Yu et~al\mbox{.}(2019)]%
        {lpride2019}
\bibfield{author}{\bibinfo{person}{Haining Yu}, \bibinfo{person}{Jiangang Shu}, \bibinfo{person}{Xiaohua Jia}, \bibinfo{person}{Hongli Zhang}, {and} \bibinfo{person}{Xiangzhan Yu}.} \bibinfo{year}{2019}\natexlab{}.
\newblock \showarticletitle{lpRide: Lightweight and Privacy-Preserving Ride Matching Over Road Networks in Online Ride Hailing Systems}.
\newblock \bibinfo{journal}{\emph{IEEE Transactions on Vehicular Technology}}  \bibinfo{volume}{68} (\bibinfo{year}{2019}), \bibinfo{pages}{10418--10428}.
\newblock


\bibitem[Yu et~al\mbox{.}(2021b)]%
        {yu2021}
\bibfield{author}{\bibinfo{person}{Haining Yu}, \bibinfo{person}{Hongli Zhang}, \bibinfo{person}{Xiangzhan Yu}, \bibinfo{person}{Xiaojiang Du}, {and} \bibinfo{person}{Mohsen Guizani}.} \bibinfo{year}{2021}\natexlab{b}.
\newblock \showarticletitle{PGRide: Privacy-Preserving Group Ridesharing Matching in Online Ride Hailing Services}.
\newblock \bibinfo{journal}{\emph{IEEE Internet of Things Journal}} \bibinfo{volume}{8}, \bibinfo{number}{7} (\bibinfo{year}{2021}), \bibinfo{pages}{5722--5735}.
\newblock


\bibitem[Yu et~al\mbox{.}(2024)]%
        {Yu2024EfficientPT}
\bibfield{author}{\bibinfo{person}{Yantao Yu}, \bibinfo{person}{Xiaoping Xue}, \bibinfo{person}{Jingxiao Ma}, \bibinfo{person}{Ellen~Z. Zhang}, \bibinfo{person}{Yunguo Guan}, {and} \bibinfo{person}{Rongxing Lu}.} \bibinfo{year}{2024}\natexlab{}.
\newblock \showarticletitle{Efficient Privacy-Preserving Task Allocation With Secret Sharing for Vehicular Crowdsensing}.
\newblock \bibinfo{journal}{\emph{IEEE Internet of Things Journal}}  \bibinfo{volume}{11} (\bibinfo{year}{2024}), \bibinfo{pages}{9473--9486}.
\newblock


\bibitem[Yuan et~al\mbox{.}(2022)]%
        {Yuan2022AnEO}
\bibfield{author}{\bibinfo{person}{Minghao Yuan}, \bibinfo{person}{Dongdong Wang}, \bibinfo{person}{Feng Zhang}, \bibinfo{person}{Shenqing Wang}, \bibinfo{person}{Shan Ji}, {and} \bibinfo{person}{Yongjun Ren}.} \bibinfo{year}{2022}\natexlab{}.
\newblock \showarticletitle{An Examination of Multi-Key Fully Homomorphic Encryption and Its Applications}.
\newblock \bibinfo{journal}{\emph{Mathematics}} \bibinfo{volume}{10}, \bibinfo{number}{24} (\bibinfo{year}{2022}), \bibinfo{numpages}{20}~pages.
\newblock
\showISSN{2227-7390}


\bibitem[Zafar et~al\mbox{.}(2022)]%
        {Zafar2022CarpoolingIC}
\bibfield{author}{\bibinfo{person}{Farkhanda Zafar}, \bibinfo{person}{Hasan~Ali Khattak}, \bibinfo{person}{Moayad Aloqaily}, {and} \bibinfo{person}{Rasheed Hussain}.} \bibinfo{year}{2022}\natexlab{}.
\newblock \showarticletitle{Carpooling in Connected and Autonomous Vehicles: Current Solutions and Future Directions}.
\newblock \bibinfo{journal}{\emph{ACM Computing Surveys (CSUR)}}  \bibinfo{volume}{54} (\bibinfo{year}{2022}), \bibinfo{pages}{1 -- 36}.
\newblock


\bibitem[Zhang et~al\mbox{.}(2020)]%
        {Zhang2020VerifiableAP}
\bibfield{author}{\bibinfo{person}{Chuan Zhang}, \bibinfo{person}{Liehuang Zhu}, \bibinfo{person}{Jianbing Ni}, \bibinfo{person}{Cheng Huang}, {and} \bibinfo{person}{Xuemin Shen}.} \bibinfo{year}{2020}\natexlab{}.
\newblock \showarticletitle{Verifiable and Privacy-Preserving Traffic Flow Statistics for Advanced Traffic Management Systems}.
\newblock \bibinfo{journal}{\emph{IEEE Transactions on Vehicular Technology}} \bibinfo{volume}{69}, \bibinfo{number}{9} (\bibinfo{year}{2020}), \bibinfo{pages}{10336--10347}.
\newblock


\bibitem[Zhang et~al\mbox{.}(2023b)]%
        {Zhang2023BSDPBS}
\bibfield{author}{\bibinfo{person}{Can Zhang}, \bibinfo{person}{Liehuang Zhu}, {and} \bibinfo{person}{Chang Xu}.} \bibinfo{year}{2023}\natexlab{b}.
\newblock \showarticletitle{BSDP: Blockchain-Based Smart Parking for Digital-Twin Empowered Vehicular Sensing Networks With Privacy Protection}.
\newblock \bibinfo{journal}{\emph{IEEE Transactions on Industrial Informatics}}  \bibinfo{volume}{19} (\bibinfo{year}{2023}), \bibinfo{pages}{7237--7246}.
\newblock


\bibitem[Zhang et~al\mbox{.}(2022b)]%
        {Zhang2022TPPRAT}
\bibfield{author}{\bibinfo{person}{Chuan Zhang}, \bibinfo{person}{Liehuang Zhu}, \bibinfo{person}{Chang Xu}, \bibinfo{person}{Kashif Sharif}, \bibinfo{person}{Kai Ding}, \bibinfo{person}{Ximeng Liu}, \bibinfo{person}{Xiaojiang Du}, {and} \bibinfo{person}{Mohsen Guizani}.} \bibinfo{year}{2022}\natexlab{b}.
\newblock \showarticletitle{TPPR: A Trust-Based and Privacy-Preserving Platoon Recommendation Scheme in VANET}.
\newblock \bibinfo{journal}{\emph{IEEE Transactions on Services Computing}} \bibinfo{volume}{15}, \bibinfo{number}{2} (\bibinfo{year}{2022}), \bibinfo{pages}{806--818}.
\newblock


\bibitem[Zhang et~al\mbox{.}(2019)]%
        {Zhang2019EfficientMP}
\bibfield{author}{\bibinfo{person}{En Zhang}, \bibinfo{person}{Feng-Hao Liu}, \bibinfo{person}{Qiqi Lai}, \bibinfo{person}{Ganggang Jin}, {and} \bibinfo{person}{Yu Li}.} \bibinfo{year}{2019}\natexlab{}.
\newblock \showarticletitle{Efficient Multi-Party Private Set Intersection Against Malicious Adversaries}.
\newblock \bibinfo{journal}{\emph{Proceedings of the 2019 ACM SIGSAC Conference on Cloud Computing Security Workshop}} (\bibinfo{year}{2019}).
\newblock


\bibitem[Zhang et~al\mbox{.}(2023a)]%
        {Zhang2023PrivacypreservingOR}
\bibfield{author}{\bibinfo{person}{Juyuan Zhang}, \bibinfo{person}{Licheng Wang}, \bibinfo{person}{Xiaoya Hu}, \bibinfo{person}{Rui Li}, {and} \bibinfo{person}{Shihui Zheng}.} \bibinfo{year}{2023}\natexlab{a}.
\newblock \showarticletitle{Privacy-preserving Online Ride-hailing Service System Based on Taking the Intersection of Private sets of Points of Interest}.
\newblock \bibinfo{journal}{\emph{2023 International Conference on Mobile Internet, Cloud Computing and Information Security (MICCIS)}} (\bibinfo{year}{2023}), \bibinfo{pages}{107--117}.
\newblock


\bibitem[Zhang et~al\mbox{.}(2021b)]%
        {junwei2021}
\bibfield{author}{\bibinfo{person}{Junwei Zhang}, \bibinfo{person}{Fan Yang}, \bibinfo{person}{Zhuo Ma}, \bibinfo{person}{Zhuzhu Wang}, \bibinfo{person}{Ximeng Liu}, {and} \bibinfo{person}{Jianfeng Ma}.} \bibinfo{year}{2021}\natexlab{b}.
\newblock \showarticletitle{A Decentralized Location Privacy-Preserving Spatial Crowdsourcing for Internet of Vehicles}.
\newblock \bibinfo{journal}{\emph{IEEE Transactions on Intelligent Transportation Systems}} \bibinfo{volume}{22}, \bibinfo{number}{4} (\bibinfo{year}{2021}), \bibinfo{pages}{2299--2313}.
\newblock


\bibitem[Zhang et~al\mbox{.}(2022a)]%
        {Zhang2022APP}
\bibfield{author}{\bibinfo{person}{Lu Zhang}, \bibinfo{person}{Wenhao Gao}, \bibinfo{person}{Shukai Chen}, \bibinfo{person}{Wei Ren}, \bibinfo{person}{Kim‐Kwang~Raymond Choo}, {and} \bibinfo{person}{Naixue~N. Xiong}.} \bibinfo{year}{2022}\natexlab{a}.
\newblock \showarticletitle{A Privacy-Preserving Proximity Testing Using Private Set Intersection for Vehicular Ad-Hoc Networks}.
\newblock \bibinfo{journal}{\emph{IEEE Transactions on Industrial Informatics}}  \bibinfo{volume}{18} (\bibinfo{year}{2022}), \bibinfo{pages}{7373--7383}.
\newblock


\bibitem[Zhang and Yang(2018)]%
        {adaboost}
\bibfield{author}{\bibinfo{person}{Pengbo Zhang} {and} \bibinfo{person}{Zhixin Yang}.} \bibinfo{year}{2018}\natexlab{}.
\newblock \showarticletitle{A Novel AdaBoost Framework With Robust Threshold and Structural Optimization}.
\newblock \bibinfo{journal}{\emph{IEEE Transactions on Cybernetics}}  \bibinfo{volume}{48} (\bibinfo{year}{2018}), \bibinfo{pages}{64--76}.
\newblock


\bibitem[Zhang et~al\mbox{.}(2021a)]%
        {Zhang2021PrivacyPreservingDL}
\bibfield{author}{\bibinfo{person}{Qiao Zhang}, \bibinfo{person}{Chunsheng Xin}, {and} \bibinfo{person}{Hongyi Wu}.} \bibinfo{year}{2021}\natexlab{a}.
\newblock \showarticletitle{Privacy-Preserving Deep Learning Based on Multiparty Secure Computation: A Survey}.
\newblock \bibinfo{journal}{\emph{IEEE Internet of Things Journal}}  \bibinfo{volume}{8} (\bibinfo{year}{2021}), \bibinfo{pages}{10412--10429}.
\newblock


\bibitem[Zhao et~al\mbox{.}(2019)]%
        {dpsurvey}
\bibfield{author}{\bibinfo{person}{Ping Zhao}, \bibinfo{person}{Guanglin Zhang}, \bibinfo{person}{Shaohua Wan}, \bibinfo{person}{Gaoyang Liu}, {and} \bibinfo{person}{Tariq Umer}.} \bibinfo{year}{2019}\natexlab{}.
\newblock \showarticletitle{A survey of local differential privacy for securing internet of vehicles}.
\newblock \bibinfo{journal}{\emph{The Journal of Supercomputing}}  \bibinfo{volume}{76} (\bibinfo{year}{2019}), \bibinfo{pages}{8391 -- 8412}.
\newblock


\bibitem[Zhou et~al\mbox{.}(2020)]%
        {Zhou2020ATQ}
\bibfield{author}{\bibinfo{person}{Changli Zhou}, \bibinfo{person}{Tian Wang}, \bibinfo{person}{Hui Tian}, \bibinfo{person}{Wenxian Jiang}, {and} \bibinfo{person}{Zhijian Wang}.} \bibinfo{year}{2020}\natexlab{}.
\newblock \showarticletitle{A Top-K Query Scheme With Privacy Preservation for Intelligent Vehicle Network in Mobile IoT}.
\newblock \bibinfo{journal}{\emph{IEEE Access}}  \bibinfo{volume}{8} (\bibinfo{year}{2020}), \bibinfo{pages}{81698--81710}.
\newblock


\bibitem[Zhou et~al\mbox{.}(2024)]%
        {Zhou2024SecureMC}
\bibfield{author}{\bibinfo{person}{Ian Zhou}, \bibinfo{person}{Farzad Tofigh}, \bibinfo{person}{Massimo Piccardi}, \bibinfo{person}{Mehran Abolhasan}, \bibinfo{person}{Daniel~Robert Franklin}, {and} \bibinfo{person}{Justin Lipman}.} \bibinfo{year}{2024}\natexlab{}.
\newblock \showarticletitle{Secure Multi-Party Computation for Machine Learning: A Survey}.
\newblock \bibinfo{journal}{\emph{IEEE Access}}  \bibinfo{volume}{12} (\bibinfo{year}{2024}), \bibinfo{pages}{53881--53899}.
\newblock


\bibitem[Zhou et~al\mbox{.}(2021)]%
        {Zhou2021EPNSEP}
\bibfield{author}{\bibinfo{person}{Jun Zhou}, \bibinfo{person}{Shiying Chen}, \bibinfo{person}{Kim‐Kwang~Raymond Choo}, \bibinfo{person}{Zhenfu Cao}, {and} \bibinfo{person}{Xiaolei Dong}.} \bibinfo{year}{2021}\natexlab{}.
\newblock \showarticletitle{EPNS: Efficient Privacy-Preserving Intelligent Traffic Navigation From Multiparty Delegated Computation in Cloud-Assisted VANETs}.
\newblock \bibinfo{journal}{\emph{IEEE Transactions on Mobile Computing}}  \bibinfo{volume}{22} (\bibinfo{year}{2021}), \bibinfo{pages}{1491--1506}.
\newblock
\urldef\tempurl%
\url{https://api.semanticscholar.org/CorpusID:239645137}
\showURL{%
\tempurl}


\bibitem[Zhu et~al\mbox{.}(2020)]%
        {locprivv}
\bibfield{author}{\bibinfo{person}{Liehuang Zhu}, \bibinfo{person}{Meng Li}, \bibinfo{person}{Zijian Zhang}, {and} \bibinfo{person}{Zhan Qin}.} \bibinfo{year}{2020}\natexlab{}.
\newblock \showarticletitle{ASAP: An Anonymous Smart-Parking and Payment Scheme in Vehicular Networks}.
\newblock \bibinfo{journal}{\emph{IEEE Transactions on Dependable and Secure Computing}} \bibinfo{volume}{17}, \bibinfo{number}{4} (\bibinfo{year}{2020}), \bibinfo{pages}{703--715}.
\newblock


\end{thebibliography}

\end{document}